\documentclass[11pt]{article}
\usepackage{amsmath}
\usepackage{amssymb}
\usepackage{graphicx}
\usepackage{mathabx}

\begin{document}
\begin{center}
$~~~~~$
\vskip 2cm
\noindent
{\Huge\bf SELENA}
\vskip 1cm
{\Large {\bf SE}mi-ana{\bf L}ytical Int{\bf E}grator for a lu{\bf N}ar {\bf A}rtificial satellite}
\vskip 1cm
{\Large\bf Final Report}
\vskip 1cm
{\Large\bf for the CNES R\&T R-S20/BS-0005-062 Research Activity}
\vskip 1cm
{\Large\bf ``Semi-analytical theory for the motion of lunar artificial satellites''}
\begin{figure*}[h]
\centering
\includegraphics[scale=0.3]{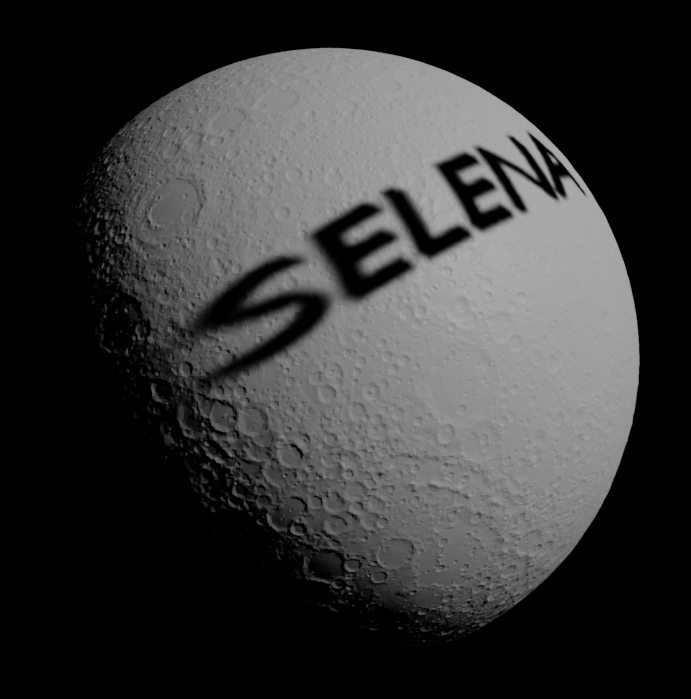}
\end{figure*}

\vskip 3cm
{\Large September 2022}
\end{center}
\vskip 1cm
\clearpage

\tableofcontents

\newpage

\section{Summary}
\label{sec:summary}
~\\
\\
\noindent
The present report summarizes the main theory and implementation steps associated with {\bf SELENA} ({\bf SE}mi-ana{\bf L}ytical Int{\bf E}grator for a lu{\bf N}ar {\bf A}rtificial satellite), i.e. the semi-analytical propagator for lunar satellite orbits developed in the framework of the the R\&T R-S20/BS-0005-062 CNES research activity in collaboration between the University of Padova (UniPd), and the  Aristotle University of Thessaloniki (AUTH), both acting as contractors with CNES.\\
\\
\noindent
A detailed account of the method, algorithms and symbolic manipulations employed in the derivation of the final theory are described in detail in sections 2-4 below: they invoke the use of canonical perturbation theory in the form of Lie series computed in `closed form', i.e., without expansions in the satellite's orbital eccentricity. These algorithms are provided in the form of a symbolic package accompanying the present report. The package contains symbolic algebra programs, as well as explicit data files containing the final Hamiltonian, equations of motion and transformations (i.e. the coefficients and exponents of each variable in each term) leading to the averaging of the short-periodic terms in the satellite's equations of motion. \\
\\
\noindent
This theory materializes in the SELENA semi-analytical propagator, which is the basic SELENA deliverable as defined by CNES in terms of practical use. The description of the software and the Graphic User Interface that accompanies the basic code is given in other deliverables accompanying the present report. The numerical validation and precision tests of the developed software materializing SELENA are described in section 5. These tests were carried out on a set of benchmarking orbits, agreed upon with CNES, as described in the same section. As a nominal error threshold, we require that most trajectories obtained by the semi-analytical propagation should exhibit an error growth rate with respect to integration under the full equations of motion smaller than $\Delta(error)/\Delta t=10$~km/year. This is roughly translated to a relative error of $10^{-6}$ per orbital revolution of the satellite. As discussed in section 5, the most severe limitation for achieving this goal for all trajectories stems from the proliferation in the number of the terms required by perturbation theory to precisely reduce the short-periodic effects due to the higher harmonics of the lunar potential. Due to the uneven distribution of the lunar mascons, for trajectories with pericentric passages lower than ~100 km, passing from a $9\times 9$ (order and degree) to the $10\times 10$ representation of the lunar potential can still introduce considerable corrections to the trajectory. On the other hand, the number of terms in perturbation theory required to cover the whole $10\times 10$ representation of the potential raises to about $10^6$. From that point on, while the symbolic packages accompanying SELENA allow in principle to produce any desired higher order corrections both in the lunar potential and in the Earth's tide, the number of terms required for computing short-periodic corrections increases to non-practical levels. \\
\\
\noindent
On the other hand, our study revealed a key property of the mean theory which renders fully open the possibility of use of SELENA beyond the present $10\times 10$ limit. Referring the reader to subsection \ref{ssec:selenamean} for details, this property can be summarized as follows:  due to the high number of terms required to compute short-periodic corrections at all timesteps along a trajectory, the benefits from the use of the mean theory appear at first sight to nearly disappear with expansions at order and degree beyond 10, both at numerical level (the time required to compute the short-periodic corrections semi-analytically becomes comparable to the time for integration of the trajectory via a fully-cartesian method), and at the theoretical level (one looses insight as regards what the secular Hamiltonian means from the point of view of the phase-space structure of the solutions). However, as will be shown in subsection \ref{ssec:selenamean}, the residuals between a full Cartesian propagation and the mean equations of motion are of the same order as of the fully-corrected equations, provided that one uses the Lie transform to properly modify \textit{just the initial condition} to the corresponding one in terms of mean elements. Hence, a \textit{single-point transformation at the beginning of the integration} is sufficient to ensure an accurate semi-analytical integration, at much higher speed. This result holds for all cases studied. The applied correction is thus a cardinal and efficient step needed to fulfill the purpose of the SELENA propagator, as a tool for fast and accurate lunar mission design. In fact, such a correction proves to be sufficient for obtaining a fast all-purpose integrator using SELENA. As a rule of thumb, with the present implementation of SELENA one should get a faster (by a factor $\sim 10 - 100$) propagator within the above stated nominal precision limits for any trajectory whose pericenter is above 100 km from the Moon's surface while the apocenter reaches a lunicentric distance as high as 10000 km.  \\
\\
\noindent
As a final note, the core of the SELENA propagator relies on a C-programmed implementation of the equations of motion and of the transformations mapping mean to osculating elements via the above averaging theory in closed form. The averaged theory was computed by a suite of symbolic programs writen in \copyright{Mathematica}. A Python-programmed Graphical User Interface allows to run SELENA in a fully graphical environment. A short description of all deliverables associated with the packages, software development and related documentation is provided in the appendix (subsection \ref{ssec:seldeliv}).\\
\\
\\
\noindent
{\bf SELENA team:} This report includes the complete collection of results by the joint work of the SELENA research team:  \\
\\
CNES:~~~~~Carlos Yanez, Space Surveillance Specialist\\
\\
UniPd:~~~~~Christos Efthymiopoulos, Associate Professor, Department of Mathematics\\
\\
AUTH:~~~~~Kleomenis Tsiganis, Professor, Department of Physics\\

~~~~~~~~~~Ioannis Gkolias, Assistant Professor, Department of Physics\\

~~~~~~~~~~M. Gaitanas, Ph.D researcher, Department of Physics\\

\clearpage
\section{Equations of Motion}
\label{sec:eqmo}

\subsection{Frame of Reference}
\label{ssec:frame}

For the purposes of the present study, similar to previous works \cite{deSae2006b,tzietal2009}, 
we adopt as a reference frame the Principal Axis Lunar Reference Frame (PALRF). 
\begin{figure}[h]
\centering
\includegraphics[scale=2.4]{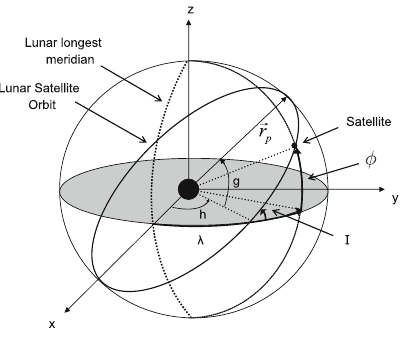}
\includegraphics[scale=0.7]{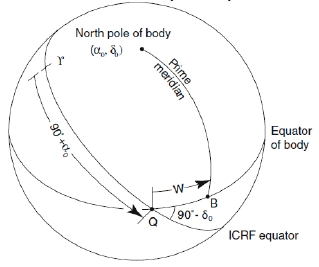}
\caption{\small Selenocentric coordinates (see \cite{tzietal2009}).} 
\label{fig:sele}
\end{figure}

\begin{figure}[!ht]
\centering
\includegraphics[width=\textwidth]{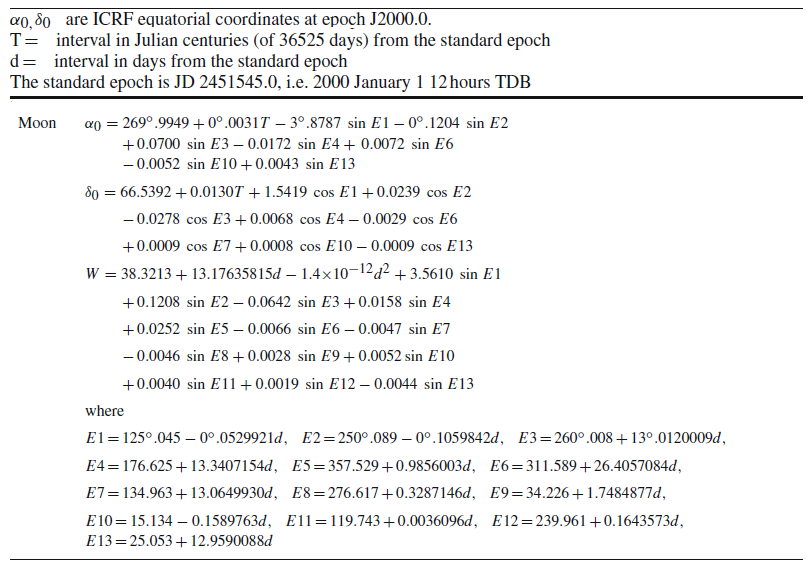}\\
\caption{\small The orientation of Moon's pole and prime meridian after IAU/IAG 2009 \cite{IAUIAG2009}.}
\label{rotation}
\end{figure}

A graphical representation of the PALRF frame is shown in Fig.~\ref{fig:sele}. The $x$ axis is taken along the longest Moon's meridian, $z$ points towards the Moon's pole. This is body-fixed frame which, therefore, changes orientation in time with respect to an inertial frame. Adopting the latter to be the International Celestial Reference Frame (ICRF, \cite{icrf}), the orientation of the PALRF with respect to the ICRF is determined by the three angles $\delta_0, a_0$ and $W$ shown in the right panel of Fig.(\ref{fig:sele}). These, in turn, are related to the set of the 3-1-3 Euler angles $\psi,\phi,\theta$ via the equations:
\begin{eqnarray}\label{euler}
\phi &= &\alpha_0 + 90^\circ \nonumber \\
\theta &= &90^\circ - \delta_0 \\
\psi &= &W~~. \nonumber
\end{eqnarray}
We adopt the recommended values for pole and prime meridian location with respect to time as specified by the IAU working group on Coordinate and Rotational Elements \cite{IAUIAG2009}. They are given by the following formulas presented in Fig.~\ref{rotation}.
\\

Using the formulas in Fig.~\ref{rotation}, the three components of the Moon's time-varying angular velocity vector $\boldsymbol{\omega}(t)$ in the Principal Axes frame (PALRF) can be computed by the following formulas:
\begin{eqnarray}\label{omet}
\omega_x(t) &= &\dot{\phi} \sin \theta \sin \psi + \dot{\theta} \cos \psi \nonumber \\
\omega_y(t) &= &\dot{\phi}  \sin \theta \cos \psi - \dot{\theta} \sin \psi \\
\omega_z(t) &= &\dot{\phi} \cos \theta + \dot{\psi} \nonumber
\end{eqnarray}

\begin{figure}
\centering
\includegraphics[width=0.31\textwidth]{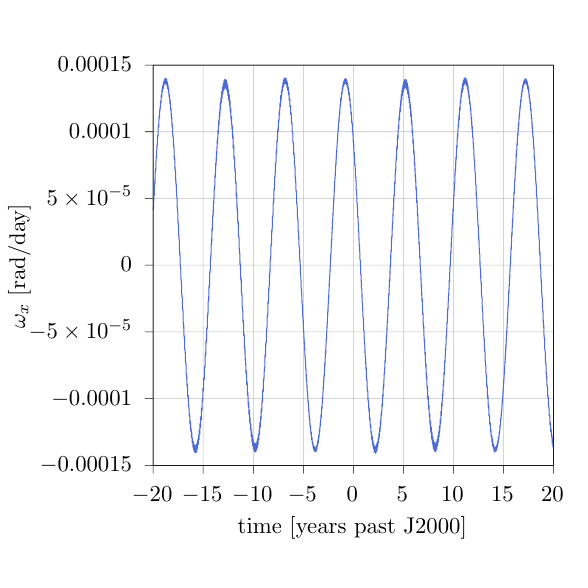}
\includegraphics[width=0.31\textwidth]{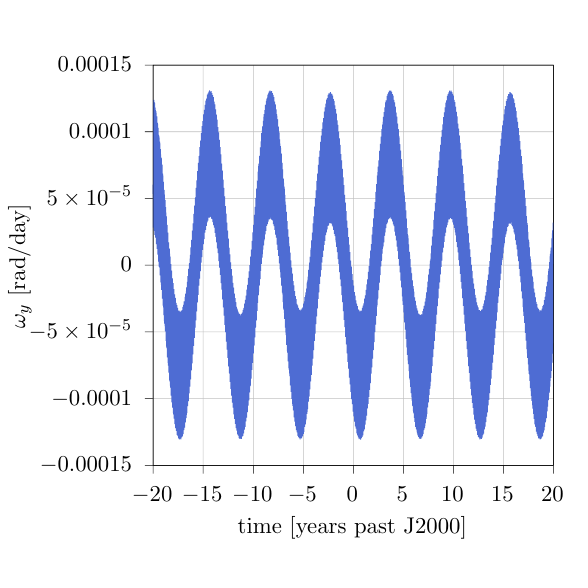}
\includegraphics[width=0.31\textwidth]{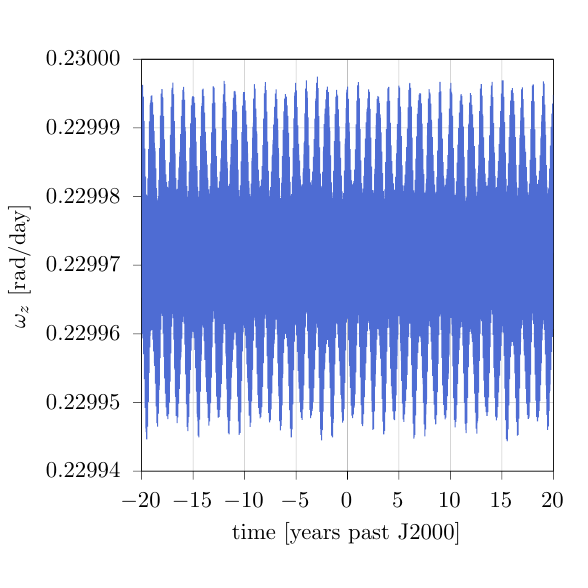}
\caption{\small Evolution of the three PALRF components of the Moon's angular velocity vector as computed by Eq.(\ref{omet})}.
\label{ometnum}
\end{figure}

Implementing the formulas (\ref{omet}), the three functions $\omega_x(t)$, $\omega_y(t)$, $\omega_z(t)$ can be computed as a function of time. The evolution of the Moon's angular velocities is given in Figure \ref{ometnum}. The time $t=0$ corresponds to the JD2000 at 12.00 Noon (UTC). 

\subsection{Hamiltonian}
\label{ssec:ham}

The Hamiltonian governing the motion of a lunar satellite in the PALRF frame has the form: 
\begin{equation}\label{ham}
{\cal H} = {1\over 2}\mathbf{p}^2 
-\boldsymbol{\omega}(t)\cdot(\mathbf{r}\times\mathbf{p}) + V(\mathbf{r},t)
\end{equation}
where: $\mathbf{r}(t)$ is the lunicentric PALRF radius vector of the satellite, and $\mathbf{p}(t)$ is the velocity vector of the satellite in a fictitious \textit{rest} frame whose axes coincide with the axes of the PALRF at the time $t$. The angular momentum vector $\boldsymbol\omega(t)$ is given by Eq.(\ref{omet}) (or (\ref{omefreq})). Solving Hamilton's equations, the satellite PALRF velocity vector at the time $t$ is given by
\begin{equation}\label{vel}
\dot{\mathbf{r}}(t) = \mathbf{p}(t)-\boldsymbol{\omega}(t)\times\mathbf{r}(t) 
\end{equation}
while the equation of motion is
\begin{equation}\label{eqmo}
\ddot{\mathbf{r}} = -\nabla V(\mathbf{r},t)-\dot{\boldsymbol{\omega}}\times\mathbf{r} 
-2\boldsymbol\omega\times\dot{\mathbf{r}} 
- \boldsymbol{\omega}\times (\boldsymbol{\omega}\times\mathbf{r})
\end{equation}

The potential $V(\mathbf{r},t)$ is given by:
\begin{equation}\label{pot}
V(\mathbf{r}(t) = V_\Moon(\mathbf{r}) + V_\Earth(\mathbf{r},t)+ V_\Sun(\mathbf{r},t)
+V_{SRP}(\mathbf{r},t)
\end{equation}
where the adopted form of the terms $V_\Moon$, $V_\Earth$, $V_\Sun$, $V_{SRP}$ in SELENA is as described in the following subsections:

\subsubsection{Lunar potential}
\label{sssec:lunpot}
The lunar potential in multipole harmonic form is
\begin{equation}\label{potmoon}
V_\Moon(\mathbf{r}) = -{\mathcal{G}M_\Moon\over r}\sum_{n=0}^\infty
\left({R_\Moon\over r}\right)^n\sum_{m=0}^n P_{nm}(\sin\phi)
[C_{nm}\cos(m\lambda)+S_{nm}\sin(m\lambda)]
\end{equation}
where $\mathcal{G}$ is Newton's gravity constant, $M_\Moon$ is the Moon's mass and $R_\Moon$ the equatorial Moon radius. The angles $\phi$ and $\lambda$ are the satellite's longitude and latitude (see Figure 1), $r$ the selenocentric satellite's distance. We have $\sin\phi = z/r$, $\cos\phi = \rho/r$ (with $\rho = (x^2+y^2)^{1/2}$), $\cos\lambda = x/\rho$, $\sin\lambda = y/\rho$. The functions $P_{nm}$ are normalized Legendre polynomials of degree $n$ and order $m$, and $C_{nm}$, $S_{nm}$ are the zonal ($m=0$) and tesseral ($m\neq 0$) coefficients of the lunar gravity potential. Accurate values of these parameters are provided by the GRAIL data \cite{grail}. 
In particular, as the origin of the PALRF coordinates is at the Moon's baricenter, we have $C_{10}=C_{11}=S_{11}=0$. The complete table of constants and $10\times 10$ set of GRAIL coefficients is given in the appendix (subsection \ref{ssec:grailcoef}).

\subsubsection{Earth's tide}
\label{sssec:earthpot}
The Earth's tidal potential term $V_\Earth(\mathbf{r},t)$ is given by
\begin{equation}\label{potearth}
V_\Earth(\mathbf{r},t) = -\mathcal{G}M_\Earth
\left({1\over\sqrt{r^2+r_\Earth^2 - 2\mathbf{r}\cdot\mathbf{r}_\Earth}}-
{\mathbf{r}\cdot\mathbf{r}_\Earth\over r_\Earth^3}\right)~~.
\end{equation}
where $\mathbf{r}_\Earth(t)$ is the lunicentric PALRF radius vector of the Earth. Multipole expansion yields:
\begin{equation}\label{potearthmpole}
V_\Earth(\mathbf{r},t) = 
 V_{\Earth,P0}\left(\mathbf{r}_\Earth(t)\right)
+V_{\Earth,P2}\left(\mathbf{r},\mathbf{r}_\Earth(t)\right)
+V_{\Earth,P3}\left(\mathbf{r},\mathbf{r}_\Earth(t)\right)
+V_{\Earth,P4}\left(\mathbf{r},\mathbf{r}_\Earth(t)\right)
\end{equation}
where

$$
V_{\Earth,P0}\left(\mathbf{r}_\Earth(t)\right)=
-{\mathcal{G}M_\Earth \over r_\Earth(t)}
$$
$$
V_{\Earth,P2}\left(\mathbf{r},\mathbf{r}_\Earth(t)\right)=
{\mathcal{G}M_\Earth \over r_\Earth(t)}
\left(
{1\over 2}{r^2\over r_\Earth^2(t)}
-{3\over 2}{(\mathbf{r}\cdot\mathbf{r}_\Earth(t))^2\over r_\Earth^4(t)}
\right)
$$
$$
V_{\Earth,P3}\left(\mathbf{r},\mathbf{r}_\Earth(t)\right)=
{\mathcal{G}M_\Earth \over r_\Earth(t)}
\left(
{3\over 2}{r^2(\mathbf{r}\cdot\mathbf{r}_\Earth(t))\over r_\Earth^4(t)}
-{5\over 2}{(\mathbf{r}\cdot\mathbf{r}_\Earth(t))^3\over r_\Earth^6(t)}
\right)
$$
$$
V_{\Earth,P4}\left(\mathbf{r},\mathbf{r}_\Earth(t)\right)=
{\mathcal{G}M_\Earth \over r_\Earth(t)}
\left(
-{3\over 8}{r^4\over r_\Earth^4(t)}
+{15\over 4}{r^2(\mathbf{r}\cdot\mathbf{r}_\Earth(t))^2\over r_\Earth^6(t)}
-{35\over 8}{(\mathbf{r}\cdot\mathbf{r}_\Earth(t))^4\over r_\Earth^8(t)}\right)
$$
The term $V_{\Earth,P0}$ does not depend on the satellite's radius vector $\mathbf{r}$, thus it can be omitted from the Hamiltonian and from the resulting equations of motion. The terms $V_{\Earth,P2}$, $V_{\Earth,P3}$ and $V_{\Earth,P4}$ are hereafter referred to as the quadrupolar, octopolar and hexapolar Earth's tidal terms respectively. 

The lunicentric PALRF radius vector of the Earth $\mathbf{r}_\Earth(t)$ can be computed, as a function of the epoch $t$, using (semi-)analytical theories (for example \cite{meeus,montenbruck}), or from state of the art ephemerides services, like the DE421 ephemeris provided by NASA's JPL or the INPOP19a ephemeris provided by IMCCE. The resulting products are high-accuracy estimations of Earth's vector to the level of one meter. 

\begin{figure}
\centering
\includegraphics[width=0.95\textwidth]{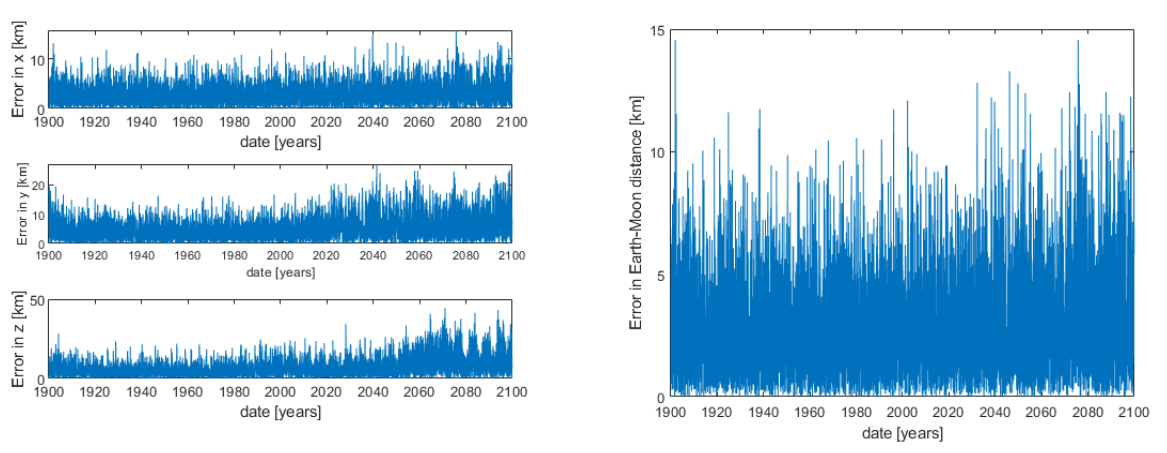}
\caption{\small Left: the difference in the three PALRF coordinates $(x_\Earth(t),y_\Earth(t),z_\Earth(t))$ as computed by the NAFF formula with the tables \ref{xenaff}, \ref{yenaff}, \ref{zenaff}, and by the INPOP19a determination of the Earth's ephemeris in the PALRF frame. Right: the corresponding error in the computation of the Earth's lunicentric distance. }
\label{fig:eartherr}
\end{figure}

In order to avoid the cumbersome and less accurate formulas of semi-analytic theory, SELENA adopts a representation of the vector $\mathbf{r}_\Earth(t)$ obtained by the NAFF algorithm (Numerical Analysis of the Fundamental Frequencies; introduced by J. Laskar in \cite{nafflaskar}). The high-accuracy ephemerides (INPOP19a) is numerically analyzed as a multiply-periodic time series involving the fundamental frequencies of the Earth-Moon orbit. Different levels of truncation in the frequency domain provide different levels of accuracy: using an expansion with the 50 leading frequencies an accuracy of the level of 10km is achieved with respect to INPOP19a. This exceeds by far the precision required for the inclusion of the Earth tide effect in the semi-analytical propagation of lunar satellite orbits. The NAFF determined lunicentric PALRF Earth radius vector $\mathbf{r}_\Earth(t)$ is then given by:
\begin{equation}\label{rbearth}
\mathbf{r}_\Earth(t) = (x_\Earth(t),y_\Earth(t),z_\Earth(t))= 
\sum_i (\mathbf{A}_i \cos\omega_i t + \mathbf{B}_i \sin\omega_i t)
\end{equation}
where $t$ is the ephemeris time calculated in seconds past J2000. The coefficients $\mathbf{A}_i$ and $\mathbf{B}_i$ and the frequencies $\omega_i$ are given in components according to the Tables 
(\ref{xenaff}), (\ref{yenaff}) and (\ref{zenaff}). Figure \ref{fig:eartherr} shows the error in the individual components $(x_\Earth(t),y_\Earth(t),z_\Earth(t))$ as computed by the above tables in comparison with the INPOP19a determination of the Earth's ephemeris in the PALRF frame. 

\begin{table}[h]
\centering
\begin{tabular}{|c|c|c|c|c|c|c|c|}
\hline
$i$ & $\omega_i$ $[rad/s]$ & $A_i^{(x)}$ $[km]$  & $B_i^{(x)}$ $[km]$ & $i$ & $\omega_i$ $[rad/s]$ & $A_i^{(x)}$ $[km]$  & $B_i^{(x)}$ $[km]$\\
\hline\hline
1   &    0	            &   382469.63   &    0          &   26  &    2.99241e-6	    &   17.11       &    3.27       \\                     
2   &    3.32012e-8	    &   0.99        &    -1.        &   27  &    3.05881e-6	    &   -1.09       &    14.94      \\    
3   &    6.64022e-8	    &   0.35        &    3.14       &   28  &    4.3729e-6	    &   -2.19       &    0.97       \\    
4   &    1.76602e-7	    &   -6.71       &    2.05       &   29  &    4.52701e-6	    &   4.05        &    -7.22      \\    
5   &    1.99097e-7	    &   39.58       &    1.25       &   30  &    4.572e-6	    &   -24.13      &    9.49       \\    
6   &    3.53204e-7	    &   -189.3      &    128.62     &   31  &    4.7261e-6	    &   90.03       &    -145.28    \\        
7   &    3.98194e-7	    &   2.13        &    0.18       &   32  &    4.9252e-6	    &   1369.26     &    -2010.23   \\        
8   &    4.19607e-7	    &   -35.43      &    -41.03     &   33  &    5.07931e-6	    &   0.35        &    8.3        \\    
9   &    5.52301e-7	    &   -7.12       &    4.4        &   34  &    5.1018e-6	    &   -1.84       &    5.93       \\    
10  &    6.18704e-7	    &   -1.38       &    -1.75      &   35  &    5.1243e-6	    &   -10.26      &    13.76      \\    
11  &    1.8878e-6	    &   1.29        &    4.59       &   36  &    5.27841e-6	    &   -0.75       &    584.51     \\    
12  &    1.93279e-6	    &   0.81        &    -1.86      &   37  &    5.34481e-6	    &   -1285.72    &    148.03     \\        
13  &    2.0869e-6	    &   34.53       &    146.51     &   38  &    5.4775e-6	    &   0.27        &    -6.03      \\    
14  &    2.2196e-6	    &   -11.79      &    0.86       &   39  &    7.0121e-6	    &   1.54        &    0.59       \\    
15  &    2.286e-6	    &   730.56      &    3836.11    &   40  &    7.2112e-6	    &   12.92       &    5.64       \\        
16  &    2.44011e-6	    &   41.09       &    37.63      &   41  &    7.36531e-6	    &   5.61        &    -1.31      \\    
17  &    2.4626e-6	    &   50.8        &    95.97      &   42  &    7.3878e-6	    &   3.07        &    0.35       \\    
18  &    2.4851e-6	    &   -1.39       &    -9.46      &   43  &    7.5644e-6	    &   71.25       &    -13.41     \\    
19  &    2.57282e-6	    &   1.32        &    -1.06      &   44  &    7.63081e-6	    &   2.26        &    7.26       \\    
20  &    2.6392e-6	    &   14796.88    &    14794.17   &   45  &    7.91761e-6	    &   16.81       &    -16.75     \\            
21  &    2.6617e-6	    &   -6.98       &    -14.67     &   46  &    7.98401e-6	    &   28.08       &    22.31      \\    
22  &    2.6724e-6	    &   16.34       &    -0.81      &   47  &    9.8504e-6	    &   -1.79       &    -4.57      \\    
23  &    2.68417e-6	    &   -0.22       &    -2.32      &   48  &    0.0000102036    &   -5.36       &    -3.67      \\    
24  &    2.70561e-6	    &   -24.25      &    30.57      &   49  &    0.00001027	    &   1.19        &    -2.26      \\    
25  &    2.8383e-6	    &   -17.17      &    -18.66     &   50  &    0.0000106232    &   -0.17       &    -1.46      \\    
\hline
\end{tabular}
\caption{The Fourier decomposition of the x-coordinate of Earth's position in the Moon's principal axis frame.}
\label{xenaff}
\end{table}
\clearpage
\begin{table}[h]
\centering
\begin{tabular}{|c|c|c|c|c|c|c|c|}
\hline
$i$ & $\omega_i$ $[rad/s]$ & $A_i^{(y)}$ $[km]$  & $B_i^{(y)}$ $[km]$ & $i$ & $\omega_i$ $[rad/s]$ & $A_i^{(y)}$ $[km]$  & $B_i^{(y)}$ $[km]$\\
\hline\hline
 1      &  0	               &    -124.77  &  0           &   26     &  2.7056e-6	      &    -34.63   &  -27.48   \\
 2      &  3.3201e-8	       &    10.81    &  6.42        &    27     &  2.8383e-6	  &    -119.38  &  109.38   \\
 3      &  6.63896e-8	       &    -34.31   &  3.92        &    28     &  2.99241e-6	  &    -2.58    &  13.47    \\
 4      &  6.86868e-8	       &    3.37     &  2.53        &    29     &  3.05881e-6	  &    2.99     &  0.21     \\
 5      &  1.24541e-7	       &    -2.17    &  -0.33       &    30     &  4.52701e-6	  &    -11.99   &  -6.74    \\
 6      &  1.54108e-7	       &    -4.07    &  -5.47       &    31     &  4.572e-6	      &    -4.89    &  -12.44   \\
 7      &  1.76603e-7	       &    -11.93   &  -38.76      &    32     &  4.7261e-6	  &    -237.3   &  -147.17  \\
 8      &  1.99097e-7	       &    60.59    &  -1403.61    &    33     &  4.7486e-6	  &    -2.06    &  -2.58    \\
 9      &  2.20511e-7	       &    -1.84    &  1.73        &    34     &  4.9252e-6	  &    -3262.64 &  -2223.9  \\
 10     &  3.53204e-7	       &    -219.21  &  -322.38     &    35     &  5.07931e-6	  &    -4.61    &  0.19     \\
 11     &  3.98194e-7	       &    1.23     &  -14.25      &    36     &  5.1018e-6	  &    -3.39    &  -1.05    \\
 12     &  4.19607e-7	       &    66.83    &  -57.74      &    37     &  5.1243e-6	  &    28.26    &  21.11    \\
 13     &  5.52301e-7	       &    -7.61    &  -12.3       &    38     &  5.27841e-6	  &    -283.15  &  -0.51    \\
 14     &  6.18704e-7	       &    2.47     &  -1.95       &    39     &  5.34481e-6	  &    148.26   &  1287.76  \\
 15     &  1.8878e-6	       &    12.98    &  -3.65       &    40     &  5.4775e-6	  &    3.87     &  0.17     \\
 16     &  2.0869e-6	       &    367.49   &  -86.34      &    41     &  7.2112e-6	  &    -6.76    &  15.5     \\
 17     &  2.2196e-6	       &    1.23     &  17.01       &    42     &  7.36531e-6	  &    1.33     &  5.72     \\
 18     &  2.24101e-6	       &    2.78     &  -3.29       &    43     &  7.3878e-6	  &    0.29     &  -2.58    \\
 19     &  2.286e-6	           &    8406.78  &  -1596.08    &    44     &  7.5644e-6	  &    14.06    &  74.71    \\
 20     &  2.44011e-6	       &    161.4    &  -175.79     &    45     &  7.63081e-6	  &    7.79     &  -2.43    \\
 21     &  2.4626e-6	       &    204.84   &  -108.29     &    46     &  7.91761e-6	  &    12.88    &  12.93    \\
 22     &  2.4851e-6	       &    -44.92   &  6.53        &    47     &  7.98401e-6	  &    22.36    &  -28.14   \\
 23     &  2.6392e-6	       &    29773.72 &  -29749.79   &    48     &  9.8504e-6	  &    4.       &  -1.57    \\
 24     &  2.6617e-6	       &    -29.94   &  14.23       &    49     &  0.0000102036   &    3.17     &  -4.63    \\
 25     &  2.68419e-6	       &    -4.62    &  0.47        &    50     &  0.00001027	  &    -2.31    &  -1.21    \\
\hline
\end{tabular}
\caption{The Fourier decomposition of the y-coordinate of Earth's position in the Moon's principal axis frame.}
\label{yenaff}
\end{table}
\clearpage
\begin{table}[h]
\centering
\begin{tabular}{|c|c|c|c|c|c|c|c|}
\hline
$i$ & $\omega_i$ $[rad/s]$ & $A_i^{(z)}$ $[km]$  & $B_i^{(z)}$ $[km]$ & $i$ & $\omega_i$ $[rad/s]$ & $A_i^{(z)}$ $[km]$  & $B_i^{(z)}$ $[km]$\\
\hline\hline
1   &    0	          	&   145.68      &   0          & 26  &    3.09201e-6	 	&   4.63        &   4.78    \\
2   &    2.66761e-9	 	&   1.92        &   -5.83      & 27  &    3.22471e-6	 	&   0.87        &   -0.61   \\
3   &    1.06614e-8	 	&   1.31        &   -3.1       & 28  &    4.5388e-6	    	&   6.07        &   3.07    \\
4   &    2.2419e-8	  	&   0.06        &   1.12       & 29  &    4.6929e-6	  	    &   1.48        &   -0.26   \\
5   &    3.32011e-8	 	&   -2579.36    &   2895.76    & 30  &    4.75931e-6	 	&   3.66        &   12.35   \\
6   &    1.65896e-7	 	&   10.16       &   12.45      & 31  &    4.892e-6	  	    &   28.87       &   -3.75   \\
7   &    1.8731e-7	  	&   0.26        &   -2.95      & 32  &    4.9584e-6	    	&   65.69       &   263.14  \\
8   &    2.09803e-7	 	&   -7.71       &   16.85      & 33  &    5.11251e-6	 	&   7.01        &   5.74    \\
9   &    2.32298e-7	 	&   -14.02      &   14.43      & 34  &    5.135e-6	  	    &   3.53        &   5.84    \\
10  &    3.86406e-7	 	&   -81.19      &   619.48     & 35  &    5.1575e-6	    	&   -0.36       &   -1.74   \\
11  &    4.52808e-7	 	&   1.26        &   0.02       & 36  &    5.24521e-6	 	&   0.81        &   -0.72   \\
12  &    5.85502e-7	 	&   -4.67       &   26.64      & 37  &    5.31161e-6	 	&   920.11      &   821.45  \\
13  &    1.8546e-6	  	&   -1.47       &   2.29       & 38  &    5.3341e-6	  	    &   -0.47       &   -0.86   \\
14  &    2.0537e-6	  	&   -39.8       &   56.65      & 39  &    5.37801e-6	 	&   2.73        &   -3.87   \\
15  &    2.2528e-6	  	&   -825.59     &   1073.41    & 40  &    5.5107e-6	    	&   -4.67       &   -4.55   \\
16  &    2.286e-6	  	&   0.29        &   1.46       & 41  &    5.66481e-6	 	&   -1.51       &   -0.2    \\
17  &    2.3192e-6	  	&   -7.07       &   -4.24      & 42  &    7.178e-6	    	&   -0.66       &   -2.     \\
18  &    2.45189e-6	 	&   15.51       &   -18.45     & 43  &    7.2444e-6	    	&   2.07        &   -0.96   \\
19  &    2.47331e-6	 	&   -28.71      &   2.89       & 44  &    7.39851e-6	 	&   2.96        &   -5.45   \\
20  &    2.606e-6	  	&   -0.96       &   16.29      & 45  &    7.5312e-6	  	    &   -1.         &   -0.77   \\
21  &    2.6392e-6	  	&   5.52        &   3.88       & 46  &    7.59761e-6	   	&   45.54       &   -75.74  \\
22  &    2.66171e-6	 	&   -7.83       &   8.29       & 47  &    7.95081e-6	 	&   -2.83       &   -50.53  \\
23  &    2.6724e-6	  	&   -44649.42   &   2554.55    & 48  &    9.8836e-6	  	    &   -1.69       &   -0.63   \\
24  &    2.8715e-6	  	&   25.79       &   -0.36      & 49  &    0.0000102368	    &   -8.41       &   2.09    \\
25  &    3.02561e-6	 	&   20.84       &   -15.99     & 50  &    0.00001059	 	&   -1.63       &   1.83    \\
\hline
\end{tabular}
\caption{The Fourier decomposition of the z-coordinate of Earth's position in the Moon's principal axis frame.}
\label{zenaff}
\end{table}

\clearpage

\begin{figure}[!h]
\centering
\includegraphics[width=1.0\textwidth]{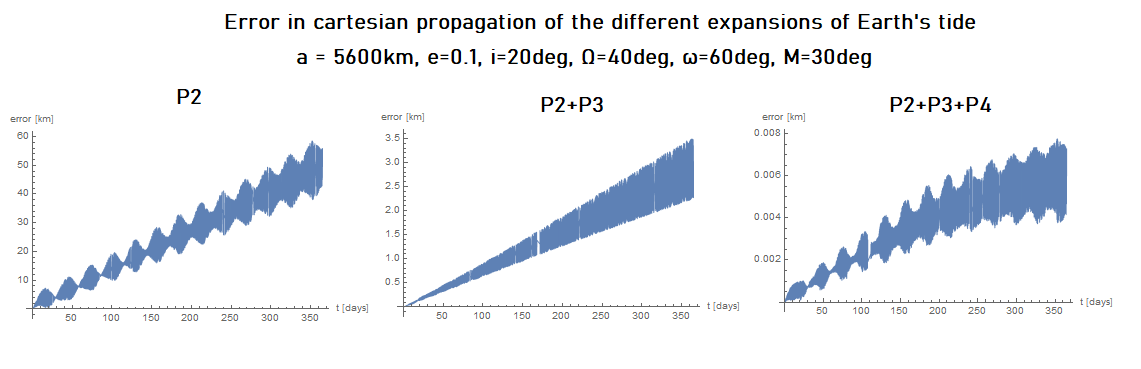}
\includegraphics[width=1.0\textwidth]{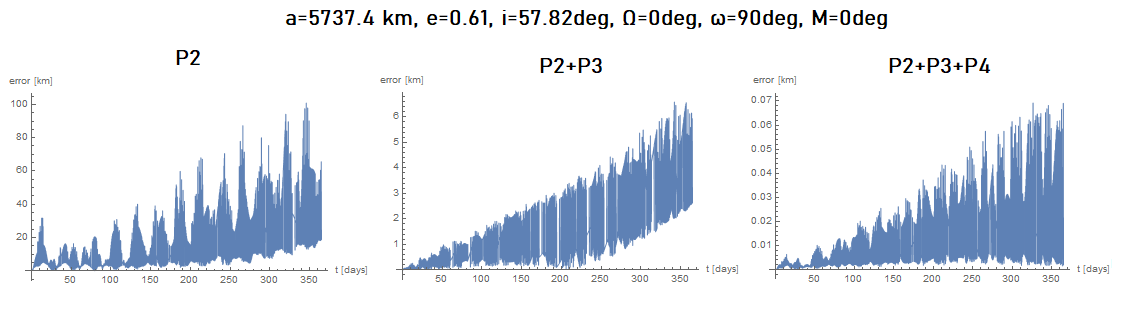}
\caption{\small The growth in time of the difference in km between Cartesian integration of a trajectory under the potential $V=-GM_\Moon/r +V_\Earth$, with $V_\Earth$ given by the exact formula of Eq.(\ref{potearth}), or by the multipole expansion truncated at $P_2$, $P_3$ or $P_4$. Two different cases are considered, with the initial conditions in each trajectory as indicated in the panels. }
\label{fig:p2p3p4}
\end{figure}
The relative importance of the Earth's tidal perturbation increases with altitude, roughly proportionally to $r^2/r_\Earth^2$. Figure \ref{fig:p2p3p4} shows a comparision between a Cartesian integration under the potential $V=-GM_\Moon/r +V_\Earth$, with $V_\Earth$ given by the exact formula of Eq.(\ref{potearth}), or by the multipole expansion truncated at $P_2$, $P_3$ or $P_4$, for two trajectories with semimajor axis $a\approx 5600$km, and eccentricities $e=0.1$ (top) and $e=0.6$ (bottom). The initial conditions of the second trajectory are indicative of the anticipated trajectory of the Lunar Pathfinder orbiter. We note that even for apocentric distances as high as $~10000 km$ a truncation of the Earth's tidal potential at $P_3$ generates trajectories well below the nominal SELENA 10~km/yr error, while the inclusion of the P4 term reduces the error to meters. In order to avoid the computational cost of short-periodic corrections associated with the averaging of the $P_4$ terms, all semi-analytical computations are truncated to the $P_3$ term (instead, $P_2^2$ square corrections produced by the averaging procedure turn to be important, see section 4 below). 

\subsubsection{Solar tide}
\label{sssec:sunpot}
The term $V_\Sun(\mathbf{r},t)$, due to the Sun's tidal force on the satellite, is taken as
\begin{equation}\label{potsun}
V_\Sun(\mathbf{r},t) = -\mathcal{G}M_\Sun
\left({1\over\sqrt{r^2+r_\Sun^2(t) - 2\mathbf{r}\cdot\mathbf{r}_\Sun(t)}}-
{\mathbf{r}\cdot\mathbf{r}_\Sun(t)\over r_\Sun^3(t)}\right)
\end{equation}
where $\mathbf{r}_\Sun(t)$ is the lunicentric PALRF radius vector of the Sun. Analogously to the 
Earth's tidal potential, we can omit the $P_0$ term in the multipole expansion of the Sun's tidal 
potential. Here, however, we already reach sufficient precision by including in the Hamiltonian only the quadrupole term:
\begin{equation}\label{potsunP2}
V_\Sun(\mathbf{r},t) = V_{\Sun,P2}(\mathbf{r},\mathbf{r}_\Sun(t))+\ldots=
{\mathcal{G}M_\Sun \over r_\Sun(t)}
\left(
{1\over 2}{r^2\over r_\Sun^2(t)}
-{3\over 2}{(\mathbf{r}\cdot\mathbf{r}_\Sun(t))^2\over r_\Sun^4(t)}
\right)+\ldots
\end{equation}
The Sun's lunicentric PALRF radius vector $\mathbf{r}_\Sun(t)$ can be computed as a function of the epoch $t$ by NAFF-decomposing the ephemeris data analogously as for the Earth's vector $\mathbf{r}_\Earth(t)$. This leads to
\begin{equation}\label{rbsun}
\mathbf{r}_\Sun(t) = (x_\Sun(t),y_\Sun(t),z_\Sun(t))= 
\sum_i (\mathbf{C}_i \cos\Omega_i t + \mathbf{D}_i \sin\Omega_i t)
\end{equation}
where $t$ is the ephemeris time calculated in seconds past J2000. The coefficients $\mathbf{C}_i$ and $\mathbf{D}_i$ and the frequencies $\Omega_i$ are given in components according to the Tables 
(\ref{xsnaff}), (\ref{ysnaff}) and (\ref{zsnaff}). Note that only 22 terms are included before reaching machine precision in the NAFF decomposition of the $x$ and $y$ components of the Sun's radius vector, while 47 terms are required for the $z-$component. 
\begin{table}[h]
\centering
\begin{tabular}{|c|c|c|c|c|c|c|c|}
\hline
$i$ & $\omega_i$ $[rad/s]$ & $C_i^{(x)}$ $[km]$  & $D_i^{(x)}$ $[km]$\\
\hline\hline
1  	&	 0			 	&	377859.4        &      0           		\\
2  	&	 1.76602e-7		&	7953.07         &      -2447.89         \\
3  	&	 2.06441e-6		&	-8293.43        &      -12860.          \\
4  	&	 2.1094e-6		&	-397.01	        &      3574.68          \\
5  	&	 2.2635e-6		&	-614574.37	    &      -1.049770.27     \\
6  	&	 2.286e-6		&	955.96	        &      5024.01          \\
7  	&	 2.39622e-6		&	-5636.91        &      2218.71          \\
8  	&	 2.4294e-6		&	1291.01	        &      -2109.38         \\
9  	&	 2.45186e-6		&	-1530.52        &      2403.09          \\
10 	&	 2.46117e-6		&	-2377.63        &      1086.91          \\
11 	&	 2.4626e-6		&	-69917136.93    &      -132200853.79  	\\
12 	&	 2.47335e-6		&	-2946.55        &      -22.03           \\
13 	&	 2.4958e-6		&	2434.	        &      100.6            \\
14 	&	 2.52898e-6		&	-4995.4         &      3403.35          \\
15 	&	 2.63921e-6		&	13743.48        &      13736.12         \\
16 	&	 2.64491e-6		&	-3781.51        &      310.12           \\
17 	&	 2.6617e-6		&	1593848.82      &	   3356568.64		\\
18 	&	 2.71168e-6		&	604.68          &      3143.74          \\
19 	&	 2.81581e-6		&	-3320.38        &      -1759.89         \\
20 	&	 2.86079e-6		&	-1747.4         &      -4127.73         \\
21 	&	 2.88221e-6		&	10081.43        &      -25881.65        \\
22 	&	 5.1018e-6		&	1645.19	        &      -5331.99         \\
\hline
\end{tabular}
\caption{The Fourier decomposition of the x-coordinate of Sun's position in the Moon's principal axis frame.}
\label{xsnaff}
\end{table}
\clearpage
\begin{table}[h]
\centering
\begin{tabular}{|c|c|c|c|c|c|c|c|}
\hline
$i$ & $\omega_i$ $[rad/s]$ & $C_i^{(y)}$ $[km]$  & $D_i^{(y)}$ $[km]$ \\
\hline\hline
1      &    1.76602e-7   &       1410.13            &   4580.96         \\
2      &    2.06441e-6   &       -12859.39          &   8293.04         \\
3      &    2.1094e-6	 &       3591.79            &   398.93          \\
4      &    2.2635e-6	 &       -1049755.65        &   614565.8        \\
5      &    2.286e-6	 &       9540.17            &   -1811.2         \\
6      &    2.39622e-6   &       2212.04            &   5619.77         \\
7      &    2.4294e-6	 &       -2083.44           &   -1281.11        \\
8      &    2.45186e-6   &       2403.07            &   1530.51         \\
9      &    2.46117e-6   &       1087.07            &   2377.74         \\
10     &    2.4626e-6	 &       -132200640.78      &   69917024.34     \\
11     &    2.47335e-6   &       -22.3              &   2946.26         \\
12     &    2.4958e-6	 &       102.38             &   -2440.63        \\
13     &    2.52898e-6   &       3403.53            &   4995.65         \\
14     &    2.6392e-6	 &       28536.31           &   -28517.21       \\
15     &    2.64491e-6   &       310.12             &   3781.45         \\
16     &    2.6617e-6	 &       3356557.04         &   -1593843.25     \\
17     &    2.71168e-6   &       3143.77            &   -604.66         \\
18     &    2.81581e-6   &       -1610.17           &   3037.82         \\
19     &    2.86079e-6   &       -4127.71           &   1747.41         \\
20     &    2.88221e-6   &       -25763.1           &   -10035.28       \\
21     &    4.9252e-6	 &       -3252.65           &   -2217.08        \\
22     &    5.1018e-6	 &       -5376.66           &   -1659.          \\
\hline
\end{tabular}
\caption{The Fourier decomposition of the y-coordinate of Sun's position in the Moon's principal axis frame.}
\label{ysnaff}
\end{table}
\clearpage
\begin{table}[h]
\centering
\begin{tabular}{|c|c|c|c|c|c|c|c|}
\hline
$i$ & $\omega_i$ $[rad/s]$ & $C_i^{(z)}$ $[km]$  & $D_i^{(z)}$ $[km]$ & $i$ & $\omega_i$ $[rad/s]$ & $C_i^{(z)}$ $[km]$  & $D_i^{(z)}$ $[km]$\\
\hline\hline
1   &  0	        &   336.87        &    0                &  26  &  2.4294e-6	   &   -25357.64     &    68078.38   \\       
2   &  1.07065e-8   &   -37953.3      &    93444.33         &  27  &  2.45191e-6   &   100.73        &    -120.46    \\   
3   &  2.74895e-8   &   103.09        &    2.5              &  28  &  2.45993e-6   &   -1229.68      &    1270.48    \\   
4   &  3.32006e-8   &   -2548.69      &    2856.04          &  29  &  2.4626e-6	   &   -26465.47     &    -50489.93  \\       
5   &  3.92743e-8   &   -11.63        &    -85.74           &  30  &  2.46527e-6   &   -4108.53      &    702.27     \\   
6   &  7.4563e-8	&   55.55         &    -34.82           &  31  &  2.47348e-6   &   -134.15       &    21.77      \\   
7   &  8.52636e-8   &   -55.75        &    -16.1            &  32  &  2.4958e-6	   &   17098.87      &    4166.31    \\       
8   &  1.43399e-7   &   -182.05       &    -108.72          &  33  &  2.6285e-6	   &   663.59        &    -1571.29   \\       
9   &  1.65558e-7   &   96.08         &    -67.65           &  34  &  2.6617e-6	   &   595.88        &    1288.54    \\       
10  &  1.74442e-7   &   -110.16       &    -126.53          &  35  &  2.66437e-6   &   102.77        &    -13.05     \\   
11  &  1.76617e-7   &   651.08        &    301.38           &  36  &  2.6724e-6	   &   -44100.91     &    2522.99    \\       
12  &  1.88467e-7   &   32.98         &    98.71            &  37  &  2.6949e-6	   &   -428.24       &    -124.2     \\       
13  &  1.99106e-7   &   -7445.27      &    -4239.07         &  38  &  2.84901e-6   &   -427.3        &    158.36     \\   
14  &  2.09803e-7   &   1675546.42    &    -3663438.74      &  39  &  4.7154e-6	   &   108.26        &    18.55      \\       
15  &  3.86405e-7   &   -79.76        &    606.91           &  40  &  4.93591e-6   &   45.05         &    67.79      \\   
16  &  3.98187e-7   &   -59.79        &    -37.58           &  41  &  4.9584e-6	   &   64.92         &    260.05     \\       
17  &  4.089e-7	    &   15313.19      &    -29987.48        &  42  &  5.135e-6	   &   4176.46       &    6915.09    \\   
18  &  6.07997e-7   &   207.94        &    -367.19          &  43  &  5.31161e-6   &   908.72        &    811.28     \\   
19  &  1.8771e-6	&   -51.14        &    39.7             &  44  &  5.3341e-6	   &   -103.02       &    -188.48    \\       
20  &  2.0537e-6	&   -39.84        &    56.7             &  45  &  7.421e-6	   &   73.82         &    -66.2      \\   
21  &  2.0762e-6	&   -1303.33      &    924.34           &  46  &  7.59761e-6   &   46.33         &    -77.07     \\   
22  &  2.2303e-6	&   -212.31       &    655.16           &  47  &  7.77421e-6   &   112.2         &    -453.64    \\   
23  &  2.2528e-6	&   -861.71       &    1120.4           & & & &\\                   
24  &  2.2635e-6	&   -230.43       &    -402.58          & & & &\\                               
25  &  2.29671e-6   &   121.59        &    24.09            & & & &\\           
\hline
\end{tabular}
\caption{The Fourier decomposition of the z-coordinate of Sun's position in the Moon's principal axis frame.}
\label{zsnaff}
\end{table}
\clearpage

\subsubsection{Solar radiation pressure}
\label{sssec:srp}
For the solar radiation pressure force, SELENA adopts a cannonball model \cite{kaula1962}:
\begin{equation}\label{potsrp}
V_{SRP}(\mathbf{r},t) = 
{C_r P_r (A/m) a_\Sun^2 \over r_\Sun(t)^3}
\left(\mathbf{r}\cdot\mathbf{r}_\Sun(t)\right)
\end{equation}
In these formulas, $C_r$ is the satellite's average reflectivity coefficient, $P_r$ the solar radiation pressure constant, $(A/m)$ the satellite's effective area-to-mass ratio, $a_\Sun$ the semi-major axis of the Earth-Moon barycenter's heliocentric orbit. The Sun's lunicentric PALRF radius vector $\mathbf{r}_\Sun(t)$ is computed as in subsection \ref{sssec:sunpot}. 

\begin{figure}[!ht]
\centering
\includegraphics[width=0.5\textwidth]{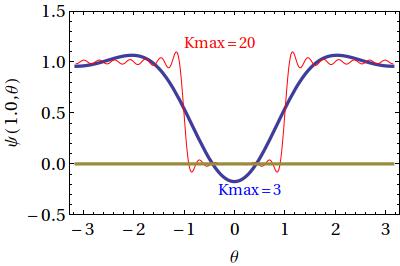}
\includegraphics[width=0.45\textwidth]{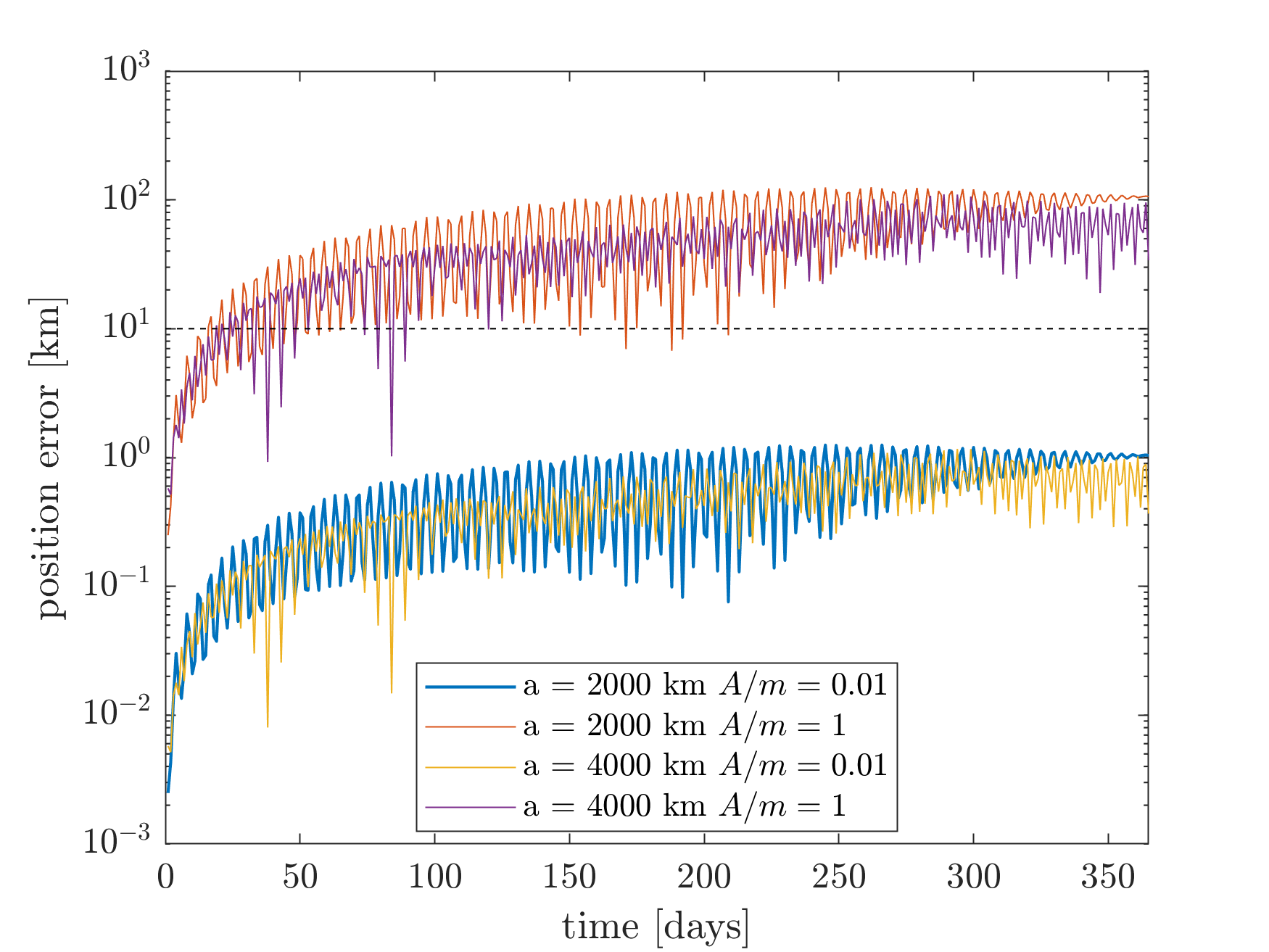}
\caption{\small (a) The growth in time of the distance between a trajectory computed with the model Kepler+SRP, and another in which the window function $\psi$ (Eq.(\ref{window})) is added as a factor to the SRP term. Four different trajectories are examined, with initial conditions corresponding to a circular orbit at distance and A/M ratio: i)$a=2000$~km, $A/M=0.01$, ii)$a=2000$~km, $A/M=1$, iii)$a=4000$~km, $A/M=0.01$, iv)$a=4000$~km, $A/M=1$. (b) The window function $\psi$ for $\phi_0=1$ rad as a function of the angle $\theta$ when $K_{max}=3$ or $K_{max}=20$. }
\label{fig:shadow}
\end{figure}
Eclipse effects can be included to the above model via multiplying the SRP potential with a shadow function $\psi$
\begin{equation}\label{potsrp}
V_{SRP}(\mathbf{r},t) = 
\psi\left(\mathbf{r},\mathbf{r}_\Sun(t)\right)
{C_r P_r (A/m) a_\Sun^2 \over r_\Sun(t)^3}
\left(\mathbf{r}\cdot\mathbf{r}_\Sun(t)\right)~~.
\end{equation}
However, we find that the eclipse effects are negligible on lunar satellite trajectories, while they become relevant only for objects well outside the satellite range, i.e., with $A/M>0.1$. Figure \ref{fig:shadow} shows the growth in time of the difference (in distance) between some initially circular trajectories computed with or without the shadow effect over an integration time of one year. The model used for the shadow function is the one proposed by Feraz-Mello in \cite{ferazmello}
\begin{eqnarray}\label{window}
\psi\left(\mathbf{r},\mathbf{r}_\Sun(t)\right) &=& 
1-\left({1\over\pi}\right)\arcsin\left({R_\Moon}\over r\right) \\
&-&\sum_{K=1}^{K_{max}}\Bigg[\left({2\over K\pi}\right)
\sin\left(K\arcsin\left({R_\Moon\over r}\right)\right)
\cos\left(K \arccos\left(\frac{\mathbf{r}\cdot\mathbf{r}_\Sun}{r r_\Sun}\right) \right)\bigg] \nonumber~~,
\end{eqnarray}
in which the window function tends to a inverse top-hut window when $K_{max}\rightarrow\infty$.  
\begin{equation}\label{psilim}
\lim_{K_{max}\rightarrow\infty} \psi(\phi_0,\theta)=
\left\{
\begin{array}{ll}
0~~~~~~~~~~~~~~~~~ &\mid\theta\mid\leq\phi_0 \\
1~~~~~~~~~~~~~~~~~ &\mid\theta\mid>\phi_0
\end{array}
\right.
\end{equation}
with $\theta$ equal to the angle between the vectors $\mathbf{r}$ and $\mathbf{r}_\Sun$ and $\phi_0 = \arcsin(R_\Moon/r)$. As shown in \cite{ferazmello}, the advantage of using such a window function $\psi$ is that it is readily expandable in Keplerian orbital elements of the satellite, while even low order truncations of the sum (say, with $K_{max}=3$ provide a satisfactory representation of the eclipse effect (see figure \ref{fig:shadow}a). 
 
Figure \ref{fig:shadow}b shows the growth in time of the distance between a trajectory computed with the model Kepler+SRP, and another in which the window function $\psi$ (Eq.(\ref{window})) is added as a factor to the SRP term. Four different trajectories are examined, with initial conditions corresponding to a circular orbit at distance and A/M ratio: i)$a=2000$~km, $A/M=0.01$, ii)$a=2000$~km, $A/M=1$, iii)$a=4000$~km, $A/M=0.01$, iv)$a=4000$~km, $A/M=1$. We note that the error in distance between trajectories computed with or without the shadow effect has only a weak dependence on the altitude of the orbit. This is because the solid angle with respect to the satellite subtended by the eclipse surface decreases as $~1/a$, but this is compensated by the fact that the size of the trajectory increases proportionally to $a$. We thus obtain an approximate estimate $\Delta(Error_{shadow})/\Delta t\approx (A/M) 100km/yr$.   

\subsection{Relative importance of the various forces}
\label{ssec:forcecompare}

In the choice of the appropriate splitting of the equations of motion in terms of different orders of smallness, a crucial criterion is the relative importance of the accelerations induced on a satellite by different sources (lunar potential harmonics, Earth and solar tide, apparent forces, SRP). 

\begin{figure}
\centering
\includegraphics[scale=0.43]{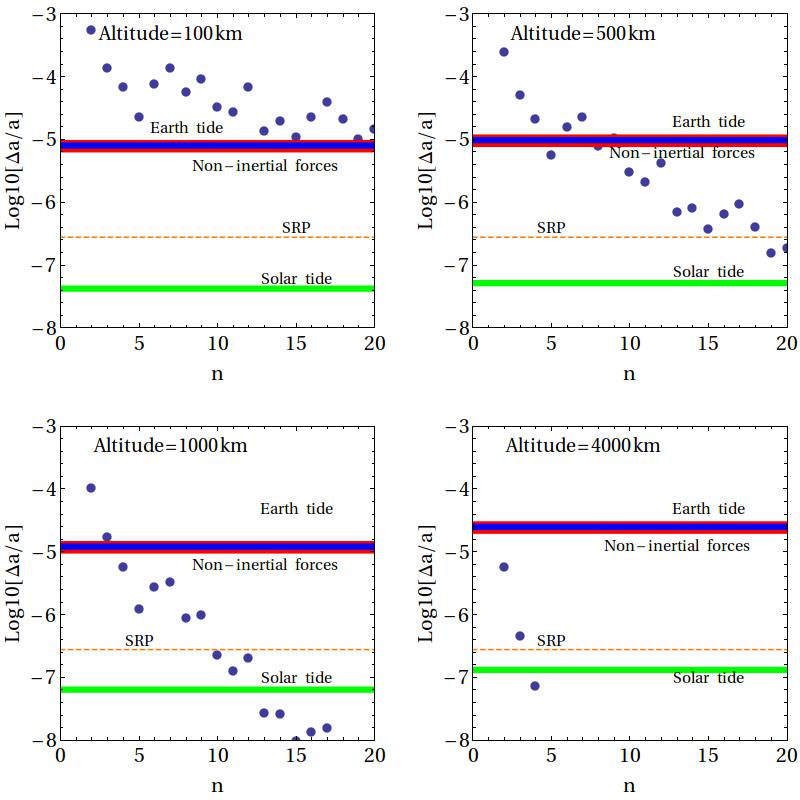}
\caption{\small Blue dots show the cumulative size (zonal+tesseral) of the relative (with respect to the Keplerian term) acceleration generated by the order $n$ terms of the lunipotential GRAIL values are used, see text). The blue line corresponds to the acceleration generated by the Earth's tide, green line the acceleration by the Solar tide, orange solid line by apparent forces, dashed line by the Solar Radiation Pressure to an object with reflectivity coefficient $C_r=1$ and area-to-mass ratio $A/M = 0.1$ meter/Kg$^2$. The altitude from the Lunar surface is equal to (top left) 100 km, (top right) 500 km, (bottom left) 1000 km, (bottom right) 4000 km.}
\label{fig:acce}
\end{figure}
Figure \ref{fig:acce} shows an estimate of the various non-Keplerian perturbations to the acceleration at four different altitudes, namely $r-R_\Moon=100$~km, $r-R_\Moon=500$~km,  $r-R_\Moon=1000$~km, and $r-R_\Moon=4000$~km from the lunar surface, where $r$ is the lunicentric distance of the satellite. An estimate of the relative (with respect to the Keplerian) acceleration generated by the n-th order harmonics is given by the formula:
\begin{equation}\label{eq:accein}
{\Delta a(n)\over a_{Kep}}\sim \sum_{m=0}^n {(n+1) R_\Moon^n\over r^n}
\left(C_{nm}^2+S_{nm}^2\right)^{1/2}~~.
\end{equation}
Here ${\cal G} M_\Moon$ is the mass parameter of the Moon, and $C_{nm},S_{nm}$ are the unnormalized 
coefficients of the harmonic expansion of the lunar potential. Using GRAIL values \cite{grail}, we obtain the points in Fig.\ref{fig:acce}, as a function of $n$. 
The straight lines show estimates of the relative acceleration generated by the Earth's tide, the Sun's tide, and non-inertial forces in a rotating frame fixed on the Moon:
\begin{equation}\label{eq:deltaa}
{\Delta a_\Earth\over a_{Kep}}\sim 
{M_\Earth r^3\over M_\Moon a_\Earth^3},~~~~~~
{\Delta a_\Sun\over a_{Kep}}\sim 
{M_\Sun r^3\over M_\Moon a_\Sun^3},~~~~~~
{\Delta a_{NI}\over a_{Kep}}\sim {n_\Moon^2 r^3\over {\cal G} M_\Moon} = 
{(M_\Earth+M_\Moon) r^3\over M_\Moon a_\Earth^3}   
\end{equation}
where $n_\Moon=\sqrt{{\cal G}(M_\Moon+M_\Earth)/a_\Earth^3}$ the mean orbital frequency of the Moon-Earth Keplerian ellipse, taken (for estimates) equal to the rotation frequency of the Moon. Since $M_\Moon<<M_\Earth$, the synchronous Moon's rotation implies that non-inertial forces due to the Moon's rotation are always of similar size as the Earth's tidal force, independently of the altitude (i.e. of $r$). \\
\\
\noindent
Some key remarks regarding Figure \ref{fig:acce} are: \\
\\
\noindent
(i) For low-altitude ($<100$ km) orbits, several harmonics of the lunar potential (most notably $J_2$, $C_{22}$, $J_7$ and several others up to $n=10$ compete in size, as a consequence of the  effect of the inhomogeneous lunar mass concentrations (mascons). The coefficients of the zonal harmonics $J_n$, $n=3,4,\ldots$ and of several tesseral harmonics $C_{nm},S_{nm}$, $n=2,3,\ldots$ decay with $n$ way more slowly than in the case of the geopotential. It has been suggested that zonal terms up to $n=30$ still influence the secular behavior of orbits such as the low-altitude high inclination ones (\cite{laretal2020}). This implies that, contrary to  the Earth's case, for low altitude orbits several harmonics besides $J_2$ need to be normalized up two second order in the process of averaging.\\
\\
\noindent
(ii) The Moon's rotational period is very long compared to the satellite's short orbital period. The latter is of order of hours, compared to the rotation period $~27.4$ days. Thus, the rotation of the Moon introduces one more secular period to the problem (about 100 to 1000 times longer than the orbital period). \\
\\
\noindent
(iii) The Earth's tidal acceleration (and corresponding apparent accelerations) dominate beyond the altitude $\sim 3000$ km.\\
\\
\noindent
(iv) No dissipative forces enter in the calculation - the solar radiation pressure force can be modelled as a conservative force which is time periodic when shadow effects are taken into account. Thus, closed form averaging can be performed to all force terms taking benefit of the canonical structure of the equations of motion.\\

\clearpage
\section{Closed-form representation of the Hamiltonian}
\label{sec:hamclosed}

\subsection{Elements and coordinate transformations}
\label{ssec:elecoord}
Standard theory (transformation to polar-nodal and then to action-angle variables) yields a symplectic transformation connecting the set of canonical variables $(x,y,z,p_x,p_y,p_z)$ appearing in the Hamiltonian (\ref{ham}) to the set of Delaunay action-angle variables: 
\begin{eqnarray}\label{delaunay}
L &=&\sqrt{{\cal G}M_\Moon a},~~~~~~~~~~~~~~~~~~~~~~~~ M \nonumber\\
G &=&\sqrt{{\cal G}M_\Moon a}\sqrt{1-e^2},~~~~~~~~~~~~~~g \\
H &=&\sqrt{{\cal G}M_\Moon a}\sqrt{1-e^2}\cos i,~~~~~~~~h\nonumber
\end{eqnarray}
where $(a,e,i,M,g,h)$ are the osculating Keplerian elements of the orbit (semi-major axis, eccentricity, inclination, mean anomaly, argument of the perilune, longitude of the ascending node). 

It is important to recall that, by the definition of the Hamiltonian (\ref{ham}), the elements $(a,e,i,M,g,h)$ of Eq.(\ref{delaunay}) must be physically interpreted as the elements of an osculating Keplerian ellipse corresponding to a satellite's radius vector $(x,y,z)$ and velocity vector $(p_x,p_y,p_z)$ \textit{as measured by an inertial observer whose frame instantaneously coincides with the PALRF frame at the time $t$}. However, for a solidal observer permanently attached to and co-rotating with PALRF, the radius and velocity vectors at time $t$ are given by:
\begin{eqnarray}\label{radvelpalrf}
x^{(PALRF)}&=&x \nonumber\\
y^{(PALRF)}&=&y \nonumber\\
z^{(PALRF)}&=&z \\
v_x^{(PALRF)}&=&p_x-(\omega_yz-\omega_zy)   \nonumber\\ 
v_y^{(PALRF)}&=&p_y-(\omega_zx-\omega_xz)   \nonumber\\
v_z^{(PALRF)}&=&p_z-(\omega_xy-\omega_yx)   \nonumber 
\end{eqnarray}
In order to keep all benefits stemming from preserving the canonical character of the theory, it turns convenient that the semi-analytical theory implemented by SELENA operate directly on the inertial state vector 
\begin{equation}\label{zetavec}
\zeta\equiv(x,y,z,p_x,p_y,p_z)~~, 
\end{equation} 
or its equivalent state vector in osculating Keplerian elements
\begin{equation}\label{zvec}
z\equiv(a,e,i,M,g,h)=E(\zeta)~~.
\end{equation}
On the other hand, to keep the processing of input and output data as simple as possible, SELENA adopts the convention that all input or output data in the SELENA propagator be expressed in terms of the PALRF state vector 
\begin{equation}\label{zetapalrf}
\zeta^{(PALRF)}\equiv(x^{(PALRF)},y^{(PALRF)},z^{(PALRF)},v_x^{(PALRF)},v_y^{(PALRF)},v_z^{(PALRF)})~~.  
\end{equation}
or the element state vector $z^{(PALRF)}=E(\zeta^{(PALRF)})$. To this end, using the compact notation  
\begin{equation}\label{qutra}
\zeta^{(PALRF)}=Q(\zeta),~~~\zeta=Q^{-1}(\zeta^{(PALRF)})
\end{equation}
for the transformation (\ref{radvelpalrf}) and for its inverse, and denoting by 
\begin{equation}\label{ftra}
z=F(z_{mean}),~~~z_{mean}=F^{-1}(z)
\end{equation}
the canonical transformation mapping mean to osculating elements and vice versa, as provided by the semi-analytical theory discussed in the next section, the SELENA procedure for the semi-analytic propagation of the trajectories can be summarized as follows:
$$
\mbox{Input}~\left(\zeta^{(PALRF)}(t=t_0)~\mbox{or}~z^{(PALRF)}=E(z^{(PALRF)})(t=t_0))\right)
\longrightarrow
$$
$$
\mbox{Compute}~z_0
=E\left(Q^{-1}(\zeta^{(PALRF)(t=t_0)})\right)
=E\left(Q^{-1}(E^{-1}(z^{(PALRF)(t=t_0)}))\right)
\longrightarrow
$$
$$
\mbox{Compute}~z_{mean,0}=F^{-1}(z_0)
\longrightarrow
$$
$$
\mbox{Propagate by SELENA and 
compute the vector}~z_{mean}(t;z_{mean,0})
\longrightarrow
$$
$$
\mbox{Compute and output}~z(t;z_0)=F(z_{mean}(t;z_{mean,0}))
\longrightarrow
$$
$$
\mbox{Compute and output}~\zeta(t;\zeta_0) = Q^{-1}(z(t;z_0))~~.
$$
Keeping in mind the above conversions of variables, in the sequel we will exclusively focus on the manipulation of the canonical state vectors $\zeta$ and $z$ in the framework of the closed-form averaging theory adopted by SELENA. In particular, in the next section we will provide all necessary definitions and steps which conclude the process of canonical averaging by giving the form of the transformation (\ref{ftra}). As standard in secular perturbation theory, the purpose of the transformation (\ref{ftra}) adopted by SELENA can be defined as follows: compute the Hamiltonian function
\begin{equation}\label{zetanf}
Z(z_{mean}) = {\cal H}(z=F(z_{mean}))~~. 
\end{equation}
We aim to determine the appropriate form of the transformation 
$$
F:~(a,e,i,M,g,h)\rightarrow(a_{mean},e_{mean},i_{mean},M_{mean},g_{mean},h_{mean})
$$ 
such that the function $Z$ does not depend on the fast variable $M_{mean}$. Then, $a_{mean}$ becomes an integral of motion (constancy of the semi-major axis). Furthermore, the equations of motion in the mean elements, hereafter called the \textit{secular} equations of motion, are just Hamilton's equations under the Hamiltonian $Z$, called the secular Hamiltonian. Since no fast degree of freedom is embedded in $Z$, the integration of the secular equations of motion can be done with large timesteps, of order $\sim 10^{-2}$ or $\sim 10^{-3}\times T_{sec}$, where $T_{sec}$ denotes the shortest secular period associated with the Hamiltonian $Z$. In the Moon's case, this implies integration by a timestep about 1000 times larger than the timestep used in a Cartesian integrator. On the other hand, all the corrections to the state vector $z$ due to short-periodic effects are now absorbed in and can be recovered at any required time by the transformation $F$.     

\subsection{Closed form representation of the Hamiltonian terms}
\label{ssec:hamtermsclosed}
As discussed in the next section, for the purpose of performing canonical averaging in closed form, in the expression of the Hamiltonian (\ref{ham}) in terms of canonical action-angle variables, it turns convenient to keep the dependence on the latter implicit, i.e., via the Keplerian elements $(a,e,i,M,g,h)$. By the equation (\ref{delaunay}) one has that $a=a(L)$, $e=e(L,G)$, $i=i(L,G,H)$. The various terms in the Hamiltonian (\ref{ham}) then obtain the following form: 

\subsubsection{Keplerian term}
\label{sssec:keplerclosed}
\begin{equation}\label{hkep}
{\cal H}_{kep} = {1\over 2}(p_x^2+p_y^2+p_z^2) - {{\cal G}M_\Moon\over r} = 
- {{\cal G}M_\Moon\over 2a} = -{1\over 2}a^2n_s^2
\end{equation}
where $n_s = ({\cal G}M_\Moon/a^3)^{1/2}$ is the satellite's mean motion.

\subsubsection{Rotation term}
\label{sssec:releclosed}
\begin{equation}\label{hamni}
{\cal H}_{NI} = -\boldsymbol{\omega}(t)\cdot(\mathbf{r}\times\mathbf{p}) = 
-n_\Moon(t) H  
- \omega_x(t)G\sin(i)\sin(h)
+ \omega_y(t)G\sin(i)\cos(h)
\end{equation}
where $n_\Moon(t)=\omega_z(t)$. The change of symbol is to emphasize that the component $n_\Moon$ is dominant in the Moon's angular velocity vector (see subsection 2.1). 

\subsubsection{Internal terms}
\label{sssec:intclosed}
All terms $n\geq 2$ in the lunar potential $V_\Moon$ will be hereafter called to cause an \textit{internal perturbation} with respect to the Keplerian orbit, i.e., a perturbation due to force sources located at the interior of the satellite's orbit. The associated terms are called internal terms. In order that a closed-form expression of the canonical averaging of the internal terms becomes possible, it is required that the true anomaly $f=f(M,e(L,G))$ be used instead of the mean anomaly $M$ in the corresponding trigonometric arguments of these terms. To this end, use is made of the expressions 
\begin{eqnarray}\label{xyzf}
x &= &r\cos(f+g)\cos h-r\sin(f+g)\cos i\sin h \nonumber\\
y &= &r\cos(f+g)\sin h+r\sin(f+g)\cos i\cos h \\
z &= &r\sin(f+g)\sin i ~~.\nonumber
\end{eqnarray}
For example, substituting (\ref{xyzf}) into the terms
$$
V_{J2} = -J_2{{\cal G}M_\Moon R_\Moon^2\over 2r^3
}\left(1-3{z^2\over r^2}\right)~~,~~
V_{C22} = -C_{22}{3{\cal G}M_\Moon R_\Moon^2\over r^5}
\left(x^2-y^2\right)~~,
$$
$$
V_{C31} = C_{31}{{3\cal G}M_\Moon R_\Moon^3x\over 2r^5}
\left(1-5{z^2\over r^2}\right)~,~~
V_{S31} = S_{31}{{3\cal G}M_\Moon R_\Moon^3y\over 2r^5}
\left(1-5{z^2\over r^2}\right)~~.
$$
and reducing all expressions to trigonometric polynomials we arrive at:\begin{equation}\label{j2ele}
V_{J2} = J_2{{\cal G}M_\Moon R_\Moon^2\over 4 r^3}
\left[-2+3 s^2 -3s^2\cos(2f+2g)\right]
\end{equation}
\begin{eqnarray}\label{c22ele}
V_{C22}&=&C_{22}{3{\cal G}M_\Moon R_\Moon^2\over 4 r^3}
\Bigg[
-2s^2\cos 2h
+3(-2+2c+s^2)\cos(2f+2g-2h)
\nonumber\\
&~&
~~~~~~~~~~~~~~~~~~-3(2+2c-s^2)\cos(2f+2g+2h)
\Bigg]
\end{eqnarray}
\begin{eqnarray}\label{c31ele}
V_{C31}&=&C_{31}{3{\cal G}M_\Moon R_\Moon^3\over 16 r^4}
\Bigg[
(-4-11c+5s^2+15c^3)\cos(f+g-h)
\nonumber\\
&~&~~~~~~~~~~~
+(-4+11c+5s^2-15c^3)\cos(f+g+h) \\
&~&~~~~~~~~~~~
+15(c-s^2-c^3)\cos(3f+3g-h) \nonumber\\
&~&~~~~~~~~~~~
-15(c+s^2-c^3)\cos(3f+3g+h) 
\Bigg] \nonumber
\end{eqnarray}
\begin{eqnarray}\label{s31ele}
V_{S31}&=&S_{31}{3{\cal G}M_\Moon R_\Moon^3\over 16 r^4}
\Bigg[
(4+11c-5s^2-15c^3)\sin(f+g-h)
\nonumber\\
&~&~~~~~~~~~~~
+(-4+11c+5s^2-15c^3)\sin(f+g+h) \\
&~&~~~~~~~~~~~
-15(c-s^2-c^3)\sin(3f+3g-h) \nonumber\\
&~&~~~~~~~~~~~
-15(c+s^2-c^3)\sin(3f+3g+h) 
\Bigg] \nonumber
\end{eqnarray}
where $c=\cos i$, $s=\sin i$. 

\subsubsection{External terms}
\label{sssec:extclosed}
All pertubations (whether gravitational or not) stemming from force sources located at the exterior of the satellite's orbit, i.e., the Earth's tide, the solar tide and the solar radiation pressure will be hereafter called \textit{external}. In order that a closed-form expression of the canonical averaging of the external terms becomes possible, it is required that the eccentric anomaly $u=u(M,e(L,G))$ be used instead of the mean anomaly $M$ in the corresponding trigonometric arguments of these terms. To this end, using the equations  
\begin{eqnarray}\label{cosfu}
\cos f &= &{a(\cos u - e)\over r}~~\\
\sin f &= &{a\eta \sin u\over r}~~,\nonumber
\end{eqnarray}
where 
\begin{equation}\label{eta}
\eta=\sqrt{1-e^2}~
 \end{equation}
is the \textit{eccentricity function}, from the expressions (\ref{xyzf}) we arrive at
\begin{eqnarray}\label{xyzu}
x &=&{a\over 4}\bigg[-2e(1-\cos i)\cos(g-h)-2e(1+\cos i)\cos(g+h) \nonumber\\
~ &~&~~~+(1-\eta)(1-\cos i)\cos(g-h-u)+(1+\eta)(1-\cos i)\cos(g-h+u) \nonumber\\
~ &~&~~~+(1-\eta)(1+\cos i)\cos(g+h-u)+(1+\eta)(1+\cos i)\cos(g+h+u)\bigg] \nonumber\\
y &=&{a\over 4}\bigg[2e(1-\sin i)\sin(g-h)-2e(1+\cos i)\sin(g+h) \\
~ &~&~~~+(1-\eta)(1-\cos i)\sin(g-h-u)-(1+\eta)(1-\cos i)\sin(g-h+u) \nonumber\\
~ &~&~~~+(1-\eta)(1+\cos i)\sin(g+h-u)+(1+\eta)(1+\cos i)\sin(g+h+u)\bigg] \nonumber\\
z &=&{a\over 4}\bigg[-2e\cos(g-h)+2e\cos(g+h) \nonumber\\
~ &~&~~~+(1-\eta)\cos(g-h-u)+(1+\eta)\cos(g-h+u) \nonumber\\
~ &~&~~~-(1-\eta)\cos(g+h-u)-(1+\eta)\cos(g+h+u)\bigg]~~. \nonumber
\end{eqnarray}
Substituting, now, Eq.(\ref{xyzu}), as well as 
\begin{equation}\label{ru}
r=a(1-\cos u) 
\end{equation}
into an external perturbation term leads to a purely trigonometric expression in terms of the angles $(u,g,h)$. For example, for the Earth tidal term $V_{\Earth,P2}$ we find:
\begin{equation}\label{potetideele}
V_\Earth(a,e,i,u,g,h;\mathbf{r}_\Earth(t)) =  {{\cal G}M_\Earth a^2\over 64 r_\Earth^3(t)}
\sum_{k,l,m}\Phi_{k,l,m}(e,i)\cos(kg+lh+mu)
\end{equation}
where
$$
\Phi_{0,0,0}=-16-8e^2+24s^2+12e^2s^2
$$
$$
\Phi_{0,0,1}=32e-48es^2
$$
$$
\Phi_{0,0,2}=-8e^2+12e^2s^2
$$
$$
\Phi_{0,2,0}=-24s^2-12e^2s^2
$$
$$
\Phi_{2,0,0}=-36e^2s^2
$$
$$
\Phi_{2,2,0}=-36e^2-36ce^2+18e^2s^2
$$
$$
\Phi_{2,-2,0}=-36e^2+36ce^2+18e^2s^2
$$
$$
\Phi_{2,0,1}=24es^2+24e\eta s^2
$$
$$
\Phi_{2,0,-1}=24es^2-24e\eta s^2
$$
$$
\Phi_{2,0,2}=-12s^2+6e^2s^2-12\eta s^2
$$
$$
\Phi_{2,0,-2}=-12s^2+6e^2s^2+12\eta s^2
$$
$$
\Phi_{0,2,1}=\Phi_{0,2,-1}=24e s^2
$$
$$
\Phi_{0,2,2}=\Phi_{0,2,-2}=-6 e^2 s^2
$$
$$
\Phi_{2,2,1}=24e+24ce+24e\eta+24ce\eta-12es^2-12e\eta s^2
$$
$$
\Phi_{2,2,-1}=24e+24ce-24e\eta-24ce\eta-12es^2+12e\eta s^2
$$
$$
\Phi_{2,-2,1}=24e-24ce+24e\eta-24ce\eta-12es^2-12e\eta s^2
$$
$$
\Phi_{2,-2,-1}=24e-24ce-24e\eta+24ce\eta-12es^2+12e\eta s^2
$$
$$
\Phi_{2,2,2}=-12-12c+6e^2+6ce^2-12\eta-12c\eta+6s^2-3e^2s^2+6\eta s^2
$$
$$
\Phi_{2,2,-2}=-12-12c+6e^2+6ce^2+12\eta+12c\eta+6s^2-3e^2s^2-6\eta s^2
$$
$$
\Phi_{2,-2,2}=-12+12c+6e^2-6ce^2-12\eta+12c\eta+6s^2-3e^2s^2+6\eta s^2
$$
$$
\Phi_{2,-2,-2}=-12+12c+6e^2-6ce^2+12\eta-12c\eta+6s^2-3e^2s^2-6\eta s^2
$$

\clearpage
\section{SELENA canonical averaging procedure}
\label{sec:selenaave}

\subsection{General}
In the present section we illustrate the steps in the construction of a canonical averaging theory for the Hamiltonian (\ref{ham}) in closed form, i.e., without expansions in powers of the satellite's orbital eccentricity.\\ 
\\
Closed-form canonical averaging has proven to be a powerful tool for the treatment of a wide-class perturbed near-Keplerian problems (see \cite{lar2021} for a comprehensive introduction). Besides presenting its use in the formulation of the secular theory behind the SELENA propagator, the purpose of the present section is to provide a complete and autonomous set of formulas explaining how to implement the core feature of the method, i.e., \textit{Delaunay normalization} with the help of a symbolic manipulator. \\
\\
Two key remarks regarding the implementation of the method are in order: \\
\\
\noindent
1. As discussed in detail below, for the averaging method to lead to formulas in closed-form, it is required that the so-called kernel of the homological equation by which the Lie generating function of the normalizing transformations is computed be composed entirely by the Kepler Hamiltonian ${\cal H}_{kep}$ (Eq.(\ref{hkep})). While this algebraic property is naturally satisfied in the case of the averaging of the central body's zonal harmonics (which do not depend on the orientation of the orbital apsides with respect to the body), it is no longer satisfied as regards the averaging of neither the central body's tesseral harmonics, nor the tidal pertubations of any third body. For such bodies, instead, the kernel of the homological equation contains the combination ${\cal H}_{kep}- \Omega_{CB} L_\perp$, where $\Omega_{CB}$ is the rotational velocity of the central body and $L_\parallel$ the component of the satellite's angular momentum along the axis of rotation of the central body. In order to face this problem, two strategies can be adopted:
\begin{itemize}
\item
Expand the tesseral harmonic and tidal terms in powers of the orbital eccentricity. This leads to a mixed theory, which is in closed form as regards zonal harmonics and in series expansion in the eccentricity for tesseral harmonics (see, for example, \cite{cofalf1981}\cite{wnu1988}\cite{metetal1993}\cite{metexe1995}\cite{deletal2006}\cite{deletal2011}\cite{laretal2018}) in the case of Earth's satellites; we note that this strategy is implemented in both the DSST and STELA semi-analytical propagators for Earth-satellite orbits (see \cite{cefetal2019}\cite{stela} and references therein). 
\item
\textit{Relegation} (\cite{pal1992}\cite{segcof2000}\cite{depetal2001}\cite{pal2002}\cite{laretal2013}\cite{mahalf2019}): according to the relegation method, for averaging both tesseral and tidal terms we keep in the kernel of the homological equation only the Keplerian term ${\cal H}_{kep}$. This allows to obtain closed-form formulas for the Lie canonical transformation, at the cost, however, of introducing a sequence of additional averaging steps,  the i-th step caused by the Poisson bracket between the generating function of the (i-1)-th step and the $\Omega_{CB} L_\perp$. The speed of convergence of this sequence depends on how smaller than unity is the ratio $n/\Omega_{CB}$, where $n$ is the satellite's orbital frequency. This is because the method's i-th step progresses geometrically as $(n/\Omega_{CB})^i$.\\
\end{itemize}
\noindent
In the case of Earth satellites, relegation is an option only for LEO or lower MEO orbits ($n/\Omega_{CB}<1$). On the other hand, in the case of the Moon, relegation is the natural strategy to select since $n/\Omega_{CB}< 10^{-2}$ both for low and high altitude orbits. Furthermore, in the implementation of SELENA it turns convenient from the symbolic-computational point of view to adopt an alternative of the relegation algorithm (see \cite{laretal2013}\cite{caveft2022}), which introduces in the formulas powers of the eccentricity instead of powers of the quantity $n/\Omega_{CB}$, but without any expansion of the original tesseral or tidal terms (see subsection \ref{sssec:step7} below). Note also that, owing to the small value of the Moon's angular velocity, the relegation terms where ignored altogether in the lunar theory of \cite{deSae2006a}\cite{desaehen2006}. \\
\\
\noindent
2. Apart from the use of formulas stemming directly from the definition of Keplerian elements, in a number of substeps of the averaging method it is required to perform a sequence of algebraic reductions and simplifications whose understanding and correct programming is crucial for a manipulator being able to efficiently implement the method. Partial accounts for these algebraic operations can be found in \cite{jef1971}\cite{depcof1982}\cite{dep1982}\cite{kel1989}\cite{met1991}\cite{hea2000}\cite{danetal1995}\cite{deSae2006b}, and in the recent book \cite{lar2021}. We are aware of no reference text where a full account is given. Thus, in the present and in the following section we aim to provide a unified account of the complete set of formulas, algebraic simplifications and functional operations of closed-form perturbation theory as implemented in the framework of SELENA's semi-analytic theory. 

\subsection{Canonical averaging by Lie series}
\label{ssec:lieseries}
\subsubsection{Composition of Lie Series}
\label{sssec:liecompo}
The canonical averaging of the Hamiltonian in closed form produced by the SELENA symbolic package works by the method of the composition of Lie series. See \cite{eft2012} for a comprehensive introduction. In the organization of the Lie series the `book-keeping' scheme proposed in \cite{eft2012} is adopted. The scheme serves the purpose of averaging any Hamiltonian of the form $H(\mathbf{q},\mathbf{p},t)$, where $\mathbf{z}=(\mathbf{q},\mathbf{p})$ is a set of $N_c$ canonical pairs of variables $\mathbf{q}=(q_1,\ldots,q_{N_c})$, $\mathbf{p}=(p_1,\ldots,p_{N_c})$. It is realised by the following steps:\\
\\
{\bf Step 1: (Hamiltonian preparation and book-keeping):} introduce a book-keeping symbol, say $\epsilon$, with numerical value equal to $\epsilon = 1$, and split the Hamiltonian according to:
\begin{equation}\label{hamgenbk}
{\cal H} = Z_0 + \epsilon {\cal H}_1 + \epsilon^2 {\cal H}_2 + \ldots +\epsilon^N {\cal H}_N = Z_0+\sum_{s=1}^N\epsilon^s {\cal H}_s.
\end{equation}
The integer $N$ is called the maximum truncation order. The splitting is subjective, and reflects one's perceived hierarchy of the size of the effect that the various perturbations have on the trajectories with respect to those of the basic dynamical model, given by the term $Z_0$.  The model $Z_0$ has to be already in average form. The index $s$ in ${\cal H}_s$ indicates that the perturbation to the dynamics induced by ${\cal H}_s$ is considered as `of $s-$th order of smallness'. \\
\\
{\bf Step 2 (Lie derivative operations):} Let ${\cal D}$ be the space of real functions $f(\mathbf{q},\mathbf{p},t)$ being $C^\infty$ in $(\mathbf{q},\mathbf{p})~\forall t\in\mathbb{R}$. For any two functions $f,g\in {\cal D}$ define the Poisson bracket operator:
\begin{equation}\label{poissonfg}
{\cal L}_gf=\{f,g\}=\sum_{j=1}^N
{\partial f\over\partial q_j}{\partial g\over\partial p_j} -
{\partial f\over\partial p_j}{\partial g\over\partial q_j}~~.
\end{equation}
Let $S({\cal D})\subset{\cal D}$ be a subset of ${\cal D}$ with the properties: i) ${\cal H}\in S({\cal D})$, ii) $\forall f,g \in S({\cal D}), {\cal L}_g f\in S({\cal D})$. Then, it can be easily demonstrated that  $S({\cal D})$, supplemented with the addition operation between functions and the multiplication of a function with a real number, is a vector space. Define now the orthogonal null and range spaces ${\cal N}_{L_{Z0}}\in S({\cal D})$, ${\cal R}_{L_{Z0}}\in S({\cal D})$ of the operator ${\cal L}_{Z_0}: S({\cal D})\rightarrow S({\cal D})$ by the properties: i) $S({\cal D})={\cal N}_{L_{Z0}}\oplus{\cal R}_{L_{Z0}}$, ii) $\forall f\in{\cal N}_{L_{Z0}}$ we have that ${\cal L}_{Z0}f=0$, iii) $\forall f\in{\cal R}_{L_{Z0}}$ we have that ${\cal L}_{Z_0}f\in{\cal R}_{L_{Z0}}$, ${\cal L}_{Z_0}^{-1}f\in{\cal R}_{L_{Z0}}$. Establish an explicit method to symbolically compute ${\cal L}_{Z_0}^{-1}f$ for all functions $f\in{\cal R}_{L_{Z0}}$. \\
\\
{\bf Step 3 (computation of Lie generating functions):}. For any function $\chi\in S({\cal D})$ the Lie exponential operator of $\chi$ is defined as:
\begin{equation}\label{lieexpchi}
\exp\left({\cal L}_\chi\right): S({\cal D})\rightarrow S({\cal D}),~~~  
\exp\left({\cal L}_\chi\right):=\sum_{j=0}^\infty{1\over j!}\left({\cal L}_\chi\right)^j~~.
\end{equation}
Provide an algorithm by which to determine $K$ integers $S_1,S_2,\ldots,S_K$ and $K$ corresponding suitably defined functions $\chi^{(1)},\chi^{(2)},\ldots,\chi^{(K)}$, called the \textit{Lie generating functions}, such that the function
\begin{equation}\label{hamknorm}
{\cal H}^{(K)}= 
\exp\left({\cal L}_{\epsilon^{S_K}\chi^{(K)}}\right)
\circ
\exp\left({\cal L}_{\epsilon^{S_{K-1}}\chi^{(K-1)}}\right)
\circ\ldots\circ
\exp\left({\cal L}_{\epsilon^{S_1}\chi^{(1)}}\right){\cal H}
\end{equation}
truncated at order $N$ in the book-keeping parameter obtains the form
\begin{equation}\label{hamnfbk}
\left({\cal H}^{(K)}\right)^{\leq N} 
= Z^{(K)}+R^{(K)}=Z_0 + \epsilon Z_1 + \epsilon^2 Z_2 + \ldots +\epsilon^N Z_N + R^{(K)} 
\end{equation}
where all the functions $Z_s$, $s=0,\ldots N$ have the desired averaging properties (for example: they are independent of fast-periodic trigonometric terms), while omission of the terms $R^{(K)}$ from Hamilton's equations of motion leads to an acceptable error in the trajectories. 

The generating functions $\chi_1,\ldots,\chi_K$ are specified recursively. The $r-$th step proceeds as follows: assume the functions $\chi_1,\ldots,\chi_{r-1}$ have been specified. The Hamiltonian
\begin{equation}\label{hamrnorm}
{\cal H}^{(r-1)}= 
\exp\left({\cal L}_{\epsilon^{S_{r-1}}\chi^{(r-1)}}\right)
\circ
\exp\left({\cal L}_{\epsilon^{S_{r-2}}\chi^{(r-2)}}\right)
\circ\ldots\circ
\exp\left({\cal L}_{\epsilon^{S_1}\chi^{(1)}}\right){\cal H}
\end{equation}
has the N-truncated form
\begin{equation}\label{hamrm1bk}
\left({\cal H}^{(r-1)}\right)^{\leq N} = 
Z_0 + \epsilon {\cal H}^{(r-1)}_1 + \epsilon^2{\cal H}^{(r-1)}_2 + \ldots +\epsilon^N {\cal H}^{(r-1)}_N~~. 
\end{equation}
Choose \textit{one} of the terms ${\cal H}_s^{(r-1)}$ in $\left({\cal H}^{(r-1)}\right)^{\leq N}$, corresponding to one of the integer indices $s=\{1,\ldots,N\}$ on which the next partial averaging is to be performed. For a valid choice ${\cal H}_s^{(r-1)}$ must satisfy the properties: i) ${\cal L}_{Z_0}{\cal H}^{(r-1)}_s\neq 0$, ii) omission of the term ${\cal L}_{Z_0}^{-1}{\cal L}_{Z_0}{\cal H}^{(r-1)}_s$ from Hamilton's equations of motion leads to a non-acceptable error in the trajectories. Set the integer $S_r$ equal to the $s$ chosen. Split ${\cal H}^{(r-1)}_{S_r}$ according to
\begin{equation}\label{hrm1s}
{\cal H}^{(r-1)}_{S_r}=h^{(r-1)}_{S_r}+R^{(r-1)}_{S_r} 
\end{equation}
where: i) ${\cal L}_{Z_0}h^{(r-1)}_{S_r}\neq 0$, and ii) $R^{(r-1)}_{S_r}$ is an \textit{acceptable remainder}. By acceptable it is meant that $R^{(r-1)}_{S_r}$ can be decomposed as $R^{(r-1)}_{S_r}=X^{(r-1)}_{S_r}+{\cal Y}^{(r-1)}_{S_r}+\tilde{R}^{(r-1)}_{S_r}$, where:

- the function $X^{(r-1)}_{S_r}$ has the desired averaging properties, 

- the function ${\cal Y}^{(r-1)}_{S_r}$ is to be normalized at a succeeding step,

- omission of the function $\tilde{R}^{(r-1)}_{S_r}$ from Hamilton's equations of motion leads to an acceptable error in the trajectories. \\
\\
Define 
\begin{equation}\label{zetadef}
\zeta^{(r-1)}_{S_r}=h^{(r-1)}_{S_r}-{\cal L}_{Z_0}^{-1}{\cal L}_{Z_0}h^{(r-1)}_{S_r}=
<h^{(r-1)}_{S_r}>_M={1\over 2\pi}\int_0^{2\pi}h^{(r-1)}_{S_r}dM~~.   
\end{equation}
Solve the homological equation
\begin{equation}\label{homor}
{\cal L}_{Z_0}\chi^{(r)}=h^{(r-1)}_{S_r}-\zeta^{(r-1)}_{S_r}
\end{equation}
considered as a linear partial differential equation with unknown function $\chi^{(r)}$. The solution comes to the form
\begin{equation}\label{chisol}
\chi^{(r)}= {\cal L}_{Z_0}^{-1}{\cal L}_{Z_0}(h^{(r-1)}_{S_r}-\zeta^{(r-1)}_{S_r})+\chi^{(r)}_{\cal N}
\end{equation}
where the part $\chi^{(r)}_{\cal N}$ belongs to the nucleus ${\cal N}_{L_{Z0}}$, and can be varied at will. \\
\\
Compute the next N-truncated Hamiltonian in the normalizing sequence:
\begin{equation}\label{hamrbk}
\left({\cal H}^{(r)}\right)^{\leq N}
=\left(\exp\left({\cal L}_{\epsilon^{S_r}\chi^{(r)}}\right)\left({\cal H}^{(r-1)}\right)^{\leq N}\right) ^{\leq N}
\end{equation}
This has the form
\begin{equation}\label{hamrm1bk}
\left({\cal H}^{(r)}\right)^{\leq N} = Z_0 + \epsilon {\cal H}^{(r)}_1 + \epsilon^2{\cal H}^{(r)}_2 + \ldots +\epsilon^N {\cal H}^{(r)}_N~~, 
\end{equation}
where ${\cal H}^{(r)}_{S_r}=Z^{(r)}_{S_r}+R^{(r)}_{S_r}$ with $Z^{(r)}_{S_r}=\zeta^{(r-1)}_{S_r}+X^{(r-1)}_{S_r}$, $R^{(r)}_{S_r}=R^{(r-1)}_{S_r}$. \\
\\
{\bf Step 4 (Equations of motion and normalizing transformation:)} Calling
\begin{equation}\label{hamave}
Z=Z_0+Z_1+...+Z_N=\left(Z^{(K)}\right)^{\leq N} 
\end{equation}
the \textit{averaged Hamiltonian} (or the \textit{normal form}), (\ref{hamnfbk}) leads to the averaged equations of motion
\begin{equation}\label{hamnfbkeq}
\dot{\mathbf{Q}}={\partial Z\over\partial\mathbf{P}},~~
\dot{\mathbf{P}}=-{\partial Z\over\partial\mathbf{Q}},~~
\end{equation}
for a set of new canonical variables $(\mathbf{Q},\mathbf{P})$, called the $\textit{mean variables}$, related to the original variables $(\mathbf{q},\mathbf{p})$ via the N-truncated transformation given by:
\begin{eqnarray}\label{qptra}
(\mathbf{q},\mathbf{p})&=&F(\mathbf{Q},\mathbf{P})~~\\
~&=&
\bigg[\left(
\exp\left({\cal L}_{\epsilon^{S_K}\chi^{(K)}}\right)
\circ
\exp\left({\cal L}_{\epsilon^{S_{K-1}}\chi^{(K-1)}}\right)
\circ\ldots\circ
\exp\left({\cal L}_{\epsilon^{S_1}\chi^{(1)}}\right)\right)_{(\mathbf{Q},\mathbf{P})}
(\mathbf{Q},\mathbf{P})\bigg]^{\leq N}~~\nonumber
\end{eqnarray}
\begin{eqnarray}\label{qptrainv}
(\mathbf{Q},\mathbf{P})&=&F^{-1}(\mathbf{q},\mathbf{p})~~\\
~&=&
~~~~~\bigg[\left(
\exp\left({\cal L}_{\epsilon^{S_1}\chi^{(1)}}\right)
\circ
\exp\left({\cal L}_{\epsilon^{S_{2}}\chi^{(2)}}\right)
\circ\ldots\circ
\exp\left({\cal L}_{\epsilon^{S_K}\chi^{(K)}}\right)\right)_{(\mathbf{q},\mathbf{p})}
(\mathbf{q},\mathbf{p})\bigg]^{\leq N}~~\nonumber
\end{eqnarray}
Compute the direct and inverse transformations $(\mathbf{q},\mathbf{p})\rightarrow(\mathbf{Q},\mathbf{P})$, $(\mathbf{Q},\mathbf{P})\rightarrow(\mathbf{q},\mathbf{p})$. Equations (\ref{hamnfbkeq}) provide the theory for the semi-analytical propagator, while Eqs.(\ref{qptra}) and (\ref{qptrainv}) provide the transformations back and forth to the new variables on which the theory operates. In particular, for any set of initial conditions $(\mathbf{q}_0,\mathbf{p}_0)$, semi-analytical propagation of the corresponding trajectory proceeds according to the following scheme:
\begin{itemize}
\item 
Use Eq.(\ref{qptrainv}) to compute the initial condition $(\mathbf{Q}_0,\mathbf{P}_0)=F^{-1}(\mathbf{q}_0,\mathbf{p}_0)$.
\item
Use the averaged equations of motion (\ref{hamnfbkeq}) to propagate the initial condition $(\mathbf{Q}_0,\mathbf{P}_0)$ and obtain $(\mathbf{Q}(t),\mathbf{P}(t))$.
\item 
Use Eq.(\ref{qptra}) to compute the trajectory in the original variables $(\mathbf{q}(t),\mathbf{p}(t))=F(\mathbf{Q}(t),\mathbf{P}(t))$.
\end{itemize}

A key remark as regards the above averaging scheme is that the choice of the composition of Lie series, in conjunction with book-keeping, as SELENA's normalizing scheme provides a wider flexibility in the choice of which terms are to be averaged and in what order, with respect to the use of the Lie triangle \cite{hor1966}\cite{dep1969}\cite{dep1982}, which is the most commonly chosen averaging scheme in the framework of near-Keplerian satellite motions. In particular, note that the finite sequence $(S_1,S_2,\ldots,S_K)$ need not be ordered, i.e. no condition of the form $S_j\leq S_{j+1}$ is required along the sequence. Since everyone of the Lie transformations generated by the functions $\chi^{(r)}$, $r=1,2,\ldots,K$ is, by definition, canonical, one can interchange the order, or even the number of the averaging steps performed. In general, permutations of the exponentials in Eqs.(\ref{hamnfbk}) or (\ref{qptra}) do not commute, a fact implying that different sequences $(S_1,S_2,\ldots,S_K)$ lead to different definitions of the mean variables and of the thereby produced averaged equations of motion. In addition, the definition of what condition defines the `averaging property' satisfied by the functions $X^{(r)}$ in the acceptable remainder can change on the go. As an example, in the actual theory implemented by SELENA, three different definitions are adopted for the properties of $X^{(r)}$ in three distinct groups of individual steps referring to i) the `elimination of the parallax' (\cite{dep1981}\cite{laraparallax}\cite{lar2021}), ii) terms generated by relegation, or iii) Delaunay normalization. 

We now report the implementation of Steps 1-4 of Section 3 in the semi-analytical theory of the SELENA propagator, giving the explicit formulas used in the symbolic programming of the associated averaging Lie transformations.  

\subsection{Step 1: Hamiltonian preparation and book-keeping}
\label{ssec:hamprepbk}

\subsubsection{Functional algebraic operations - CS-, Parallax- and RM- reduction}  

For the purposes of normalization in closed form, the following algebraic operations, acting on the Hamiltonian (\ref{ham})  are of use in the preparation of the initial Hamiltonian function of Eq.(\ref{hamgenbk}):\\
\\
\textbf{Cosine to Sine (CS-)reduction:} This operation should be applied to \textit{both internal and external terms}, and results in a reduction of the size of all obtained expressions by about $20\%$. We have
\begin{eqnarray}\label{csred}
CS[expr] &= &\mbox{Substitute all factors} \\
~&~&\cos(i)^p~\mbox{within $expr$ with:}  
\left\{\begin{array}{ll}
        (1-\sin^2(i))^{p/2}~~&p~\mbox{even}\\
        \cos(i)(1-\sin^2(i))^{(p-1)/2}~~&p~\mbox{odd}\\
       \end{array}
\right. \nonumber
\end{eqnarray}
~\\
\\
\textbf{Parallax reduction:} This operation should be applied only to \textit{internal terms}. For any expression of the form
\begin{eqnarray}\label{finte}
F_I &= &\sum_{p\geq 3}{1\over r^p}
\sum_{k,l,m}Q_{I,k,l,m}(a,e,i)\cos(kg+lh+mf) \\
~&+&\sum_{p\geq 3}{1\over r^p}
\sum_{k,l,m}K_{I,k,l,m}(a,e,i)\sin(kg+lh+mf) \nonumber 
\end{eqnarray}
introduce the operation
\begin{eqnarray}\label{pfinte}
P[F_I] &= &{1\over r^2}\sum_{p\geq 3}
\left({1+e\cos f\over a\eta^2}\right)^{p-2}
\sum_{k,l,m}Q_{I,k,l,m}(a,e,i)\cos(kg+lh+mf) \\
~&+&{1\over r^2}\sum_{p\geq 3}
\left({1+e\cos f\over a\eta^2}\right)^{p-2}
\sum_{k,l,m}K_{I,k,l,m}(a,e,i)\sin(kg+lh+mf) \nonumber 
\end{eqnarray}
Due to the identity $1/r^p = (1/r^2)(1/r^{p-2})$, the function $P[F_I]$ is 
an equivalent representation of the function $F_I$. Furthermore, since 
$p-2\geq 1$, after trigonometric reduction Eq.(\ref{pfinte}) yields a sum 
of purely trigonometric monomials $\cos(kg+lh+mf)$, $\sin(kg+lh+mf)$
\begin{eqnarray}\label{pfintered}
P[F_I] &= &{1\over r^2}
\sum_{k,l,m}B_{I,k,l,m}(a,e,i)\cos(kg+lh+mf) \\
~&+&{1\over r^2}
\sum_{k,l,m}D_{I,k,l,m}(a,e,i)\sin(kg+lh+mf) \nonumber 
\end{eqnarray}
~\\
\\
\textbf{$r-$to the minus one (RM-)reduction: } This operation should be applied only to \textit{external terms}, and plays in the normalization process a role analogous to the one of parallax reduction for the internal terms. For any expression of the form
\begin{eqnarray}\label{fexte}
F_E &= &\sum_{k,l,m}Q_{I,k,l,m}(a,e,i)\cos(kg+lh+mu) \\
   ~&+ &\sum_{k,l,m}K_{I,k,l,m}(a,e,i)\sin(kg+lh+mu) \nonumber 
\end{eqnarray}
introduce the operation
\begin{eqnarray}\label{pfexte}
RM[F_E] &= &{a\over r}
\left(1-e\cos u\right)
\sum_{k,l,m}Q_{I,k,l,m}(a,e,i)\cos(kg+lh+mu) \\
~&+&{a\over r}
\left(1-e\cos u\right)
\sum_{k,l,m}K_{I,k,l,m}(a,e,i)\sin(kg+lh+mu) \nonumber 
\end{eqnarray}
After trigonometric reduction Eq.(\ref{pfexte}) yields a sum of trigonometric monomials $\cos(kg+lh+mu)$, $\sin(kg+lh+mu)$
\begin{eqnarray}\label{pfextered}
RM[F_E] &= &{1\over r}
\sum_{k,l,m}B_{I,k,l,m}(a,e,i)\cos(kg+lh+mu) \\
~&+&{1\over r}
\sum_{k,l,m}D_{I,k,l,m}(a,e,i)\sin(kg+lh+mu) \nonumber 
\end{eqnarray}

\subsubsection{Book-keeping}  
\label{sssec:bookkeeping}
We set the truncation order equal to $N=4$, and book-keep the Hamiltonian according to 
\begin{equation}\label{hamselbk}
{\cal H} =Z_0+\epsilon {\cal H}_1+\epsilon^2{\cal H}_2+\epsilon^3{\cal H}_3+\epsilon^4{\cal H}_4
\end{equation}
where
\begin{eqnarray}\label{hamselbk234}
Z_0&=&{\cal H}_{Kep} \nonumber\\
~&~&\nonumber\\
{\cal H}_1&=&-n_\Moon(t)H \nonumber\\
~&~&\nonumber\\
{\cal H}_2&=&CS[P[V_{C20}]]+CS[RM[V_{\Earth,P2}]] \\
~&~&\nonumber\\
{\cal H}_3&=&CS\bigg[P\bigg[V_{C21}+V_{C22}+V_{S21}+V_{S22}
+\sum_{n=3}^{10}\sum_{m=0}^n(V_{Cnm}+V_{Snm})\bigg]\bigg] 
+CS[RM[V_{\Earth,P3}]] \nonumber \\
~&~&\nonumber\\
{\cal H}_4&=&- \omega_x(t)G\sin(i)\sin(h) + \omega_y(t)G\sin(i)\cos(h)
+CS[RM[V_{\Sun,P2}]]+CS[RM[V_{SRP}]]~~. \nonumber
\end{eqnarray}
We denote ${\cal H}_{rot,xy}=-\omega_x(t)G\sin(i)\sin(h) + \omega_y(t)G\sin(i)\cos(h)$.

\subsection{Step 2: Lie derivative operations}
\label{ssec:lieder}
To normalize both interior and exterior terms of the Hamiltonian, use will be made of the Poisson bracket between functions of the form $F(L,G,H,M,g,h)$ whose explicit expressions are, however, given instead in terms of the elements $(a,e,i,f,g,h)$, for functions arising throughout the normalization of `internal' terms, or $(a,e,i,u,g,h)$ for functions arising throughout the normalization of `external' terms. Whenever needed, Poisson brackets are to be computed by use of the following formulas:

\subsubsection{Poisson bracket for functions involving internal terms}
\label{sssec:poissonint}
Let $A(a,e,i,f,g,h)$, $B(a,e,i,f,g,h)$ be two such functions. The Poisson bracket $\{A,B\}$ is defined by
\begin{eqnarray}\label{poiss}
\{A,B\}_I&=&CS\Bigg[
\left({\partial A\over\partial M}\right)_I
\left({\partial B\over\partial L}\right)_I+ 
\left({\partial A\over\partial g}\right)_I
\left({\partial B\over\partial G}\right)_I+ 
\left({\partial A\over\partial h}\right)_I
\left({\partial B\over\partial H}\right)_I \\
&-& 
\left({\partial A\over\partial L}\right)_I
\left({\partial B\over\partial M}\right)_I- 
\left({\partial A\over\partial G}\right)_I
\left({\partial B\over\partial g}\right)_I- 
\left({\partial A\over\partial H}\right)_I
\left({\partial B\over\partial h}\right)_I\Bigg]   \nonumber
\end{eqnarray}
where the subscript $I$ means `derivative computed for an internal term'. This has the following meaning: consider first the quantity called `equation of the center' $\phi$, defined by 
\begin{equation}\label{phi}
\phi = f-M~~.
\end{equation}
Any function $F$ (say $A$, or $B$) representing an `internal' term will depend on the Delaunay action-variables either directly, or through some combination of the quantities $(a,e,i,f,g,h)$, and their composite functions $r$,$\phi$,$n$, $\eta$,$\cos i$, $\sin i$. Then, the derivatives of the function $F$ are computed by the following formulas:
$$
\left({\partial F\over\partial M}\right)_I = 
{\partial F\over\partial M}+
{\partial F\over\partial f}{\partial f\over\partial M}+
{\partial F\over\partial r}{\partial r\over\partial M}+
{\partial F\over\partial\phi}{\partial\phi\over\partial M}
$$
$$
\left({\partial F\over\partial g}\right)_I = 
{\partial F\over\partial g}
$$
$$
\left({\partial F\over\partial h}\right)_I = 
{\partial F\over\partial h}
$$
$$
\left({\partial F\over\partial L}\right)_I = 
{\partial F\over\partial L}+
{\partial F\over\partial a}{\partial a\over\partial L}+
{\partial F\over\partial e}{\partial e\over\partial L}+
{\partial F\over\partial n_s}{\partial n_s\over\partial L}+
{\partial F\over\partial \eta}{\partial \eta\over\partial L}+
{\partial F\over\partial f}{\partial f\over\partial L}+
{\partial F\over\partial r}{\partial r\over\partial L}+
{\partial F\over\partial \phi}{\partial \phi\over\partial L}
$$
$$
\left({\partial F\over\partial G}\right)_I = 
{\partial F\over\partial G}+
{\partial F\over\partial e}{\partial e\over\partial G}+
{\partial F\over\partial \eta}{\partial \eta\over\partial G}+
{\partial F\over\partial f}{\partial f\over\partial G}+
{\partial F\over\partial r}{\partial r\over\partial G}+
{\partial F\over\partial \phi}{\partial \phi\over\partial G}+
{\partial F\over\partial \cos i}{\partial \cos i\over\partial G}+
{\partial F\over\partial \sin i}{\partial \sin i\over\partial G}
$$
$$
\left({\partial F\over\partial H}\right)_I = 
{\partial F\over\partial \cos i}{\partial \cos i\over\partial H}+
{\partial F\over\partial \sin i}{\partial \sin i\over\partial H}
$$
The intermediate derivatives have to be computed according to a set of formulas which allows for various simplifications to be carried out automatically as the normalization process moves on. Thus we have:
$$
{\partial a\over\partial L}={2\over a n_s}~~,~~
{\partial n_s\over\partial L}=-{3\over a^2}
$$
$$
{\partial e\over\partial L}={\eta^2\over a^2 e n_s}~~,~~
{\partial e\over\partial G}=-{\eta\over a^2 e n_s}
$$
$$
{\partial \eta\over\partial L}=-{\eta\over a^2 n_s}~~,~~
{\partial \eta\over\partial G}={1\over a^2 n_s}
$$
$$
{\partial \cos i\over\partial G}=-{\cos i\over a^2 \eta n_s}~~,~~
{\partial \cos i\over\partial H}={1\over a^2 \eta n_s}
$$
$$
{\partial \sin i\over\partial G}=-{1-\sin^2i\over a^2 \eta n_s \sin i}~~,~~
{\partial \sin i\over\partial H}=-{\cos i\over a^2 \eta n_s \sin i}
$$
$$
{\partial f\over\partial M}={a^2\eta\over r^2}~~,~~
{\partial f\over\partial L}=
{2\sin f\over a^2 e n_s}+{\sin 2f\over 2 a^2 n_s}~~,~~
{\partial f\over\partial G}=
-{2\sin f\over a^2 e \eta n_s}-{\sin 2f\over 2 a^2 \eta n_s}
$$
$$
{\partial \phi\over\partial M}={a^2\eta\over r^2}-1~~,~~
{\partial \phi\over\partial L}=
{2\sin f\over a^2 e n_s}+{\sin 2f\over 2 a^2 n_s}~~,~~
{\partial \phi\over\partial G}=
-{2\sin f\over a^2 e \eta n_s}-{\sin 2f\over 2 a^2 \eta n_s}
$$
$$
{\partial r\over\partial M}={a e\sin f\over \eta}~~,~~
{\partial r\over\partial L}={2r\over a^2n_s}-{\eta^2\cos f\over a e n_s}~~,~~
{\partial r\over\partial G}={\eta\cos f\over a e n_s}
$$
\subsubsection{Poisson bracket for functions involving external terms}
\label{sssec:poissonext}
Let $A(a,e,i,u,g,h)$, $B(a,e,i,u,g,h)$ be two such functions. We define
\begin{eqnarray}\label{poissb}
\{A,B\}_E&=&CS\Bigg[
\left({\partial A\over\partial M}\right)_E
\left({\partial B\over\partial L}\right)_E+ 
\left({\partial A\over\partial g}\right)_E
\left({\partial B\over\partial G}\right)_E+ 
\left({\partial A\over\partial h}\right)_E
\left({\partial B\over\partial H}\right)_E \\
&-& 
\left({\partial A\over\partial L}\right)_E
\left({\partial B\over\partial M}\right)_E- 
\left({\partial A\over\partial G}\right)_E
\left({\partial B\over\partial g}\right)_E- 
\left({\partial A\over\partial H}\right)_E
\left({\partial B\over\partial h}\right)_E\Bigg]   \nonumber
\end{eqnarray}
where the subscript $E$ means `derivative computed for an external term'. Defining the corresponding `equation of center' quantity
\begin{equation}\label{phiu}
\phi_u = u-M~~,
\end{equation}
as well as the radial distance as function of the eccentric anomaly 
$$
r_u = a(1-e\cos u)
$$
any function $F$ (say $A$, or $B$) representing an `enternal' term will depend on the Delaunay action-variables either directly, or through some combination of the quantities $(a,e,i,u,g,h)$, and their composite functions $r_u$,$\phi_u$,$n$, $\eta$,$\cos i$, $\sin i$. Then, the derivatives of the function $F$ are computed by the following formulas:
$$
\left({\partial F\over\partial M}\right)_E = 
{\partial F\over\partial M}+
{\partial F\over\partial u}{\partial u\over\partial M}+
{\partial F\over\partial r_u}{\partial r_u\over\partial M}+
{\partial F\over\partial\phi_u}{\partial\phi_u\over\partial M}
$$
$$
\left({\partial F\over\partial g}\right)_E = 
{\partial F\over\partial g}
$$
$$
\left({\partial F\over\partial h}\right)_E = 
{\partial F\over\partial h}
$$
$$
\left({\partial F\over\partial L}\right)_E = 
{\partial F\over\partial L}+
{\partial F\over\partial a}{\partial a\over\partial L}+
{\partial F\over\partial e}{\partial e\over\partial L}+
{\partial F\over\partial n_s}{\partial n_s\over\partial L}+
{\partial F\over\partial \eta}{\partial \eta\over\partial L}+
{\partial F\over\partial u}{\partial u\over\partial L}+
{\partial F\over\partial r_u}{\partial r_u\over\partial L}+
{\partial F\over\partial \phi_u}{\partial \phi_u\over\partial L}
$$
$$
\left({\partial F\over\partial G}\right)_E = 
{\partial F\over\partial G}+
{\partial F\over\partial e}{\partial e\over\partial G}+
{\partial F\over\partial \eta}{\partial \eta\over\partial G}+
{\partial F\over\partial u}{\partial u\over\partial G}+
{\partial F\over\partial r_u}{\partial r_u\over\partial G}+
{\partial F\over\partial \phi_u}{\partial \phi\over\partial G}+
{\partial F\over\partial \cos i}{\partial \cos i\over\partial G}+
{\partial F\over\partial \sin i}{\partial \sin i\over\partial G}
$$
$$
\left({\partial F\over\partial H}\right)_E = 
{\partial F\over\partial \cos i}{\partial \cos i\over\partial H}+
{\partial F\over\partial \sin i}{\partial \sin i\over\partial H}
$$
The formulas for the intermediate derivatives are the same as for internal terms. For the new quantities $u$,$\phi_u$,$r_u$, we have
$$
{\partial u\over\partial M}={a \over r_u}~~,~~
{\partial u\over\partial L}={\eta^2\sin u\over a e n_s r_u}~~,~~
{\partial u\over\partial G}=-{\eta\sin u\over a e n_s r_u}
$$
$$
{\partial \phi_u\over\partial M}={a \over r_u}-1~~,~~
{\partial \phi_u\over\partial L}={\eta^2\sin u\over a e n_s r_u}~~,~~
{\partial \phi_u\over\partial G}=-{\eta\sin u\over a e n_s r_u}
$$
$$
{\partial r_u\over\partial M}={a^2\eta\sin u \over r_u}~~,~~
{\partial r_u\over\partial L}=
{2r\over a^2n_s}-{\eta^2(-e+\cos u) \over e n_s r_u}~~,~~
{\partial r_u\over\partial G}={\eta(-e+\cos u) \over e n_s r_u}
$$

\subsection{Step 3: determination of the Lie generating functions}
\label{ssec:liegen}
The normalization of the hamiltonian involves the computation of 10 Lie generating functions by the following sequence:
\begin{itemize}
\item
Function $\chi^{(1)}, S_1=2$: elimination of the parallax for the $C_{20}$ (i.e. $J_2$) internal term
\item
Function $\chi^{(2)}, S_2=3$: elimination of the parallax for the remaining $C_{nm}$ and $S_{nm}$ internal terms
\item
Function $\chi^{(3)}, S_3=4$: elimination of the parallax for the $C_{20}^2$ (i.e. $J_2^2$) internal term
\item
Function $\chi^{(4)}, S_4=2$: Delaunay normalization of the $C_{20}$ internal term
\item
Function $\chi^{(5)}, S_5=3$: Delaunay normalization of the remaining $C_{nm}$ and $S_{nm}$ internal terms. 
\item
Function $\chi^{(6)}, S_6=4$: Delaunay normalization of the $C_{20}^2$ internal term
\item
Function $\chi^{(7)}, S_7=4$: normalization of the $n_\Moon C_{nm}$ and $n_\Moon S_{nm}$ internal relegation terms for $m\neq 0$. 
\item
Function $\chi^{(8)}, S_8=2$: Delaunay normalization of the Earth $P_2$ external term
\item
Function $\chi^{(9)}, S_9=3$: Delaunay normalization of the Earth $P_3$ external term and of the Earth $n_\Moon P_2$ external relegation term
\item
Function $\chi^{(10)}, S_{10}=4$: Delaunay normalization of the Earth $P_2^2$, Sun $P_2$ and SRP external terms and of the Earth $n_\Moon P_3$ external relegation term. 
\end{itemize}

\subsubsection{Normalization of internal terms (steps 1-7): definitions} 
\label{sssec:steps1to7def}
In the normalization of internal terms (steps 1-7 above) use is made of the following definitions:\\
\\
{\it Nucleus monomial term} is called a term of the form:
\begin{equation}
F_N = b_{m_1,m_2}(a,e,i)\small{{\cos\atop\sin}}(m_1g+m_2h)
\end{equation}
{\it  Range monomial term} is called a term of the form:
\begin{equation}
F_R = b_{k,m_1,m_2}(a,e,i)\small{{\cos\atop\sin}}(kf+m_1g+m_2h)
\end{equation}
{\it $\phi-$Nucleus monomial term} is called a term of the form:
\begin{equation}
F_{\phi N} = \phi^s b_{m_1,m_2}(a,e,i)\small{{\cos\atop\sin}}(m_1g+m_2h)
,~~~s=1,2,\ldots
\end{equation}
{\it  $\phi-$Range monomial term} is called a term of the form:
\begin{equation}
F_{\phi R} = \phi^s b_{k,m_1,m_2}(a,e,i)\small{{\cos\atop\sin}}(kf+m_1g+m_2h)
,~~~s=1,2,\ldots
\end{equation}
{\it $r-$Nucleus monomial term} is called a term of the form:
\begin{equation}
F_{rN} = {1\over r^2}b_{m_1,m_2}(a,e,i)\small{{\cos\atop\sin}}(m_1g+m_2h)
\end{equation}
{\it  $r-$Range monomial term} is called a term of the form:
\begin{equation}
F_{rR} = {1\over r^2}b_{k,m_1,m_2}(a,e,i)\small{{\cos\atop\sin}}(kf+m_1g+m_2h)
\end{equation}
{\it $r\phi$-Nucleus monomial term} is called a term of the form:
\begin{equation}
F_{r\phi N} = \phi^s{1\over r^2}b_{m_1,m_2}(a,e,i)\small{{\cos\atop\sin}}(m_1g+m_2h)
,~~~s=1,2,\ldots
\end{equation}
{\it  $r\phi$-Range monomial term} is called a term of the form:
\begin{equation}
F_{r\phi R} = 
\phi^s{1\over r^2}b_{k,m_1,m_2}(a,e,i)\small{{\cos\atop\sin}}(kf+m_1g+m_2h)
,~~~s=1,2,\ldots
\end{equation}
~\\
Of use are the following important equations, whose demonstration is straightforward:
\begin{eqnarray}\label{poissphinuc}
\{Z_0,\phi^s\cos(kg+lh)\} &= 
&n_ss\phi^{s-1}\left(1-{a^2\eta\over r^2}\right)\cos(kg+lh) \\
\{Z_0,\phi^s\sin(kg+lh)\} &= 
&n_ss\phi^{s-1}\left(1-{a^2\eta\over r^2}\right)\sin(kg+lh) \nonumber
\end{eqnarray}
\begin{eqnarray}\label{poissphirag}
\{Z_0,\phi^s\cos(kg+lh+mf)\} &= 
&n_ss\phi^{s-1}\left(1-{a^2\eta\over r^2}\right)\cos(kg+lh+mf) \nonumber\\
&+&n_s\phi^s{a^2\eta m\over r^2}\sin(kg+lh+mf)  \\
\{Z_0,\phi^s\sin(kg+lh+mf)\} &= 
&n_ss\phi^{s-1}\left(1-{a^2\eta\over r^2}\right)\sin(kg+lh+mf) \nonumber\\
&-&n_s\phi^s{a^2\eta m\over r^2}\cos(kg+lh+mf)  \nonumber
\end{eqnarray}

We also have the following rules, valid for all integers $(k,m_1,m_2)$ and useful in the computation of averages of the above functions \cite{koz1962}\cite{kel1989}: 
\begin{eqnarray}\label{aveint}
<{\cos(kf+m_1g+m_2h)\over r^2}>_M&=& 0 \nonumber\\ 
<{\sin(kf+m_1g+m_2h)\over r^2}>_M&=& 0          \\
<\cos(kf+m_1g+m_2h)>_M          &=& 
{-e^{|k|}(1+|k|\eta)\over (1+\eta)^{|k|}}\cos(m_1g+m_2h)  \nonumber\\
<\sin(kf+m_1g+m_2h)>_M          &=& 
{-e^{|k|}(1+|k|\eta)\over (1+\eta)^{|k|}}\sin(m_1g+m_2h)  \nonumber\\
<\phi>_M                        &=&<f-M>_M=0 \nonumber
\end{eqnarray}

\subsubsection{Normalization step 1: elimination of the parallax for the $J_2$ 
internal term}
\label{sssec:step1}
After P-reduction, the $J_2$ term (Eq.(\ref{j2ele})) obtains the form
\begin{eqnarray}\label{j2ele2}
CS[P[V_{C20}]]&=&{a^2n_s^2 R_\Moon^2 C_{20}\over 2 \eta^2 r^2}
\bigg(1-{3\over 2}s^2+e(1-{3\over 2}s^2)\cos(f) \\
&+&{3\over 4}es^2\cos(f+2g)+{3\over 2}s^2\cos(2f+2g)+{3\over 4}es^2\cos(3f+2g)\bigg) 
\nonumber
\end{eqnarray}
thus it contains only r-Nucleus and r-Range terms. Following \cite{dep1981}, the purpose of the elimination of the parallax process is to eliminate from the Hamiltonian the r-Range terms. To this end, we invoke the following general rules:\\
\\
{\bf Rule for the elimination of the parallax in internal terms:} The homological equation
\begin{equation}\label{homoelpar}
{\cal L}_{Z_0}\chi = {1\over r^2}b_{k,m_1,m_2}(a,e,i)\small{{\cos\atop\sin}}(kf+m_1g+m_2h)
\end{equation}
admits the solution
\begin{equation}\label{homoelparsol}
\chi = {1\over a^2\eta}{b_{k,l,m}(a,e,i)\over k n_s}\small{{~~\sin\atop -\cos}}(kf+m_1g+m_2h)
\equiv S_{HrR}[{1\over r^2}b_{k,m_1,m_2}(a,e,i)\small{{\cos\atop\sin}}(kf+m_1g+m_2h)]~~,
\end{equation}
where, we denote by $S_{HrR}[\cdot]$ the solution to the homological equation for r-Range terms. \\

We then have for the first normalization step ($r=1$): 
\begin{eqnarray}\label{defstep1}
S_1 &=& 2   \nonumber \\
h^{(0)}_2 &=& \mbox{r-Range terms of~} CS[P[V_{C20}]]\nonumber \\
X^{(0)}_2 &=& \mbox{r-Nucleus terms of~} CS[P[V_{C20}]]\\
{\cal Y}^{(0)}_2 &=& CS[RM[V_{\Earth,P2}]]\nonumber \\
\tilde{R}^{(0)}_2 &=& 0\nonumber 
\end{eqnarray}
We then readily find (from Eq.(\ref{aveint}):
\begin{equation}\label{j2elezeta}
\zeta^{(0)}_2=<h^{(0)}_2>_M=0 
\end{equation}
\begin{equation}\label{chi1homosol}
{\cal L}_{Z0}\chi^{(1)}=h^{(0)}_2-\zeta^{(0)}_2=h^{(0)}_2\Longrightarrow 
\chi^{(1)}=S_{HrR}(h^{(0)}_2)~~,
\end{equation}
that is
$$
\chi^{(1)}={n_s R_\Moon^2 C_{20}\over 2 \eta^3}
\bigg(e(1-{3\over 2}s^2)\sin(f) \\
+{3\over 4}es^2\sin(f+2g)+{3\over 4}s^2\sin(2f+2g)+{1\over 4}es^2\sin(3f+2g)\bigg)~.
$$
The 4-truncated Hamiltonian ${\cal H}^{(1)}$ reads
\begin{equation}\label{ham1nf}
\left({\cal H}^{(1)}\right)^{\leq 4}=
\left(\exp\left({\cal L}_{\epsilon^2\chi^{(1)}}\right){\cal H}^{(0)}\right)^{\leq 4}=
Z_0+\sum_{s=1}^4\epsilon^s{\cal H}^{(1)}_s 
\end{equation}
with
$$
{\cal H}^{(1)}_1={\cal H}^{(0)}_1
$$
$$
{\cal H}^{(1)}_2=X^{(0)}_2+{\cal Y}^{(0)}_2=
{n_s^2a^2 R_\Moon^2 C_{20}\over 2 \eta^2 r^2}
(1-{3\over 2}s^2)+CS[RM[V_{\Earth,P2}]]
$$
$$
H^{(1)}_3=H^{(0)}_3
$$
$$
{\cal H}^{(1)}_4={\cal H}^{(0)}_4+\{{\cal H}^{(0)}_2,\chi^{(1)}\}+{1\over 2}\{\{Z_0,\chi^{(1)}\},\chi^{(1)}\}
$$
~\\
\\
Three remarks are in order:\\
\\
- We have $\{-n_\Moon H,\chi^{(1)}\}=0$, that is, the elimination of the parallax for the $J_2$ term produces no relegation terms of book-keeping order 3. Physically, since the $J_2$ term is axi-symmetric, the terms remains invariant under rotation of the Moon. \\
\\
- The Poisson bracket $\{H^{(0)}_2,\chi^{(1)}\}$ generates two terms:
$$
\{H^{(0)}_2,\chi^{(1)}\}= \{CS[P[V_{C20}]],\chi^{(1)}\}+\{CS[RM[V_{\Earth,P2}]],\chi^{(1)}\}
$$
The second term admits no normalization in closed form. However, this term is neglected by SELENA (see normalization step 6), since it is very small for trajectories at both small and large lunicentric distances. Its omission may, however, have some influence to very eccentric trajectories (see section 5). \\
\\
- For the next step in the elimination of the parallax to be caried in closed form, it is required that the following algebraic-reduction, called \textit{$\eta-$simplification} is performed on the expression $\{CS[P[V_{C20}]],\chi^{(1)}\}+{1\over 2}\{\{Z_0,\chi^{(1)}\},\chi^{(1)}\}$:\\
\\
\textbf{$\eta-$simplification ($S_\eta[~]$):} Let $exprint$ be an expression containing only internal terms with the same coefficient $C_{nm}^{jc}$ or $S_{nm}^{js}$, with fixed exponent $jc$ or $js$. Then:\\
\\
- Compute $exprintcs=CS[exprint]$\\
\\
- Split $exprintcs$ into terms of the form $cterm*s^{is}\cos(kf+m_1g+m_2h)$ or 
$cterm*s^{is}\sin(mf+m_1g+m_2h)$ for fixed $(is,k,m_1,m_2)$. \\
\\
- For each term, compute the maximum exponent of $1/\eta$ in $cterm$, denoted by 
$E_{\eta,max}$. \\
\\
- The term has the form of a sum of NT subterms:
$$
cterm= subterm_1+subterm_2+\ldots+subterm_{NT}
$$
By D'Alembert rules, all the subterms contain powers of the eccentricity $e$ of the same parity, i.e., only even or only odd. 

- If the parity is even, compute:
$$
S_\eta[cterm]=\left({1\over\eta}\right)^{E_{\eta,max}}
\mbox{Simplify}\bigg[\eta^{E_{\eta,max}}\bigg(subterm_1+subterm_2+\ldots
+subterm_{NT}\bigg)_{e\rightarrow\sqrt{1-\eta^2}}\bigg]
$$
- If the parity is odd, compute:
$$
S_\eta[cterm]=e\left({1\over\eta}\right)^{E_{\eta,max}}
\mbox{Simplify}\bigg[{\eta^{E_{\eta,max}}\over e}\bigg(subterm_1+subterm_2+\ldots
+subterm_{NT}\bigg)_{e\rightarrow\sqrt{1-\eta^2}}\bigg]
$$
where Simplify[] means to make the algebraic reduction of common terms. \\
\\
Implementing the above procedure to the terms $\{CS[P[V_{C20}]],\chi^{(1)}\}+{1\over 2}\{\{Z_0,\chi^{(1)}\},\chi^{(1)}\}$ we thus get:
\begin{equation}\label{etaredh4}
{\cal H}^{(1)}_4={\cal H}^{(0)}_4+
S_\eta\bigg[P\bigg[
\{CS[P[V_{C20}]],\chi^{(1)}\}+{1\over 2}\{\{Z_0,\chi^{(1)}\},\chi^{(1)}\}\bigg]\bigg]
+ \tilde{R}^{(1)}_4
\end{equation}
where $\tilde{R}^{(1)}_4=\{CS[RM[V_{\Earth,P2}]],\chi^{(1)}\}$ will be omitted from further computations. 

\subsubsection{Normalization step 2: elimination of the parallax for the remaining 
internal terms $C_{nm}$, $S_{nm}$}
\label{sssec:step2}
This proceeds analogously as in the $J_2$ case. Thus we have:
\begin{eqnarray}\label{defstep2}
S_2 &=& 3   \nonumber \\
h^{(1)}_3 &=& \mbox{r-Range terms of~} CS[P[V_{Cnm}+V_{Snm}]]~\mbox{for all $(n,m)$: 
$V_{Cnm}\in H^{(1)}_3$ or $V_{Snm}\in H^{(1)}_3$}\nonumber \\
X^{(1)}_3 &=& \mbox{r-Nucleus terms of~} CS[P[V_{Cnm}+V_{Snm}]]\\
{\cal Y}^{(1)}_3 &=& CS[RM[V_{P3}]]\nonumber \\
\tilde{R}^{(1)}_3 &=& 0\nonumber 
\end{eqnarray}
\begin{equation}\label{csnmelezeta}
\zeta^{(1)}_3=<h^{(1)}_3>_M=0 
\end{equation}
\begin{equation}\label{chi2homosol}
{\cal L}_{Z0}\chi^{(2)}=h^{(1)}_3\Longrightarrow 
\chi^{(2)}=S_{HrR}(h^{(1)}_3)~~.
\end{equation}
We then obtain the Hamiltonian after the second normalization step:
\begin{equation}\label{ham2nf}
\left({\cal H}^{(2)}\right)^{\leq 4}=
\left(\exp\left({\cal L}_{\epsilon^3\chi^{(2)}}\right){\cal H}^{(1)}\right)^{\leq 4}=
Z_0+\sum_{s=1}^4\epsilon^s{\cal H}^{(2)}_s 
\end{equation}
with
$$
{\cal H}^{(2)}_1={\cal H}^{(1)}_1={\cal H}^{(0)}_1=-n_\Moon H
$$
$$
{\cal H}^{(2)}_2={\cal H}^{(1)}_2=X^{(0)}_2+{\cal Y}^{(0)}_2=
{n_s^2 a^2 R_\Moon^2 C_{20}\over 2 \eta^2 r^2}
(1-{3\over 2}s^2)+CS[RM[V_{\Earth,P2}]]
$$
$$
{\cal H}^{(3)}_3=\mbox{r-Nucleus terms of~} CS[P[V_{Cnm}+V_{Snm}]]~+CS[RM[V_{\Earth,P3}]],~~(n,m)\neq(2,0)
$$
$$
{\cal H}^{(2)}_4={\cal H}^{(0)}_4
+S_\eta[P[\{{\cal H}^{(0)}_2,\chi^{(1)}\}+{1\over 2}\{\{Z_0,\chi^{(1)}\},\chi^{(1)}\}]]
+P[\{-n_\Moon H,\chi^{(2)}\}]~~.
$$
We note the appearance, at book-keeping order 4, of the \textit{relagation term} $P[\{-n_\Moon H,\chi^{(2)}\}]$. This term gets a non-null contribution from every normalized harmonic $C_{nm}$ or $S_{nm}$ with $m\neq 0$, since these are the harmonics depending trigonometrically on the angle $h$. Physically, this expresses the fact that the sine-cosine distribution of the values of all tesseral harmonics depends on the orientation of the Moon.  

\subsubsection{Normalization step 3: elimination of the parallax for the  
internal term $J_2^2$}
\label{sssec:step3}
The term $J_2^2$ in the normalized Hamiltonian $H^{(2)}$ stems from the Poisson brackets 
$$
{\cal H}_{J_2^2}=S_\eta[P[\{H^{(0)}_2,\chi^{(1)}\}+{1\over 2}\{\{Z_0,\chi^{(1)}\},\chi^{(1)}\}]]
$$
in $H^{(2)}_4$. By construction, ${\cal H}_{J_2^2}$ contains only r-Range and r-Nucleus terms. 
Thus, the elimination of the parallax for ${\cal H}_{J_2^2}$ proceeds in the same way as at the previous steps:
\begin{eqnarray}\label{defstep3}
S_3 &=& 4   \nonumber \\
h^{(2)}_4 &=& \mbox{r-Range terms of~} {\cal H}_{J_2^2}\nonumber\\ 
X^{(2)}_4 &=& \mbox{r-Nucleus terms of~} {\cal H}_{J_2^2}+{\cal H}_{rot,xy}\\
{\cal Y}^{(2)}_4 &=& P[\{-n_\Moon H,\chi^{(2)}\}]
+CS[RM[V_{\Sun,P2}]]+CS[RM[V_{SRP}]]\nonumber \\
\tilde{R}^{(2)}_4 &=& \tilde{R}^{(1)}_4\nonumber 
\end{eqnarray}
\begin{equation}\label{j2sqelezeta}
\zeta^{(2)}_4=<h^{(2)}_4>_M=0 
\end{equation}
\begin{equation}\label{chi3homosol}
{\cal L}_{Z0}\chi^{(3)}=h^{(2)}_4\Longrightarrow 
\chi^{(3)}=S_{HrR}(h^{(2)}_4)~~.
\end{equation}
We then obtain the Hamiltonian after the third normalization step:
\begin{equation}\label{ham3nf}
\left({\cal H}^{(3)}\right)^{\leq 4}=
\left(\exp\left({\cal L}_{\epsilon^4\chi^{(3)}}\right){\cal H}^{(2)}\right)^{\leq 4}=
Z_0+\sum_{s=1}^4\epsilon^s{\cal H}^{(3)}_s 
\end{equation}
with
$$
{\cal H}^{(3)}_1={\cal H}^{(2)}_1=-n_\Moon H
$$
$$
{\cal H}^{(3)}_2={\cal H}^{(2)}_2=
{n_s^2 a^2 R_\Moon^2 C_{20}\over 2 \eta^2 r^2}
(1-{3\over 2}s^2)+CS[RM[V_{\Earth,P2}]]
$$
$$
{\cal H}^{(3)}_3={\cal H}^{(2)}_3=\mbox{r-Nucleus terms of~} CS[P[V_{Cnm}+V_{Snm}]]~+CS[RM[V_{\Earth,P3}]],~~(n,m)\neq(2,0)
$$
$$
{\cal H}^{(2)}_4=\mbox{r-Nucleus terms of}~{\cal H}_{J_2^2}
+P[\{-n_\Moon H,\chi^{(2)}\}]+CS[RM[V_{\Sun,P2}]]+CS[RM[V_{SRP}]]
+{\cal H}_{rot,xy}+\tilde{R}^{(1)}_4
$$
This concludes the process of Deprit's elimination of the parallax for all the internal terms.  

\subsubsection{Normalization step 4: Delaunay normalization for the internal r-Nucleus term $J_2$}
\label{sssec:step4}
We return to the terms of second order in book-keeping in the Hamiltonian ${\cal H}^{(3)}$. 
These contain the r-Nucleus term ${n_s^2 a^2 R_\Moon^2 C_{20}\over 2 \eta^2 r^2}
(1-{3\over 2}s^2)$. To Delaunay-normalize it we invoke the following\\
\\
{\bf Rule for the Delaunay-normalization of r-Nucleus internal terms:} The homological equation
\begin{eqnarray}\label{homodelnuk}
{\cal L}_{Z_0}\chi &=& {1\over r^2}b_{m_1,m_2}(a,e,i)\small{{\cos\atop\sin}}(m_1g+m_2h)-
<{1\over r^2}b_{m_1,m_2}(a,e,i)\small{{\cos\atop\sin}}(m_1g+m_2h)>_M \\
~&=& {1\over r^2}b_{m_1,m_2}(a,e,i)\small{{\cos\atop\sin}}(m_1g+m_2h)-
 {1\over a^2\eta}b_{m_1,m_2}(a,e,i)\small{{\cos\atop\sin}}(m_1g+m_2h) \nonumber
\end{eqnarray}
admits the solution
\begin{equation}\label{homornuksol}
\chi = {b_{m_1,m_2}(a,e,i)\over n_s a^2\eta}\small{{\cos\atop\sin}}(m_1g+m_2h)\phi
\equiv S_{HrN}[{1\over r^2}b_{m_1,m_2}(a,e,i)\small{{\cos\atop\sin}}(m_1g+m_2h)]~~
\end{equation}
where $S_{HrN}[~]$ stands for the solution of the homological equation for r-Nucleus terms. 
We then have:
\begin{eqnarray}\label{defstep4}
S_4 &=& 2   \nonumber \\
h^{(3)}_2 &=& \mbox{r-Nucleus terms in $J_2$} \nonumber\\ 
X^{(3)}_2 &=& 0\\
{\cal Y}^{(3)}_2 &=& CS[RM[V_{\Earth,P2}]]\nonumber \\
\tilde{R}^{(3)}_2 &=& 0\nonumber 
\end{eqnarray}
\begin{equation}\label{j2delzeta}
\zeta^{(3)}_2=<h^{(3)}_2>_M={n_s^2 R_\Moon^2 C_{20}\over 2 \eta^3} 
\end{equation}
\begin{equation}\label{chi3homosol}
{\cal L}_{Z0}\chi^{(4)}=h^{(3)}_4\Longrightarrow 
\chi^{(4)}=S_{HrN}[h^{(3)}_4]~~.
\end{equation}
We then obtain the Hamiltonian after the 4-th normalization step:
\begin{equation}\label{ham4nf}
\left({\cal H}^{(4)}\right)^{\leq 4}=
\left(\exp\left({\cal L}_{\epsilon^2\chi^{(4)}}\right){\cal H}^{(3)}\right)^{\leq 4}=
Z_0+\sum_{s=1}^4\epsilon^s{\cal H}^{(4)}_s 
\end{equation}
with
$$
{\cal H}^{(4)}_1={\cal H}^{(3)}_1=-n_\Moon H
$$
$$
{\cal H}^{(4)}_2=\zeta^{(3)}_2+CS[RM[V_{\Earth,P2}]]=
{n_s^2 R_\Moon^2 C_{20}\over 2 \eta^3}(1-{3\over 2}s^2)
+CS[RM[V_{\Earth,P2}]]
$$
$$
{\cal H}^{(4)}_3={\cal H}^{(3)}_3=\mbox{r-Nucleus terms of~} CS[P[V_{Cnm}+V_{Snm}]]~+CS[RM[V_{\Earth,P3}]],~~(n,m)\neq(2,0)
$$
$$
{\cal H}^{(4)}_4={\cal H}^{(3)}_4+\{{\cal H}^{(3)}_2,\chi^{(4)}\}
+{1\over 2}\{\{Z_0,\chi^{(4)}\},\chi^{(4)}\}
$$
$$
=\mbox{r-Nucleus terms of}~{\cal H}_{J_2^2}
+S_\eta\bigg[\left\{{n_s^2 a^2 R_\Moon^2 C_{20}\over 2 \eta^2 r^2}
(1-{3\over 2}s^2),\chi^{(4)}\right\}
+{1\over 2}\{\{Z_0,\chi^{(4)}\},\chi^{(4)}\}\bigg]
$$
$$
+P[\{-n_\Moon H,\chi^{(2)}\}]+CS[RM[V_{\Sun,P2}]]+CS[RM[V_{SRP}]]
+{\cal H}_{rot,xy}+\left\{CS[RM[V_{\Earth,P2}]],\chi^{(2)}+\chi^{(4)}\right\}
$$
Note that, similarly to the case  $\tilde{R}^{(2)}_4=\{CS[RM[V_{\Earth,P2}]],\chi^{(1)}\}$, here as well the term of book-keeping order 4 produced by $\{CS[RM[V_{\Earth,P2}]],\chi^{(2)}+\chi^{(4)}\}$ admits no normalization in closed form. However, due to its small value both for close and distant trajectories, this term can be included to the acceptable remainder (see error analysis in section 5). 
\subsubsection{Normalization step 5: Delaunay normalization for the internal r-Nucleus 
terms $C_{nm}$, $S_{nm}$ with $(n,m)\neq (2,0)$}
\label{sssec:step5}
Using the same rule (Eq.\ref{homodelnuk}) for the normalization of the r-Nucleus terms produced by the elimination of the parallax (step 2) we find:
\begin{eqnarray}\label{defstep5}
S_5 &=& 3   \nonumber \\
h^{(4)}_3 &=& \mbox{r-Nucleus terms of~} CS[P[V_{Cnm}+V_{Snm}]]
~\mbox{for all $(n,m)\neq(2,0)$}\nonumber \\
X^{(4)}_3 &=& 0\\
{\cal Y}^{(4)}_3 &=& CS[RM[V_{P3}]]\nonumber \\
\tilde{R}^{(4)}_3 &=& 0\nonumber 
\end{eqnarray}
\begin{equation}\label{csnmdelzeta}
\zeta^{(4)}_3=<h^{(4)}_3>_M
\end{equation}
\begin{equation}\label{chi5homosol}
{\cal L}_{Z0}\chi^{(5)}=h^{(4)}_3\Longrightarrow 
\chi^{(5)}=S_{HrN}(h^{(4)}_3)~~.
\end{equation}
We then obtain the Hamiltonian after the fifth normalization step:
\begin{equation}\label{ham5nf}
\left({\cal H}^{(5)}\right)^{\leq 4}=
\left(\exp\left({\cal L}_{\epsilon^3\chi^{(5)}}\right){\cal H}^{(4)}\right)^{\leq 4}=
Z_0+\sum_{s=1}^4\epsilon^s{\cal H}^{(5)}_s 
\end{equation}
with
$$
{\cal H}^{(5)}_1=-n_\Moon H
$$
$$
{\cal H}^{(5)}_2={\cal H}^{(4)}_2=\zeta^{(3)}_2+CS[RM[V_{\Earth,P2}]]=
{n_s^2 R_\Moon^2 C_{20}\over 2 \eta^3}(1-{3\over 2}s^2)
+CS[RM[V_{\Earth,P2}]]
$$
$$
{\cal H}^{(5)}_3=\zeta^{(4)}_3+{\cal Y}^{(4)}_3=\zeta^{(4)}_3+CS[RM[V_{\Earth,P3}]]
$$
where we recall that $\zeta{(4)}_3$ contains all averaged terms $C_{nm}$, $S_{nm}$ in closed form for $(n,m)\neq(2,0)$.
\begin{eqnarray*}
{\cal H}^{(5)}_4&=&{\cal H}^{(4)}_4+\{-n_\Moon H,\chi^{(5)}\} \\
&=&\mbox{r-Nucleus terms of}~{\cal H}_{J_2^2}
+S_\eta\bigg[\left\{{n_s^2 a^2 R_\Moon^2 C_{20}\over 2 \eta^2 r^2}
(1-{3\over 2}s^2),\chi^{(4)}\right\}
+{1\over 2}\{\{Z_0,\chi^{(4)}\},\chi^{(4)}\}\bigg]\\
&+&P[\{-n_\Moon H,\chi^{(2)}+\chi^{(5)}\}]+CS[RM[V_{\Sun,P2}]]+CS[RM[V_{SRP}]]\\
&~&~\\
&+&{\cal H}_{rot,xy}+\left\{CS[RM[V_{\Earth,P2}]],\chi^{(2)}+\chi^{(5)}\right\} 
\end{eqnarray*}
As in the second normalization step, here as well we note the appearance, at book-keeping order 4, of the \textit{relagation term} $P[\{-n_\Moon H,\chi^{(2)}+\chi^{(5)}\}]$, due to the non-invariance of the the sine-cosine distribution of the values of all tesseral harmonics with respect to the orientation of the Moon.  

\subsubsection{Normalization step 6: Delaunay normalization for the $J_2^2$ terms}
\label{sssec:step6}
After step 5, the $J_2^2$ terms produced in ${\cal H}^{(5)}_4$ are:
\begin{eqnarray}\label{j2sqdel}
{\cal H}^{(5)}_{4,J_2^2}&=&\mbox{r-Nucleus terms of}~{\cal H}_{J_2^2} \\
&+&
S_\eta\bigg[\left\{{n_s^2 a^2 R_\Moon^2 C_{20}\over 2 \eta^2 r^2}
(1-{3\over 2}s^2),\chi^{(4)}\right\}
+{1\over 2}\{\{Z_0,\chi^{(4)}\},\chi^{(4)}\}\bigg] \nonumber
\end{eqnarray}
This can be easily seen to contain Nucleus, r-Nucleus and r-Range terms. To compute the generating function which normalizes the r-Nucleus and the r-Range terms, we invoke the following\\
\\
\textbf{General rule for the normalization of second order internal terms\\
$C_{nm}^2,S_{nm}^2,C_{nm}C_{n'm'},C_{nm}S_{n',m'}$:}\\
\\
Let ${\cal H}_{2nd}$ be a function containing combinations of internal terms of the form 
$C_{nm}^2$, $S_{nm}^2$, $C_{nm}C_{n'm'}$, $C_{nm}S_{n',m'}$ as they arise by Delaunay normalization at second order after the elimination of the parallax. The function ${\cal H}_{2nd}$ can always be decomposed as a sum of Nucleus, Range, $r-$Nucleus, $r-$Range and $r\phi-$Range terms:
\begin{equation}\label{ham2nddecomp}
{\cal H}_{2nd}={\cal H}_{2nd,N}+{\cal H}_{2nd,R}
+{\cal H}_{2nd,rN}+{\cal H}_{2nd,rR}+{\cal H}_{2nd,r\phi R} ~~.
\end{equation}
~\\
To Delaunay-normalize the function ${\cal H}_{2nd}$\\
\\
1. Compute a generating function $\chi_{r\phi R}$ which normalizes ${\cal H}_{2nd,r\phi R}$ by the following\\
\\
\textbf{$\boldsymbol{r\phi R}$-Rule ($R_{r\phi R}[\cdot]$)}
$$
\mbox{for every term } 
B_{k,m_1,m_2}(a,e,i){\phi\over r^2}\sin(kf+m_1g+m_2h)
\mbox{in the Hamiltonian} 
$$
$$
\mbox{add the term } 
-B_{k,l,f}(a,e,i){\phi\over a^2 \eta k n_s}\cos(kf+m_1g+m_2h)
\mbox{in the generating function} 
$$
and
$$
\mbox{for every term } 
D_{k,g,m_1,m_2}(a,e,i){\phi\over r^2}\cos(kf+m_1g+m_2h)
\mbox{in the Hamiltonian} 
$$
$$
\mbox{add the term } 
D_{k,l,f}(a,e,i){\phi\over a^2 \eta k n_s}\sin(kf+m_1g+m_2h)
\mbox{in the generating function} 
$$
According to the previous definitions, we set:
\begin{equation}
\chi_{r\phi R}=R_{r\phi R}\left[{\cal H}_{2nd,r\phi R}\right]~~. 
\end{equation}
~\\
2. Compute
\begin{equation}\label{hphifree}
{\cal H}_{\phi-free}=\{Z_0,\chi_{r\phi R}\}+{\cal H}_{2nd,r\phi R} 
\end{equation}
The function ${\cal H}_{\phi-free}$ contains range and r-Range terms:
$$
{\cal H}_{\phi-free}={\cal H}_{\phi-free,R}+{\cal H}_{\phi-free,rR}
$$
~\\
3. Compute
\begin{equation}\label{hkillint0red}
{\cal H}_{f2r}={\cal H}_{2nd,N}+R_{f2r}\bigg[{\cal H}_{2nd,R}+
{\cal H}_{\phi-free,R}\bigg] 
\end{equation}
where the operation $R_{f2r}[\cdot]$ acts on Range-functions according to:\\
\\
\\
\textbf{$\boldsymbol{R_{f2r}}$-Rule ($R_{f2r}[\cdot]$)}: Let $F_R$ be a range function. Each term in $F_R$ has one of the forms:
$$
F_{R,c}=a(a,e,i)\cos(kf+m_1g+m_2h),~~~F_{R,s}=b(a,e,i)\sin(kf+m_1g+m_2h)
$$
Expand each term as
\begin{eqnarray*}
F_{R,c}&=&a(a,e,i)\bigg[
\left(\sum_{l=0,2l+1\leq k}(-1)^l{k\choose 2l}(\cos f)^{k-2l}(\sin f)^{2l}\right)\cos(m_1g+m_2h) \\
&~&~~~~~~~~~~
-
\left(\sum_{l=0,2l+1\leq k}(-1)^l{k\choose 2l+1}(\cos f)^{k-2l-1}(\sin f)^{2l+1}\right)\sin(m_1g+m_2h)\bigg]
\end{eqnarray*}
\begin{eqnarray*}
F_{R,s}&=&b(a,e,i)\bigg[
\left(\sum_{l=0,2l+1\leq k}(-1)^l{k\choose 2l}(\cos f)^{k-2l}(\sin f)^{2l}\right)\sin(m_1g+m_2h) \\
&~&~~~~~~~~~~
+
\left(\sum_{l=0,2l+1\leq k}(-1)^l{k\choose 2l+1}(\cos f)^{k-2l-1}(\sin f)^{2l+1}\right)\cos(m_1g+m_2h)\bigg]
\end{eqnarray*}
$$
F_{R,s}=b(a,e,i)
\left(
\cos(kf)\sin(m_1g+m_2h)+\sin(kf)\cos(m_1g+m_2h)
\right)
$$
Substitute all even powers of $\sin f$ in the above expressions according to $(\sin f)^{2l}=(1-\cos^f)^l$ and expand. Substitute $\cos f$ with $\cos f\rightarrow a\eta^2/r -1$. Reduce common factors, and trigonometrically reduce the resulting expression. \\
\\
Applying now the $R_{f2r}[\cdot]$ rule to the function ${\cal H}_{2nd,R}+{\cal H}_{\phi-free,R}$ (Eq.(\ref{hkillint0red})), we find that only Nucleus or r-Nucleus terms are produced. Then
\begin{equation}\label{calhf2r}
{\cal H}_{f2r}={\cal H}_{2nd,N}+{\cal H}_{f2r,N}+{\cal H}_{f2r,rN} 
\end{equation}
~\\
4. Finally, define 
\begin{equation}\label{hkill}
{\cal H}_{Del}= {\cal H}_{\phi-free,rR}+{\cal H}_{f2r,rN}
+{\cal H}_{2nd,rN}+{\cal H}_{2nd,rR}
\end{equation}
The function ${\cal H}_{Del}$ contains only $r-$Nucleus and $r-$Range terms. The terms $r-$Nucleus can be normalized by the $S_{HrN}$ rule (Eq.(\ref{homornuksol})), while the terms $r-$Range can be normalized with the $S_{HrR}$ rule (Eq.(\ref{homoelparsol})). Thus, the Lie generating function normalizing ${\cal H}_{2nd}$ reads:
\begin{equation}\label{chi2nd}
\chi_{2nd}=
R_{r\phi R}\left[{\cal H}_{2nd,r\phi R}\right] +
R_{HrR}\left[{\cal H}_{\phi-free,rR}+{\cal H}_{2nd,rR}\right] +
R_{HrN}\left[{\cal H}_{f2r,rN}+{\cal H}_{2nd,rN}\right]
\end{equation}
while, after the normalization, we are left only with the nucleus terms
\begin{equation}\label{nf2nd}
\{Z_0,\chi_{2nd}\}+{\cal H}_{2nd}={\cal H}_{2nd,N}+{\cal H}_{f2r,N}~~. 
\end{equation}
~\\
In the case of the $J_2^2$ terms, we find that ${\cal H}_{2nd}$ contains only Nucleus, $r-$Nucleus and $r-$Range terms. Thus, we have:
\begin{eqnarray}\label{defstep6}
S_6 &=& 4   \nonumber \\
h^{(5)}_4 &=& \mbox{r-Range and r-Nucleus terms in $J_2^2$ of~} {\cal H}^{(5)}_4\nonumber\\ 
X^{(5)}_4 &=& \mbox{Nucleus terms in $J_2^2$ of~} {\cal H}^{(5)}_4\\
{\cal Y}^{(5)}_4 &=& P[\{-n_\Moon H,\chi^{(2)}+\chi^{(5)}\}]
+{\cal H}_{rot,xy}+CS[RM[V_{\Sun,P2}]]+CS[RM[V_{SRP}]]\nonumber \\
\tilde{R}^{(5)}_4 &=&\left\{CS[RM[V_{\Earth,P2}]],\chi^{(2)}+\chi^{(4)}\right\} \nonumber 
\end{eqnarray}
\begin{equation}\label{j2sqdelzeta}
\zeta^{(6)}_4=<h^{(5)}_4>_M={r^2\over a^2\eta n_s}\times 
\mbox{r-Nucleus terms in $J_2^2$ of~} {\cal H}^{(5)}_4 
\end{equation}
\begin{eqnarray}\label{chi6homosol}
{\cal L}_{Z0}\chi^{(6)}&=&h^{(5)}_4\Longrightarrow 
\chi^{(6)}=S_{HrR}\left[\mbox{r-Range terms in $J_2^2$ of~} 
{\cal H}^{(5)}_4\right] \\
&+&S_{HrN}\left[\mbox{r-Nucleus terms in $J_2^2$ of~} {\cal H}^{(5)}_4\right]~~.\nonumber
\end{eqnarray}
We then obtain the Hamiltonian after the 6-th normalization step:
\begin{equation}\label{ham6nf}
\left({\cal H}^{(6)}\right)^{\leq 4}=
\left(\exp\left({\cal L}_{\epsilon^4\chi^{(6)}}\right){\cal H}^{(5)}\right)^{\leq 4}=
Z_0+\sum_{s=1}^4\epsilon^s{\cal H}^{(6)}_s 
\end{equation}
with
$$
{\cal H}^{(6)}_1==-n_\Moon H
$$
$$
{\cal H}^{(6)}_2={\cal H}^{(5)}_2=\zeta^{(3)}_2+CS[RM[V_{\Earth,P2}]]=
{n_s^2 R_\Moon^2 C_{20}\over 2 \eta^3}(1-{3\over 2}s^2)
+CS[RM[V_{\Earth,P2}]]
$$
$$
{\cal H}^{(6)}_3={\cal H}^{(5)}_3\zeta^{(4)}_3+{\cal Y}^{(4)}_3=\zeta^{(4)}_3+CS[RM[V_{\Earth,P3}]]
$$
\begin{eqnarray*}
{\cal H}^{(6)}_4&=&X^{(5)}_4+\zeta^{(5)}_4+
+P[\{-n_\Moon H,\chi^{(2)}+\chi^{(5)}\}]+CS[RM[V_{\Sun,P2}]]+CS[RM[V_{SRP}]]\\
&+&{\cal H}_{rot,xy}+\left\{CS[RM[V_{\Earth,P2}]],\chi^{(2)}+\chi^{(4)}\right\} \nonumber
\end{eqnarray*}
This concludes the process of Delaunay normalization of all the internal terms via Deprit's elimination of the parallax. 

\subsubsection{Normalization step 7: Normalization for the $n_\Moon C_{nm}$ and 
$n_\Moon S_{nm}$ internal relegation terms}
\label{sssec:step7}
These are the terms:
\begin{equation}\label{hamrelint}
{\cal H}_{I,rel}= P[\{-n_\Moon H,\chi^{(2)}+\chi^{(5)}\}]~~.
\end{equation} By construction, the term $P[\{-n_\Moon H,\chi^{(2)}\}]$ is Range, while the term $P[\{-n_\Moon H,\chi^{(5)}\}]$ is $\phi-$Nucleus. Expanding $\phi$ as:
\begin{eqnarray}\label{phiexp}
\phi&=&2e\sin(f)
-{3\over 4}e^2\sin(2f)
+{1\over 3}e^3\sin(3f)
-e^4\left({1\over 8}\sin(2f)+{5\over 32}\sin(4f)\right) \\
&+&e^5\left({1\over 8}\sin(3f)+{3\over 40}\sin(5f)\right)
-e^6\left({3\over 64}\sin(2f)+{3\over 32}\sin(4f)+{7\over 192}\sin(6f)\right)+O(e^7)\nonumber
\end{eqnarray}
and substituting $\phi$ by the above series in the $\phi-$Nucleus terms, we arrive at an expression ${\cal H}_{I,rel}'=\left[{\cal H}_{I,rel}\right]_{\phi\rightarrow(\ref{phiexp})}$
which contains only Range terms. The normalization of Range terms can now be achieved by the following iterative algorithm, which eliminates the need for relegation (\cite{laretal2013}):\\
\\
for a pre-selected number of iterations $j=1,\ldots,N_{rel}$ define the sequence of generating functions $\chi_{I,rel,j}$ by the following\\
\\
\textbf{Rule for the normalization of Range terms $S_{HR}[\cdot,N_{rel}]$:}\\
\\
1. Set ${\cal H}_{I,rel,1}={\cal H}_{I,rel}' $\\
\\
2. Compute $\chi_{I,rel,1}$ by the following
\\
$$
\mbox{for every term } 
B_{k,m_1,m_2}(a,e,i)\sin(kf+m_1g+m_2h) \mbox{in the Hamiltonian} 
$$
$$
\mbox{add the term } 
-B_{k,l,f}(a,e,i){1\over k n_s}\cos(kf+m_1g+m_2h)
\mbox{in the generating function} 
$$
and
$$
\mbox{for every term } 
D_{k,g,m_1,m_2}(a,e,i)\cos(kf+m_1g+m_2h)
\mbox{in the Hamiltonian} 
$$
$$
\mbox{add the term } 
D_{k,l,f}(a,e,i){1\over k n_s}\sin(kf+m_1g+m_2h)
\mbox{in the generating function} 
$$
~\\
3. Repeat the procedure for $j=2,\ldots,N_{rel}$, by setting
$$
{\cal H}_{I,rel,j}=\{Z_0,\chi_{I,rel,j-1}\}
$$
and computing $\chi_{I,rel,j}$ through the function ${\cal H}_{I,rel,j}$.\\
\\
4. Set 
\begin{equation}\label{chiirel}
\chi_{I,rel}=\sum_{j=1}^{N_{rel}}\chi_{I,rel,j}~~.
\end{equation}

One can now readily see that the function $\{Z_0,\chi_{I,rel}\}+{\cal H}_{I,rel}'$ leaves only Nucleus terms after the normalization, while it leaves a $O(e^{N_{rel}+1})$ remainder which can be neglected. We are thus led to:
\begin{eqnarray}\label{defstep7}
S_7 &=& 4   \nonumber \\
h^{(6)}_4 &=& {\cal H}_{I,rel}'\nonumber\\ 
X^{(6)}_4 &=& X^{(5)}_4+\zeta^{(5)}_4\\
{\cal Y}^{(6)}_4 &=& 
{\cal H}_{rot,xy}+CS[RM[V_{\Sun,P2}]]+CS[RM[V_{SRP}]]\nonumber \\
\tilde{R}^{(6)}_4 &=&\left\{CS[RM[V_{\Earth,P2}]],\chi^{(2)}+\chi^{(4)}\right\} \nonumber 
\end{eqnarray}
\begin{equation}\label{zeta6}
\zeta^{(6)}_4=<h^{(6)}_4>_M= \mbox{computed by Eq.(\ref{aveint})} 
\end{equation}
\begin{eqnarray}\label{chi7homosol}
{\cal L}_{Z0}\chi^{(7)}&=&h^{(6)}_4-\zeta^{(6)}_4\Longrightarrow 
\chi^{(6)}=S_{HR}\left[h^{(6)}_4\right] 
\end{eqnarray}
We then obtain the Hamiltonian after the 7-th normalization step:
\begin{equation}\label{ham7nf}
\left({\cal H}^{(7)}\right)^{\leq 4}=
\left(\exp\left({\cal L}_{\epsilon^4\chi^{(7)}}\right){\cal H}^{(6)}\right)^{\leq 4}=
Z_0+\sum_{s=1}^4\epsilon^s{\cal H}^{(7)}_s 
\end{equation}
with
$$
{\cal H}^{(7)}_1==-n_\Moon H
$$
$$
{\cal H}^{(7)}_2={\cal H}^{(5)}_2=\zeta^{(3)}_2+CS[RM[V_{\Earth,P2}]]=
{n_s^2 R_\Moon^2 C_{20}\over 2 \eta^3}(1-{3\over 2}s^2)
+CS[RM[V_{\Earth,P2}]]
$$
$$
{\cal H}^{(7)}_3={\cal H}^{(5)}_3+\zeta^{(4)}_3+{\cal Y}^{(4)}_3=\zeta^{(4)}_3+CS[RM[V_{\Earth,P3}]]
$$
\begin{eqnarray*}
{\cal H}^{(7)}_4&=&X^{(5)}_4+\zeta^{(5)}_4+\zeta^{(6)}_4
+CS[RM[V_{\Sun,P2}]]+CS[RM[V_{SRP}]]\\
&+&{\cal H}_{rot,xy}+\left\{CS[RM[V_{\Earth,P2}]],\chi^{(2)}+\chi^{(4)}\right\} \nonumber
\end{eqnarray*}
Note that $\zeta^{(7)}_4=\mbox{Kernel terms of~}(\{Z_0,\chi_{I,rel}\}+{\cal H}_{I,rel}')$.
This concludes the normalization of all perturbations induced by the Moon's internal multipole harmonic terms. 

\subsubsection{Normalization of external terms: definitions}
\label{sssec:step8to10def}
In the case of external terms, the equation of the center term $\phi_u=u-M$ resumes a simple trigonometric form $u-M=e\sin u$. This implies that, in the procedure of Delaunay normalization, there is no need to carry $\phi_u$ as an independent symbol. On the other hand, the $RM[\cdot]$ operation introduces terms of the form $(1/r){\cos\atop\sin}(ku+m_1g+m_2h)$. We thus introduce the following definitions for external terms:\\
\\
{\it Nucleus external monomial term} is called a term of the form:
\begin{equation}
F^E_N = b_{m_1,m_2}(a,e,i)\small{{\cos\atop\sin}}(m_1g+m_2h)
\end{equation}
{\it  Range external monomial term} is called a term of the form:
\begin{equation}
F^E_R = b_{k,m_1,m_2}(a,e,i)\small{{\cos\atop\sin}}(ku+m_1g+m_2h)
\end{equation}
{\it $r-$Nucleus external monomial term} is called a term of the form:
\begin{equation}
F^E_{rN} = {1\over r}b_{m_1,m_2}(a,e,i)\small{{\cos\atop\sin}}(m_1g+m_2h)
\end{equation}
{\it  $r-$Range external monomial term} is called a term of the form:
\begin{equation}
F^E_{rR} = {1\over r}b_{k,m_1,m_2}(a,e,i)\small{{\cos\atop\sin}}(ku+m_1g+m_2h)
\end{equation}

Due to the relation of the differentials $dM=r du/a=(1-e\cos u)du$, the computation of averages for external terms is straightforward:
\begin{eqnarray}\label{aveexternals}
<\cos(ku+m_1g+m_2h)>_M&=& \left\{
\begin{array}{ll}
-e/2\cos(m_1g+m_2h)&~~~\mbox{if $|k|=1$}\\
\cos(m_1g+m_2h)&~~~\mbox{if $k=0$}\\
0 &~~~\mbox{if $|k|>1$}\\
\end{array}
\right. \nonumber\\
<\sin(ku+m_1g+m_2h)>_M&=& \left\{
\begin{array}{ll}
\pm e/2\sin(m_1g+m_2h)&~~~\mbox{if $k=\pm 1$}\\
\sin(m_1g+m_2h)&~~~\mbox{if $k=0$}\\
0 &~~~\mbox{if $|k|>1$}
\\
\end{array}
\right.  \\
<{1\over r}\left({\cos\atop\sin}(ku+m_1g+m_2h)\right)>_M &=& 
\left\{
\begin{array}{ll}
{1\over a}\left({\cos\atop\sin}(m_1g+m_2h)\right)
&~~~~~~\mbox{if $k=0$}\\
0 
&~~~~~~\mbox{if $k\neq 0$}\\
\end{array}
\right.   \nonumber
\end{eqnarray}

\subsubsection{Normalization step 8: Delaunay normalization of the Earth tide's $P_2$ 
external terms}
\label{sssec:step8}
We have:
\begin{eqnarray}\label{defstep8}
S_8 &=& 2   \nonumber \\
h^{(7)}_2 &=& CS[RM[V_{\Earth,P2}]]\nonumber\\ 
X^{(7)}_2 &=& {n^2 R_\Moon^2 C_{20}\over 2 \eta^3}(1-{3\over 2}s^2)\\
{\cal Y}^{(7)}_2&=&0 \nonumber\\
\tilde{R}^{(7)}_2 &=&0 \nonumber 
\end{eqnarray}
~\\
We easily verify that, by construction, $h^{(7)}_2 = CS[RM[V_{\Earth,P2}]]$ contains $r-$Nucleus and $r-$Range terms. 
\begin{equation}\label{h72}
h^{(7)}_2=h^{(7)}_{2,rN}+h^{(7)}_{2,rR}
\end{equation}
\\
Using Eq.(\ref{aveexternals}), we compute the averages
\begin{equation}\label{zeta7}
\zeta^{(7)}_2=<h^{(7)}_{2,rN}>_M+<h^{(7)}_{2,rR}>_M= \mbox{computed by Eq.(\ref{aveexternals})} 
\end{equation}

We can now solve the homological equation for the generating function $\chi^{(8)}$ using the following rules:\\
\\
\\
\textbf{Normalization of r-Range external terms: rule $S_{HrR}^E[\cdot]$}:\\
$$
\mbox{for every term } 
{1\over r}B_{k,m_1,m_2}(a,e,i)\sin(ku+m_1g+m_2h) \mbox{in the Hamiltonian} 
$$
$$
\mbox{add the term } 
-{1\over a}B_{k,l,f}(a,e,i){1\over k n_s}\cos(ku+m_1g+m_2h)
\mbox{in the generating function} 
$$
and
$$
\mbox{for every term } 
{1\over r}D_{k,g,m_1,m_2}(a,e,i)\cos(ku+m_1g+m_2h)
\mbox{in the Hamiltonian} 
$$
$$
\mbox{add the term } 
{1\over a}D_{k,l,f}(a,e,i){1\over k n_s}\sin(ku+m_1g+m_2h)
\mbox{in the generating function} 
$$\\
\\
\textbf{Normalization of r-Nucleus external terms: rule $S_{HrN}^E[\cdot]$}:\\
$$
\mbox{for every term } 
{1\over r}B_{k,m_1,m_2}(a,e,i)\sin(m_1g+m_2h) \mbox{in the Hamiltonian} 
$$
$$
\mbox{add the term } 
{\phi_u\over a n_s}B_{k,l,f}(a,e,i)\sin(ku+m_1g+m_2h)
\mbox{in the generating function} 
$$
and
$$
\mbox{for every term } 
{1\over r}D_{k,g,m_1,m_2}(a,e,i)\cos(m_1g+m_2h)
\mbox{in the Hamiltonian} 
$$
$$
\mbox{add the term } 
{\phi_u\over a n_s}D_{k,l,f}(a,e,i)\cos(ku+m_1g+m_2h)
\mbox{in the generating function} 
$$\\
\\
Using the above rules, we have:
\begin{eqnarray}\label{chi8homosol}
{\cal L}_{Z0}\chi^{(8)}&=&h^{(7)}_2-\zeta^{(7)}_2\Longrightarrow 
\chi^{(8)}=
 S_{HrR}^E\left[h^{(7)}_{2,rR}\right] 
+S_{HrN}^E\left[h^{(7)}_{2,rN}\right] ~~.
\end{eqnarray}
We then obtain the Hamiltonian after the 8-th normalization step:
\begin{equation}\label{ham8nf}
\left({\cal H}^{(8)}\right)^{\leq 4}=
\left(\exp\left({\cal L}_{\epsilon^2\chi^{(8)}}\right){\cal H}^{(7)}\right)^{\leq 4}=
Z_0+\sum_{s=1}^4\epsilon^s{\cal H}^{(8)}_s 
\end{equation}
with
$$
{\cal H}^{(8)}_1=-n_\Moon H
$$
$$
{\cal H}^{(8)}_2=\zeta^{(3)}_2+\zeta^{(7)}_2=
{n_s^2 R_\Moon^2 C_{20}\over 2 \eta^3}(1-{3\over 2}s^2)
+<CS[RM[V_{\Earth,P2}]]>_M
$$
$$
{\cal H}^{(8)}_3={\cal H}^{(7)}_3+\{-n_\Moon H,\chi^{(8)}\}=
\zeta^{(4)}_3+CS[RM[V_{\Earth,P3}]]+CS[RM[\{-n_\Moon H,\chi^{(8)}\}]]
$$
\begin{eqnarray*}
{\cal H}^{(8)}_4&=&{\cal H}^{(7)}_4
+RM[\{CS[V_{\Earth,P2}],\chi^{(8)}\}]+CS[RM[{1\over 2}\{\{Z_0,\chi^{(8)}\},\chi^{(8)}\}]] 
\nonumber\\
&=&\zeta^{(5)}_4+\zeta^{(6)}_4+CS[RM[V_{\Sun,P2}]]+CS[RM[V_{SRP}]] +{\cal H}_{rot,xy}\\
&+&RM[\{CS[V_{\Earth,P2}],\chi^{(8)}\}]+CS[RM[{1\over 2}\{\{Z_0,\chi^{(8)}\},\chi^{(8)}\}]]
\nonumber\\
&+&\left\{CS[RM[V_{\Earth,P2}]],\chi^{(2)}+\chi^{(4)}\right\} \nonumber
\end{eqnarray*}
We note the appearance of:\\
\\
- the relegation term $CS[RM[\{-n_\Moon H,\chi^{(8)}\}]]$ at book-keeping order 3. Physically, this term expresses the dependence of the forces on the angle formed between the long-axis of the ellipsoid of the Earth's tidal equipotentials and the $x-$axis of the PALRF frame. This angle is not constant and equal to zero, since it has librations proportional to the eccentricity of the Moon's geocentric orbit. Note also the operation $CS[RM[\cdot]]$ imposed on the relegation term, in order to render this term of the normalizable form $r-$Nucleus or $r-$range. \\
\\
- The $P_2^2$ term $RM[\{CS[V_{\Earth,P2}],\chi^{(8)}\}]+CS[RM[{1\over 2}\{\{Z_0,\chi^{(8)}\},\chi^{(8)}\}]]$, which expresses corrections due to influences in second order of perturbation theory of the Earth's quadrupolar tidal force. Note, now, that in order to render this term of the normalizable form $r-$Nucleus or $r-$range we revert the order of the CS- and RM-operations, i.e., we impose the operation $RM[CS[\cdot]]$ to the Poisson bracket $\{CS[V_{\Earth,P2}],\chi^{(8)}\}$. 

\subsubsection{Normalization step 9: Delaunay normalization of the relegation term 
$n_\Moon P_2$ and of the Earth's $P_3$ external term}
\label{sssec:step9}
This proceeds analogously as for the $P_2$ case:
\begin{eqnarray}\label{defstep9}
S_9 &=& 3   \nonumber \\
h^{(8)}_3 &=& CS[RM[V_{\Earth,P3}]]+RM[CS[\{-n_\Moon H,\chi^{(8)}\}]]\nonumber\\ 
X^{(8)}_3 &=& \zeta^{(4)}_3\\
{\cal Y}^{(8)}_3&=&0 \nonumber\\
\tilde{R}^{(8)}_3 &=&0 \nonumber 
\end{eqnarray}
It is easy to verify that 
\begin{equation}\label{avep2rel}
<RM[CS[\{-n_\Moon H,\chi^{(8)}\}]]>_M=0
\end{equation}
thus, only the average terms $<CS[RM[V_{\Earth,P3}]]>_M$ survive in $\zeta^{(8)}_3$, i.e.,  
\begin{equation}\label{zeta8}
\zeta^{(8)}_3=<h^{(8)}_{3,rN}>_M+<h^{(8)}_{3,rR}>_M=<CS[RM[V_{\Earth,P3}]]>_M~~. 
\end{equation}
Then
\begin{eqnarray}\label{chi9homosol}
{\cal L}_{Z0}\chi^{(9)}&=&h^{(8)}_3-\zeta^{(8)}_3\Longrightarrow 
\chi^{(9)}=
S_{HrR}^E\left[h^{(8)}_{3,rR}\right] 
S_{HrN}^E\left[h^{(8)}_{3,rN}\right] ~~.
\end{eqnarray}
We then obtain the Hamiltonian after the 9-th normalization step:
\begin{equation}\label{ham9nf}
\left({\cal H}^{(9)}\right)^{\leq 4}=
\left(\exp\left({\cal L}_{\epsilon^3\chi^{(9)}}\right){\cal H}^{(8)}\right)^{\leq 4}=
Z_0+\sum_{s=1}^4\epsilon^s{\cal H}^{(9)}_s 
\end{equation}
with
$$
{\cal H}^{(9)}_1=-n_\Moon H
$$
$$
{\cal H}^{(9)}_2={n_s^2 R_\Moon^2 C_{20}\over 2 \eta^3}(1-{3\over 2}s^2)
+<CS[RM[V_{\Earth,P2}]]>_M
$$
$$
{\cal H}^{(9)}_3=
\zeta^{(4)}_3+<CS[RM[V_{\Earth,P3}]]>_M
$$
\begin{eqnarray*}
{\cal H}^{(9)}_4
&=&
\zeta^{(5)}_4+\zeta^{(6)}_4+CS[RM[V_{\Sun,P2}]]+CS[RM[V_{SRP}]] +{\cal H}_{rot,xy}\nonumber\\
&+&
RM[\{CS[V_{\Earth,P2}],\chi^{(8)}\}]+CS[RM[{1\over 2}\{\{Z_0,\chi^{(8)}\},\chi^{(8)}\}]]\\
&+&
CS[RM[\{-n_\Moon H,\chi^{(9)}\}]]\\
&+&
\left\{CS[RM[V_{\Earth,P2}]],\chi^{(2)}+\chi^{(4)}\right\} \nonumber
\end{eqnarray*}

We now note the appearance of the new relegation term $CS[RM[\{-n_\Moon H,\chi^{(9)}\}]]$. However, due to the form of the function $h^{(8)}$ in Eq.(\ref{defstep9}), the generating function $\chi^{(9)}$ can be decomposed as:
\begin{equation}\label{chi9decomp}
\chi^{(9)}=\chi^{(9)}_{P3}+\chi^{(9)}_{n_\Moon^2 P2} 
\end{equation}
where $\chi^{(9)}_{P3}=O({\cal G}M_\Earth)$ while $\chi^{(9)}_{n_\Moon P2}=O(n_\Moon^2 {\cal G}M_\Earth)$. Thus, the term $CS[RM[\{-n_\Moon H,\chi^{(9)}\}]]$ is the sum of a term $CS[RM[\{-n_\Moon H,\chi^{(9)}_{P3}\}]]$, which reflects the dependence of the normalized Hamiltonian on the angle of the $P3$ equipotentials' principal axis with respect to the $x-$axis of the PALRF frame, and a higher order relegation term $CS[RM[\{-n_\Moon H,\chi^{(9)}_{n_\Moon^2 P2}\}]]$, which has negligible size and can be omitted from the averaged form of the Hamiltonian.  

\subsubsection{Normalization step 10: Delaunay normalization of the relegation term 
$n_\Moon P_3$, the Earth's $P_2^2$, the Sun's $P_2$ and SRP external terms}
\label{sssec:step10}
\begin{eqnarray}\label{defstep10}
S_{10} &=& 4   \nonumber \\
h^{(9)}_4 &=& CS[RM[V_{\Sun,P2}]]+CS[RM[V_{SRP}]]+RM[CS[\{-n_\Moon H,\chi^{(9)}_{P3}\}]]
\nonumber\\ 
&+&RM[\{CS[V_{\Earth,P2}],\chi^{(8)}\}]+CS[RM[{1\over 2}\{\{Z_0,\chi^{(8)}\},\chi^{(8)}\}]]\\
X^{(9)}_4 &=& \zeta^{(5)}_4+\zeta^{(6)}_4+{\cal H}_{rot,xy}\\
{\cal Y}^{(9)}_3&=&0 \nonumber\\
\tilde{R}^{(9)}_4 &=&\left\{CS[RM[V_{\Earth,P2}]],\chi^{(2)}+\chi^{(4)}\right\} 
+CS[RM[\{-n_\Moon H,\chi^{(9)}_{n_\Moon^2 P2}\}]]\nonumber 
\end{eqnarray}
We have again $<RM[CS[\{-n_\Moon H,\chi^{(9)}_{P3}\}]]>=0$ leading to 
\begin{eqnarray}\label{zeta9}
\zeta^{(9)}_4&=&<h^{(9)}_{4,rN}>_M+<h^{(9)}_{4,rR}>_M \\
             &=& <CS[RM[V_{\Sun,P2}]]>_M
+<CS[RM[V_{SRP}]]>_M=\mbox{computed by Eq.(\ref{aveexternals})} \nonumber
\end{eqnarray}
\begin{eqnarray}\label{chi10homosol}
{\cal L}_{Z0}\chi^{(10)}&=&h^{(9)}_4-\zeta^{(9)}_4\Longrightarrow 
\chi^{(10)}=
S_{HrR}^E\left[h^{(9)}_{4,rR}\right] 
S_{HrN}^E\left[h^{(9)}_{4,rN}\right] ~~.
\end{eqnarray}
\begin{equation}\label{ham10nf}
\left({\cal H}^{(10)}\right)^{\leq 4}=
\left(\exp\left({\cal L}_{\epsilon^4\chi^{(10)}}\right){\cal H}^{(9)}\right)^{\leq 4}=
Z_0+\sum_{s=1}^4\epsilon^s{\cal H}^{(10)}_s 
\end{equation}
with
$$
{\cal H}^{(10)}_1=-n_\Moon H
$$
$$
{\cal H}^{(10)}_2={n_s^2 R_\Moon^2 C_{20}\over 2 \eta^3}(1-{3\over 2}s^2)
+<CS[RM[V_{\Earth,P2}]]>_M
$$
$$
{\cal H}^{(10)}_3=
\zeta^{(4)}_3+<CS[RM[V_{\Earth,P3}]]>_M+<CS[RM[\{-n_\Moon H,\chi^{(8)}\}]]>_M
$$
\begin{eqnarray*}
{\cal H}^{(10)}_4
&=&
\zeta^{(5)}_4+\zeta^{(6)}_4+<CS[RM[V_{\Sun,P2}]]>_M+
<CS[RM[V_{SRP}]]>_M +{\cal H}_{rot,xy}\nonumber\\
&+&
<RM[\{CS[V_{\Earth,P2}],\chi^{(8)}\}]>_M+<CS[RM[{1\over 2}\{\{Z_0,\chi^{(8)}\},\chi^{(8)}\}]]>_M\\
&+&
\left\{CS[RM[V_{\Earth,P2}]],\chi^{(2)}+\chi^{(4)}\right\} \nonumber
\end{eqnarray*}

\subsection{Semi-analytical theory: final (secular) Hamiltonian}
\label{ssec:hamsecfinal}
Following Eq.(\ref{ham10nf}),  all the normalizations, the Hamiltonian is brought to a secular (free from fast-periodic terms) normal form, except for the small remainder $\left\{CS[RM[V_{\Earth,P2}]],\chi^{(2)}+\chi^{(4)}\right\}$ which will be neglected. Thus, droping now the book-keeping symbol $\epsilon=1$, the final secular model is:
\begin{equation}\label{hamsecfinal}
Z=Z_0+Z_1+Z_2+Z_3+Z_4 
\end{equation}
with
$$
Z_1=-n_\Moon H
$$
$$
Z_2={n_s^2 R_\Moon^2 C_{20}\over 2 \eta^3}(1-{3\over 2}s^2)
+<CS[RM[V_{\Earth,P2}]]>_M
$$
$$
Z_3=
\zeta^{(4)}_3+<CS[RM[V_{\Earth,P3}]]>_M
$$
\begin{eqnarray*}
Z_4
&=&
\zeta^{(5)}_4+\zeta^{(6)}_4+<CS[RM[V_{\Sun,P2}]]>_M+
<CS[RM[V_{SRP}]]>_M +{\cal H}_{rot,xy}\nonumber\\
&+&
<RM[\{CS[V_{\Earth,P2}],\chi^{(8)}\}]>_M
+<CS[RM[{1\over 2}\{\{Z_0,\chi^{(8)}\},\chi^{(8)}\}]]>_M\nonumber~~.
\end{eqnarray*}
The meaning of each term in this model is as follows:
\begin{eqnarray*}
Z_0&=&~\mbox{Keplerian}\\
~&~&~\\
Z_{rot,z}&=&-n_\Moon H=\mbox{non-inertial (Moon's principal rotation)}\\
~&~&~\\
Z_{J_2}&=&{n_s^2 R_\Moon^2 C_{20}\over 2 \eta^3}(1-{3\over 2}s^2)=\mbox{Averaged $J_2$ (closed form)}\\
~&~&~\\
Z_{mn}&=&\zeta^{(4)}_3=~\mbox{Averaged Moon's internal terms $C_{nm}$, $S_{nm}$ for $(n,m)\neq (2,0)$ (closed form)}\\
~&~&~\\
Z_{J_2^2}&=&\zeta^{(5)}_4=\mbox{Averaged $J_2^2$ (closed form)}\\
~&~&~\\
Z_{nm,rel}&=&\zeta^{(6)}_4=\mbox{Averaged relegation of tesseral harmonics (truncated series)}\\
~&~&~\\
Z_{rot,xy}&=&{\cal H}_{rot,xy}=\mbox{Non-inertial (components of the angular velocity $\omega_x$, $\omega_y$)}\\
~&~&~\\
Z_{\Earth,P_2}&=&<CS[RM[V_{\Earth,P2}]]>_M=\mbox{Averaged Earth's $P_2$ (closed form)}\\
~&~&~\\
Z_{\Earth,P_3}&=&<CS[RM[V_{\Earth,P3}]]>_M=\mbox{Averaged Earth's $P_3$ (closed form)}\\
~&~&~\\
Z_{\Earth,P_2^2}&=&<RM[\{CS[V_{\Earth,P2}],\chi^{(8)}\}]>_M+<CS[RM[{1\over 2}\{\{Z_0,\chi^{(8)}\},\chi^{(8)}\}]]>_M\\
~&=&\mbox{Averaged Earth's $P_2^2$ (closed form)}\\
~&~&~\\
Z_{\Sun,P_2}&=&<CS[RM[V_{\Sun,P2}]]>_M=\mbox{Averaged Sun's $P_2$ (closed form)}\\
~&~&~\\
Z_{SRP}&=&<CS[RM[V_{SRP}]]>_M=\mbox{Averaged SRP (closed form)}\\
\end{eqnarray*}

\subsection{Semi-analytical theory: equations of motion and transformations}
\label{ssec:eqmotrafinal}

\subsubsection{Computation of the transformations: exchange theorem}
\label{sssec:exchange}
One of the properties of the transformations defined by the composition of Lie series (Eqs.\ref{qptra} and \ref{qptrainv}) is determined by the \\
\\
\textbf{Exchange theorem:} for any composite function $f(\mathbf{q}(\mathbf{Q},\mathbf{P}),\mathbf{p}(\mathbf{Q},\mathbf{P}))$, where the transformations $\mathbf{q}(\mathbf{Q},\mathbf{P})$, $\mathbf{p}(\mathbf{Q},\mathbf{P})$ are given by Eq.(\ref{qptra}) the following identities hold
\begin{eqnarray}\label{exchange}
\left(f(\mathbf{q}(\mathbf{Q},\mathbf{P})
,\mathbf{p}(\mathbf{Q},\mathbf{P}))\right)^{\leq N}
&=&
\bigg[\bigg(
\exp\left({\cal L}_{\epsilon^{S_K}\chi^{(K)}}\right)
\circ
\exp\left({\cal L}_{\epsilon^{S_{K-1}}\chi^{(K-1)}}\right)
\circ
\ldots \nonumber\\
&~&
~~~~~\ldots\circ\exp\left({\cal L}_{\epsilon^{S_{1}}\chi^{(1)}}\right)
\bigg)_{(\mathbf{Q},\mathbf{P})}f(\mathbf{Q},\mathbf{P})\bigg]^{\leq N}~~\\
\left(f(\mathbf{Q}(\mathbf{q},\mathbf{p})
,\mathbf{P}(\mathbf{q},\mathbf{p}))\right)^{\leq N}
&=&
\bigg[\bigg(
\exp\left(-{\cal L}_{\epsilon^{S_1}\chi^{(1)}}\right)
\circ
\exp\left(-{\cal L}_{\epsilon^{S_{2}}\chi^{(2)}}\right)
\circ
\ldots \nonumber\\
&~&
~~~~~\ldots\circ\exp\left(-{\cal L}_{\epsilon^{S_{K}}\chi^{(K)}}\right)
\bigg)_{(\mathbf{q},\mathbf{p})}f(\mathbf{q},\mathbf{p})\bigg]^{\leq N}~~\nonumber
\end{eqnarray}
In words: to compute an N-truncated Lie series representation of the composision $f(\mathbf{q}(\mathbf{Q},\mathbf{P}),$ $\mathbf{p}(\mathbf{Q},\mathbf{P}))$, act with with the composition of the Lie series directly on the function $f(\mathbf{Q},\mathbf{P})$. 

\subsubsection{Decomposition of the transformations: mid-way elements}
\label{sssec:midway}
Implementing the exchange theorem with $K=10$, $\chi^{(1)}$ to $\chi^{(10)}$ computed as in the steps 1-10 above, and with the book-keeping truncation order $N=4$, allows to compute directly the transformation between any function of the form $W(z_{mean})$ where $z_{mean}$ is the osculating element state vector, and the corresponding formula for the function $w(z_mean)=W(z(z_{mean}))$, where $z$ is the osculating element state vector, and vice versa. A key point in SELENA's numerical implementation of the above formulas is that the number of terms to be computed is reduced dramatically by splitting the process in two stages, i.e., one connecting osculating elements with some \textit{mid-way} elements, reflecting only the transformation up to the elimination of the parallax of the internal terms, and another connecting the mid-way elements to mean elements. We have the following forward and backward transformation formulas: \\
\\
\textbf{Rule $z=F(z_{mean})$: tranformation from mean to osculating elements:}\\
\\
Compute
\begin{eqnarray}\label{trameantomid}
w_{mid}(z_{mean})
&=&
S_\eta\bigg[CS\bigg[W(z_{mean}) \\
&+&
\bigg(\{W,\chi^{(1)}\}+{1\over 2}\{\{W,\chi^{(1)}\},\chi^{(1)}\}
+\{W,\chi^{(2)}\}+\{W,\chi^{(3)}\}\bigg)_{z_{mean}}\bigg]\bigg] \nonumber
\end{eqnarray}
\begin{eqnarray}\label{tramdidtoosc}
w(z(z_{mid}))
&=&
S_\eta\bigg[CS\bigg[W(z_{mid}) \\
&+&
\bigg(\{W,\chi^{(4)}\}+{1\over 2}\{\{W,\chi^{(4)}\},\chi^{(4)}\}
+\{W,\chi^{(5)}\}+\{W,\chi^{(6)}\} \nonumber\\
&+&
\{W,\chi^{(7)}\}+\{W,\chi^{(8)}\}+{1\over 2}\{\{W,\chi^{(8)}\},\chi^{(8)}\}\nonumber\\
&+&
\{W,\chi^{(9)}\}+\{W,\chi^{(10)}\}\bigg)_{z_{mid}}\bigg]\bigg] \nonumber
\end{eqnarray}
\\
\textbf{Rule $z_{mean}=F^{-1}(z)$: tranformation from osculating to mean elements:}\\
\\
Compute
\begin{eqnarray}\label{traosctomid}
w_{mid}(z)
&=&
S_\eta\bigg[CS\bigg[W(z) \\
&+&
\bigg(-\{W,\chi^{(4)}\}+{1\over 2}\{\{W,\chi^{(4)}\},\chi^{(4)}\}
-\{W,\chi^{(5)}\}-\{W,\chi^{(6)}\} \nonumber\\
&-&
\{W,\chi^{(7)}\}-\{W,\chi^{(8)}\}+{1\over 2}\{\{W,\chi^{(8)}\},\chi^{(8)}\} \nonumber\\
&-&
\{W,\chi^{(9)}\}-\{W,\chi^{(10)}\}\bigg)_{z}\bigg]\bigg] \nonumber
\end{eqnarray}
\begin{eqnarray}\label{tramidtomean}
w(z_{mean})
&=&
S_\eta\bigg[CS\bigg[W(z_{mid}) \\
&+&
\bigg(-\{W,\chi^{(1)}\}+{1\over 2}\{\{W,\chi^{(1)}\},\chi^{(1)}\}
-\{W,\chi^{(2)}\}-\{W,\chi^{(3)}\}\bigg)_{z_{mid}}\bigg]\bigg] \nonumber
\end{eqnarray}

\subsubsection{Choice of integration variables: modified equinoctial elements}
\label{sssec:equinoctial}
In order to avoid apparent singularities in the equations of motion, related to a trajectory crossing the Moon's equator or polar meridian, we adopt the following six different forms of the function $W$, representing combinations of the Keplerian element state vector $z\equiv(a,e,i,M,g,h)$:\\
\\
\textbf{Modified equinoctial elements for internal terms:}
\begin{eqnarray}\label{equieleint}
W_1(a,e,i,\lambda_f,g,h)&=&~~a \nonumber\\
W_2(a,e,i,\lambda_f,g,h)&=&~~\lambda=M+g+h \nonumber\\
W_3(a,e,i,\lambda_f,g,h)&=&h_{eq}=e\cos(\varpi)=e\cos(g+h) \\
W_4(a,e,i,\lambda_f,g,h)&=&k_{eq}=e\sin(\varpi)=e\sin(g+h) \nonumber\\
W_5(a,e,i,\lambda_f,g,h)&=&q_{eq}=\sin(i/2)\cos(h)  \nonumber\\
W_6(a,e,i,\lambda_f,g,h)&=&p_{eq}=\sin(i/2)\sin(h)  \nonumber
\end{eqnarray}
where $\lambda_f=f+g+h$ is the true longitude.\\
\\
\textbf{Modified equinoctial elements for external terms:}
\begin{eqnarray}\label{equieleext}
W_1(a,e,i,\lambda_u,g,h)&=&~~a \nonumber\\
W_2(a,e,i,\lambda_u,g,h)&=&~~\lambda=M+g+h \nonumber\\
W_3(a,e,i,\lambda_u,g,h)&=&h_{eq}=e\cos(\varpi)=e\cos(g+h) \\
W_4(a,e,i,\lambda_u,g,h)&=&k_{eq}=e\sin(\varpi)=e\sin(g+h) \nonumber\\
W_5(a,e,i,\lambda_u,g,h)&=&q_{eq}=\sin(i/2)\cos(h)  \nonumber\\
W_6(a,e,i,\lambda_u,g,h)&=&p_{eq}=\sin(i/2)\sin(h)  \nonumber
\end{eqnarray}
where $\lambda_u=u+g+h$ is the eccentric longitude. \\
\\
Note that the definition of the modified equinoctial elements $(a,\lambda,h_{eq},k_{eq},p_{eq},q_{eq})$ is unique and the same throughout all terms appearing in the equations of motion or the transformations computed by the SELENA propagator. However, what changes between internal and external terms is the way that the functional dependence of these elements on the Keplerian elements is represented in internal or external terms. In particular, Hamilton's equations of motion for the mean equinoctial elements are computed using the Poisson algebra of Eqs.(\ref{poiss}) and (\ref{poissb}), by directly computing the Poisson brackets:
\begin{eqnarray}\label{eqmomean}
{dM_{mean}\over dt}
&=&
\bigg\{M_{mean},Z(a_{mean},e_{mean},i_{mean},g_{mean},h_{mean})\bigg\} 
\nonumber\\
{da_{mean}\over dt}
&=&
\bigg\{a_{mean},Z(a_{mean},e_{mean},i_{mean},g_{mean},h_{mean})\bigg\}=0 
\nonumber\\
{dh_{eq,mean}\over dt}
&=&
\bigg\{e_{mean}\cos(g_{mean}+h_{mean}),Z(a_{mean},e_{mean},i_{mean},g_{mean},h_{mean})\bigg\} 
\\
{dk_{eq,mean}\over dt}
&=&
\bigg\{e_{mean}\sin(g_{mean}+h_{mean}),Z(a_{mean},e_{mean},i_{mean},g_{mean},h_{mean})\bigg\}         
\nonumber\\
{dq_{eq,mean}\over dt}
&=&
\bigg\{\sqrt{(\cos(i_{mean})-1)/2}\cos(h_{mean}),Z(a_{mean},e_{mean},i_{mean},g_{mean},h_{mean})\bigg\}         
\nonumber\\
{dp_{eq,mean}\over dt}
&=&
\bigg\{\sqrt{(\cos(i_{mean})-1)/2}\sin(h_{mean}),Z(a_{mean},e_{mean},i_{mean},g_{mean},h_{mean})\bigg\}
\nonumber
\end{eqnarray}
with $Z$ given by Eq.(\ref{hamsecfinal}). The equations of motion (\ref{eqmomean}), with all algebraic reduction rules and simplifications adopted so far, may exhibit apparent singularities due to the presence of hidden factors of the form $(1-\eta^2)/e=e$. Such singularities can be removed by implementing the following rule to an expression $expr$ exhibiting apparent $(1/e)$ terms:\\
\\
\textbf{`$e^2-$simplify' ($S_{e2}[expr]$) rule for the simplification of factors $(1-\eta^2)/e$:}\\
\\
1. Decompose an expression $CS[expr]$ of internal terms in subterms of the form
\begin{equation}\label{sube2int}
subterm_I=factor_I P_I(\eta) e^{\ell_1}(\sin i)^{\ell_2}(\cos i){\cos\atop\sin}(kf+m_1g+m_h) 
\end{equation}
where: $factor_I$ can be one of $C_{20}$,$C_{20}^2$, $C_{nm}$, $S_{nm}$, $n_\Moon C_{nm}$, $n_{Moon}S_{nm}$ (with $(n,m)\neq (2,0)$), and $P_I(\eta)$ is a polynomial in $\eta$. Decompose an expression $expr$ of external terms in subterms of the form
\begin{equation}\label{sube2ext}
subterm_E=factor_E P_I(\eta) e^{\ell_1}(\sin i)^{\ell_2}(\cos i){\cos\atop\sin}(ku+m_1g+m_h) 
\end{equation}
where: $factor_E$ can be one of ${\cal G}M_{\Earth}$, ${\cal G}^2M_{\Earth}^2$, $n_\Moon{\cal G}M_{\Earth}$, ${\cal G}M_{\Sun}$, $n_\Moon{\cal G}M_{\Sun}$, and $P_I(\eta)$ is a polynomial in $\eta$. Note that, in both cases above, the survival of only the first power of $\cos i$ in all expressions is guaranteed by the implementation of the CS-rule before the identification of the subterms.  \\
\\
2. Use Horner's scheme to factorize $P_I(\eta)$ and $P_E(\eta)$ according to
$$
P_I(\eta)=(1-\eta^2)^{\ell_\eta}P_{I,rest}(\eta)=e^{\ell_\eta}P_{I,rest}(\eta),~~~
$$
$$
P_E(\eta)=(1-\eta^2)^{\ell_\eta}P_{E,rest}(\eta)=e^{\ell_\eta}P_{E,rest}(\eta)~~.
$$
~\\
\\
3. Control that $\ell_\eta+\ell_1\geq 0$ (this is required by the first D'Alambert rule on the symmetry of the equations of motion with respect to rotations of all bodies in space). In the polynomials $P_{I,rest}$, $P_{E,rest}$, substitute all even powers of $\eta$ with $\eta^{2\ell}=(1-e^2)^\ell$ and all odd powers with $\eta^{2\ell+1}=(1-e^2)^\ell\eta$, and expand. The resulting expressions are polynomials of the form $P(e)$ or $\eta P(e)$, where $P(e)$ is polynomial in the eccentricity $e$. \\
\\
4. make the substitutions:
$$
\cos i\rightarrow(1-2s_{1/2}^2))/2
$$
$$
(\sin i)^{\ell_2}\rightarrow
2^{\ell_2}s_{1/2}^{\ell_2}(1-s_{1/2}^2)^{\ell_2/2}~~~~~~~~~~~~
\mbox{if $\ell_2=$even}
$$
$$
(\sin i)^{\ell_2}\rightarrow
2^{\ell_2}s_{1/2}^{\ell_2}(1-s_{1/2}^2)^{(\ell_2-1)/2}c_{1/2}~~~
\mbox{if $\ell_2=$odd}
$$
and
$$
g\rightarrow\varpi-h,~~~f\rightarrow\lambda_f-\varpi,~~~u\rightarrow\lambda_u-\varpi~~~.
$$
~\\
\\
After $S_{e2}[expr]$ is performed to the Hamiltonian $Z$ (Eq.(\ref{hamsecfinal})), or to the r.h.s. of the equations of motion (\ref{eqmomean}), we arrive at expressions containing terms of the form 
$$
coef e^{k_1}s_{1/2}^{k_2}\eta^{k_3}c_{1/2}^{k_4}{\cos\atop\sin}(k\lambda_f+n_1g+n_2h)
,~~\mbox{or}~~
coef e^{k_1}s_{1/2}^{k_2}\eta^{k_3}c_{1/2}^{k_4}{\cos\atop\sin}(k\lambda_u+n_1g+n_2h)
$$ 
where the following \textit{D'Alembert rules} are satisfied:
\begin{eqnarray}\label{dalambert}
~&~&
k_3=0~\mbox{or}~1,~~~
k_4=0~\mbox{or}~1,~~~\\
~&~&
k_1\geq |n_1|,~\mod(k_1,2)=\mod(n_1,2),~~~\nonumber\\
~&~&
k_2\geq |n_2|,~\mod(k_2,2)=\mod(n_2,2),~~~\nonumber\\
~&~&
k+n_1+n_2=0\nonumber
\end{eqnarray}
The above rules guarantee that, after operating with $S_{e2}$, all resulting expressions are polynomial in the equinoctial variables $(h,k,q,p)$, having also a possible linear dependence on $\eta$ and/or $c_{1/2}$, which does not cause any apparent singularity for $e=0$ and $i=0$ or $i=\pi/2$. Note that, as demonstrated in \cite{caveft2022}, we could have reached the final expressions by writing all the way from the start the Poisson algebra of the theory (Eqs.\ref{poiss}) or $\ref{poissb}$ in modified Delaunay variables or in Poincar\'{e} canonical variables. However, this requires re-establishing the whole set of algebraic rules for simplifications inherent in the Delaunay normalization in closed form.\\
\\
As a guide to the reader, a list of the whole set of definitions of variables, symbolic functions, as well as algebraic reduction and simplification rules employed in the semi-analytical theory presented in this section is given in the appendix. 

\clearpage

\section{Numerical Results}
\label{sec:numerical}
In this section, a numerical validation of the semi-analytical theory derived in the previous sections is provided. Particularly, the implementation of SELENA is compared with a cartesian propagation of the equations of motions in their non-averaged form. A test data set of orbits is selected, including orbits at different altitudes and orientations to assess the capabilities of SELENA. The osculating orbital states computed by the Cartesian integration are compared with the mean element evolution over 365 days. Moreover, we compute the canonical transformation that allows to add the short periodic corrections to the SELENA mean propagation, confirming SELENA's capability to produce high-fidelity propagations, with noticeably reduced computational cost.

\subsection{Test data sets and integrations}
\label{ssec:testdatainte}

\subsubsection{Test data sets}
\label{sssec:testdata}
Due to the slow convergence of the Lunar potential's multipole expansion (see subsection \ref{ssec:forcecompare}), the long-term behavior of lunar trajectories depends crucially on the particular choice of initial conditions: at low altitudes ($\sim 100~km$ above the Moon's surface), the lunar mascons dominate the orbital evolution, a fact implying that only isolated slots in inclination can lead to long-term stable motion. On the other hand, at intermediate altitudes ($\sim 500~km$ above the surface), the secular dynamics changes nature close to critical inclination values where important secular resonances appear (as for example the $2g$ resonance, driven mostly by the $J_2^2$ and $J_4$ terms (see \cite{laretal2020} and references therein). At still higher altitudes we also have secular resonances driven mostly by the Earth's $P_2$ tidal term. As a typical result, secular resonances lead to Kozai cicles of coupled oscillations between the orbital eccentricity and inclination. Such oscillations, in turn, may lead to trajectories developing pericentric passages quite close to the lunar surface (i.e. below the limit of 100 km where mascons dominate) even when the semi-major axis is as long as $\sim 6000~$km. 

In order to produce tests of numerical accuracy representative of most types of orbits that would be of interest to study as regards the secular dynamics, SELENA adopts a test campaign based on two sets of trajectories, as agreed with CNES:\\
\\
\textbf{Set 1: trajectories testing the dependence of accuracy on altitude}. In order to test the dependence of the numerical performance of SELENA on altitude only, test set 1 includes 120 initially circular trajectories ($e_{initial}=0$) chosen by the following combinations of initial orbital elements:
\begin{eqnarray}
a(km)         &=& R_\Moon + 100, 200, 400, 1000, 2000, 4000 \nonumber\\
e             &=& 0 \nonumber\\
i(^\circ)     &=& 0, 30, 57.8, 63.5, 90 \\
M(^\circ)     &=& 0 \nonumber\\
g(^\circ)     &=& 0 \nonumber\\
h(^\circ)     &=& 0, 90, 180, 270 \nonumber
\end{eqnarray}
The inclinations $0$ and $90$ degrees test equatorial and polar orbits respectively, while the choice $i=30^\circ$, $57.8^\circ$ and $63.5^\circ$ are connected to the critical behavior of the trajectories close to low-order secular resonances related to the motion of the satellite's perilune and line of nodes.\\
\\
\textbf{Set 2: trajectories testing the dependence of accuracy on eccentricity and inclination}. In order now to test the numerical performance of SELENA at various altitudes against the orbital eccentricity and inclination, test 2 includes 80 trajectories arranged in two main groups: subset 2a, with $e_{initial}=0.1$, and subset 2b, with $e_{initial}=0.6$.
\begin{eqnarray}
q(km)=a(1-e) &=& R_\Moon + 200, R_\Moon + 500 \nonumber\\
e &=& 0.1~(\mbox{subset 2a}), 0.6~(\mbox{subset 2b}) \nonumber\\
i(^{\circ} &=& 0, 30, 57.8, 63.5, 90 \\
M(^\circ) &=& 0\nonumber\\
g(^\circ) &=& 0\nonumber\\
h(^\circ) &=& 0 , 90 , 180, 270 \nonumber
\end{eqnarray}
Note that the grouping of altitude now is in terms of initial pericentric distance $q=a(1-e)$, implying the values of the semi-major axis (in kilometers) $a=2.150 (q=R_\Moon+200,e=0.1)$, $a=2.480 (q=R_\Moon+500,e=0.1)$ in subset 2a, and $a=4.840 (q=R_\Moon+200,e=0.6)$, or $a=5.590 (q=R_\Moon+500,e=0.6)$ in subset 2b. \\
\\
Finally, a number of independent tests were performed for a Lunar Pathfinder type of orbit chosen by the initial conditions: $a=5737.4$~km, $e=0.61$, $i=57.82^\circ$, $h=0$, $g=90^\circ$, $M=0$. 

\subsubsection{Definitions for integrations}
\label{sssec:integrdef}
In the comparisons presented below, three different methods are used to propagate the trajectories obtained by the initial conditions of sets 1 and 2 above. These integration methods are referred to below as:\\
\\
\textbf{AUTh-Cartesian:} a full cartesian integrator developed at AUTh, in which the forces are computed by the full form of the lunar, Earth and Sun tidal and SRP potential models (Eqs.(\ref{potmoon}), (\ref{potearth}), (\ref{potsun}), (\ref{potsrp}) without expansions or approximations). \\
\\
\textbf{SELENA:} Integration of the averaged equations of motion (Eqs.(\ref{eqmomean}). In this case, the initial data, which refer to osculating elements, are first transformed to the corresponding mean elements at the time $t=0$, by numerically implementing the chain of transformations (\ref{traosctomid}) and then (\ref{tramidtomean}). As in sections \ref{sec:hamclosed} and \ref{sec:selenaave},  we collectively refer to this transformations as the `rule $F^{-1}: z_{mean}=F^{-1}(z)$. Once the initial mean elements are computed, these are propagated through the averaged equations of motion (\ref{eqmomean}), yielding a numerical time series $z_{mean}(t_i)$ for all the mean elements, with $t_i=i\Delta t$, $\Delta t=t_{int}/N_{steps}$, where $t_{int}$, the total integration time, and $N_{steps}$, the number of time data points in the series, are pre-selected by the SELENA user. Finally, the time series in the mean elements can be transformed to one in osculating elements via the direct transformation $z(t_i)=F(z_{mean}(t_i))$, which is computed again by a numerical two-step computation of the transformations (\ref{trameantomid}) and (\ref{tramdidtoosc}). Finally, from the computed time series in osculating elements $z(t_i)$ we compute the cartesian co-ordinates and velocities via the transformation $\zeta(t_i)=E^{-1}(z(t_i))$ (subsection (\ref{ssec:elecoord})).\\
\\
\textbf{SELENA-Mean:} In this method, we proceed again with the numerical integration of the averaged equations of motion (Eqs.(\ref{eqmomean}), by first transforming the initial data as in the normal SELENA procedure, i.e., according to $z_{mean}(t=0)=F^{-1}(z(t=0))$. This allows to obtain the same numerical time series $z_{mean}(t_i)$ as in the SELENA procedure. However, in the final stage, instead of performing the passage to osculating elements via the direct transformation $z(t_i)=F(z_{mean}(t_i))$, we compute the cartesian coordinates of an approximate trajectory obtained by a straightforward mapping of the \textit{mean element} time series to cartesian coordinates given by $\zeta'(t_i)=E^{-1}(z_{mean}(t_i))$, i.e., neglecting altogether the short-periodic corrections (SRC) inherent in the transformation $F$. Note, nevertheless, that a crucial point in this method is that the SPC should be computed at the passing from the initial datum to the initial value of the mean elements. As discussed below (subsection \ref{ssec:selenamean}), this last method, albeit aleviating the numerical cost of computing the lengthy SCP transformations at all data points in the series, still provides a representation of the cartesian state vector having practically the same level of precision as the one properly computed in the normal SELENA procedure. The reasons for this behavior, which practically establishes the use of SELENA as a fast computing tool able to replace a cartesian integrator of the full equations of motion, is explained in subsection \ref{ssec:selenamean}. \\
\\
We now discuss the benchmarking, precision and validation tests performed within the framework of each one of the above integration procedures. 

\subsection{Validation of the AUTh-Cartesian propagator}
\label{ssec:auth}
\begin{figure}
\centering
\includegraphics[width=0.9 \textwidth]{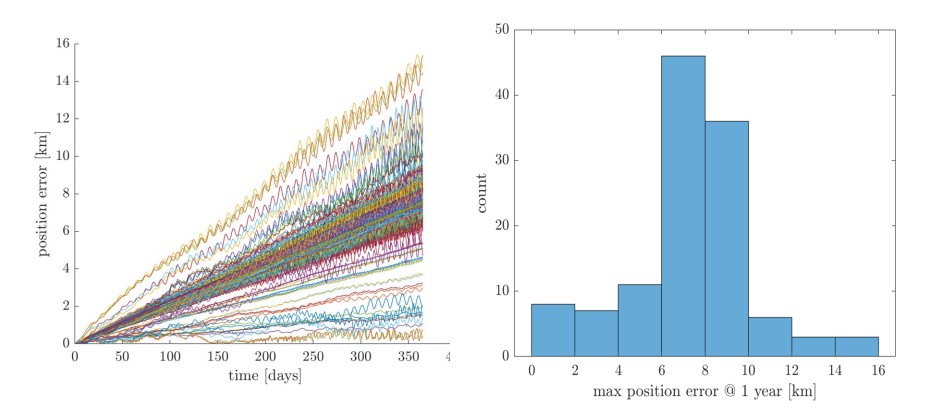}
\caption{\small (Left) The growth in time, up to $t_{int}=1~$year, of the distance $D$ in kilometers between trajectories tested for their difference when integrated in the full cartesian model by the AUTh-Cartesian or by the PATRIUS integrators. (Right) The histogram of the number of trajectories $N(D)$ whose AUTh-PATRIUS-integration distance $D$ in kilometers after one year is in one of the bins indicated in the abscissa of the figure.}
\label{fig:patriusauth}
\end{figure}
The AUTh-Cartesian integrator has served as the basis for all comparisons presented below, between trajectories computed with SELENA's averaged equations of motion and those obtained by a fully cartesian integration without averaging. In order to establish the precision levels of the AUTh-Cartesian integrator, several tests were performed with all trajectories of set 1 as well as a number of benchmarking trajectories included in the validation tests of the CNES-PATRIUS Cartesian integrator for lunar satellite orbits \cite{patrius}. Figure \ref{fig:patriusauth} summarizes the growth of the error in time for the entire set of tested trajectories, where error refers to the growth or the distance in kilometers between the trajectories computed by the same initial conditions but using the two different cartesian integrators, i.e., AUth or PATRIUS. As shown in the figure, using the same multipole truncation ($8\times 8$) for the lunar potential, the difference between the two integrations remains below the limit of 10 km/year for most trajectories, and only marginally surpassed by about $10\%$ of the trajectories in the entire test set. We note that since these are trajectories performing about $10^3 - 10^4$ orbital revolutions in one year, each revolution having a total length of $\sim10^4~km$, an absolute distance difference of $\sim 10~km$ in one year implies a relative difference of $\sim 10^{-9} - 10^{-10}$ per orbital period between the two propagators.   

\subsection{Validation of the SELENA propagator}
\label{ssec:selena}

Having established the precision levels of the AUTh full-cartesian integrator, the present section refers to the main validation tests for SELENA, i.e., a comparison of the difference in the trajectories computed by the averaged equations of motion and those under a fully-cartesian integration without approximations. 

In order to arrive at a final choice of terms to be included in the semi-analytical (averaged) model of SELENA, several tests where performed to evaluate the relative importance of various terms appearing in the semi-analytical theory in different sets of test trajectories. These partial tests can be devided in four main categories:
\begin{itemize}
\item 
Relative importance of second order internal term corrections: for example, $J_2^2$, $J_{22}C_{22}$, $C_{22}^2$ etc.
\item
Relative importance of relegation terms for the internal tesseral harmonics: $n_\Moon C_{nm}$ or 
$n_\Moon S_{nm}$.
\item 
Relative importance of second order external term corrections: $P_2^2$, $P_3^2$.
\item 
Relative importance of external relegation terms: $n_\Moon P_2$, $n_\Moon P_3$.
\end{itemize}

\subsubsection{Relative importance of second order and relegation internal terms}
\label{sssec:secondrelint}
\begin{figure}[h]
\centering
\includegraphics[width=0.9 \textwidth]{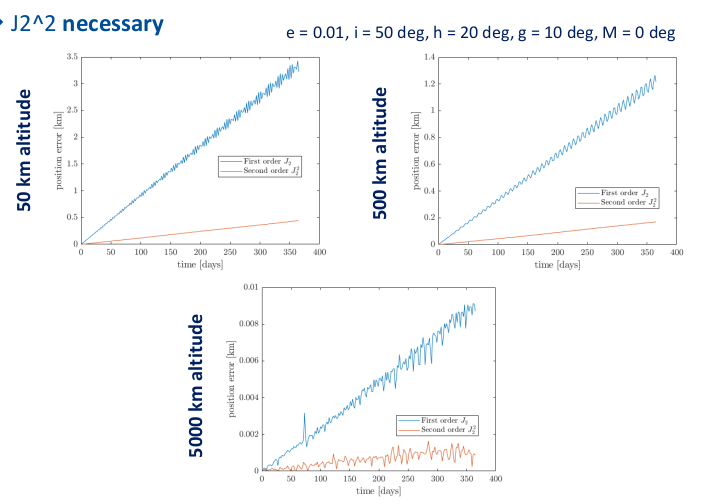}
\caption{\small The growth in time of the distance error in kilometers, estimated as the distance between the SELENA- and AUTh-Cartesian propagated trajectories, when the term $J_2^2$ is included, or not, in the SELENA-averaged equations of motion and corresponding SPCs. The test trajectories have the same initial orbital elements in all three cases, as indicated at the top of the figure, except for the value of the semi-major axis, which is equal to $a=R_\Moon+\delta$ with the altitude $\delta$ equal to (top left) $\delta=50~km$, (top right)$\delta=500~km$, and (bottom) $\delta=5000~km$. }
\label{fig:j2sqerror}
\end{figure}

\textbf{$J_2^2$-corrections:} Figure \ref{fig:j2sqerror} shows a comparison of the growth in kilometers of the distance between the SELENA- and AUTh-Cartesian propagated trajectories, when the term $J_2^2$ is included or not in the SELENA-averaged equations of motion and SPCs. The three panels correspond to trajectories with the same initial orbital elements in all three cases, except for the value of the semi-major axis, equal to $a=R_\Moon+\delta$ with the altitude $\delta$ equal to 50, 500, and 5000 km in each panel respectively. 

\begin{figure}
\centering
\includegraphics[width=0.7 \textwidth]{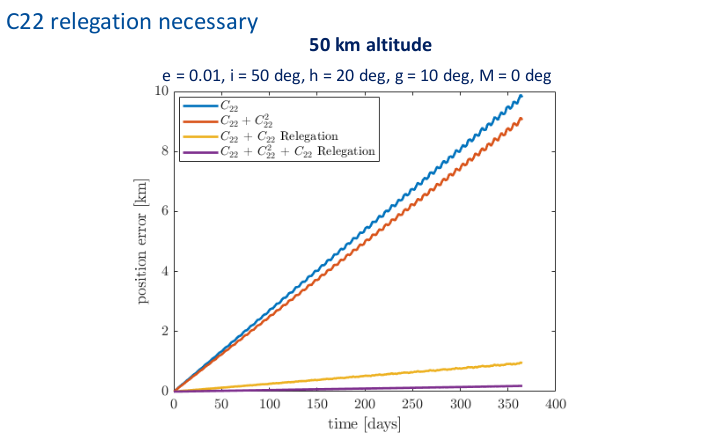}
\caption{\small Comparison of the growth in time of the distance error in kilometers (distance between the SELENA- and AUTh-Cartesian propagated trajectories), in four cases including various combinations of the $C_{22}$, $C_{22}^2$ and $n_{\Moon}C_{22}$ terms. We note the error by the omission of the $n_{\Moon}C_{22}$ term in the SPCs is dominant over all other second-order corrections. The initial conditions of the trajectory are indicated on top of the panel. }
\label{fig:c22relerror}
\end{figure}

We note that the distance error, with or without the $J_2^2$ term, grows linearly with time. A careful analysis shows that such a linear growth appears generically by the omission of any considered term in the averaged equations of motion and/or the SPC. In fact, the latter omission implies that the initial datum for the mean semi-major axis $a_{mean}(0)$, as computed by the osculating elements at time $t=0$, misses a correction of order equal to the size of the term omitted in the transformation $F$, that is, the value $a_{mean}(0)$ has an error $\Delta a_{mean}$, of order $\Delta a_{mean}=O(a_{mean}J_2^2)$ in our case. Since $a_{mean}$ is constant under the averaged equations of motion, we can obtain an estimate for the error produced in the equation for the fast frequency $\dot{M}$ as specified by the semi-analytical (averaged) equations of motion
\begin{equation}\label{dotmerror}
\dot{M}=\dot{M}_0+\Delta\dot{M}
=\left({2\over a n_s}{\partial Z\over\partial a}\right)_{a=a_{mean}}
+\left({2\over a n_s}
{\partial\tilde{R}\over\partial a}
+n_s{\Delta a\over a}\right)_{a=a_{mean}} ~~.  
\end{equation}
where the term $\tilde{R}$ represents the acceptable remainder, i.e., it is determined by the term  omitted from the equations of motion, while the term $n_s\Delta a/a$ represents an error in the determination of the Keplerian frequency (mean motion) by the error in the estimate of $a_{mean}$. By dimensional analysis, both terms are of the same order of magnitude, implying a scaling law $\dot{\Delta M}_0\sim C_{\tilde{R}}n_s$, where $C_{\tilde{R}}$ is a dimensionless coefficient of the same order as the coefficient of the term omitted from the equations of motion and/or the SPC. For a trajectory with (nearly constant) semi-major axis $a$, this implies a systematic growth of the error in kilometers of the order of 
\begin{equation}\label{errorlin}
D\sim 4\pi^2 C_{\tilde{R}} a\left({a\over R_\Moon}\right)^{3/2} t/T_{R_\Moon}~~, 
\end{equation}
where $T_{R_\Moon}\sim 0.05~days$ is the orbital period of the circular orbit with $a=R_\Moon$. This leads to the estimate
\begin{equation}\label{errorlin2}
D[km]\sim 2~C_{\tilde{R}}~a[km]~\left({a\over R_\Moon}\right)^{3/2}~t[days]~~.  
\end{equation}
Equation (\ref{errorlin2}) can be used for getting an estimate of the coefficient of the linear growth of the error distance $D$ for any of the terms excluded from a precision test. In the case of Fig.\ref{fig:j2sqerror} we have $C_{\tilde{R}}=O(J_2^2)\sim 10^{-6}$. For orbits close to the Moon's surface $a\sim 2\times 10^3~km$ we find $D[km]\sim 4\times 10^{-3} t[days]$, implying an error $\sim 1~km/year$, as shown in Fig.\ref{fig:j2sqerror}. Including the $J_2^2$ term, instead, drops the error to the scale of meters/year.\\
\\
\textbf{$C_{22}^2$ and $C_{22}-$relegation corrections:} Figure \ref{fig:c22relerror} shows now the same error growth analysis as above, but testing the inclusion, or not, of the $C_{22}^2$ term of the averaged theory in the equations of motion and in the SPCs, as well as the inclusion, or not, of the $n_\Moon C_{22}$ relegation term in the SPCs. We now observe that the second-order corrections due to relegation are far more important than the ones due to $C_{22}^2$ term.
\begin{figure}
\centering
\includegraphics[width=0.9\textwidth]{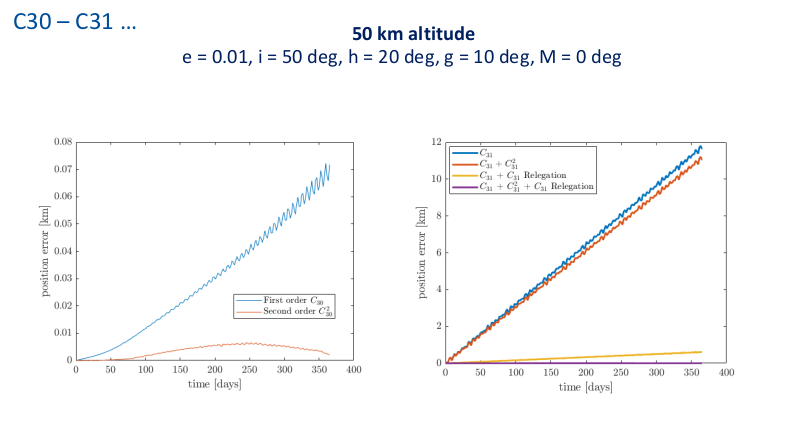}
\caption{\small (left) Same as in Fig.\ref{fig:j2sqerror}, but for the zonal harmonic $C_{30}$. (right) Same as in Fig.\ref{fig:c22relerror}, but for the tesseral harmonic $C_{31}$.}
\label{fig:c3031relerror}
\end{figure}

Taking into account the functional form of each harmonic in the lunar potential, for both zonal or tesseral harmonics $C_{nm},S_{nm}$ the coefficient $C_{\tilde{R}}$ can be estimated as
\begin{equation}\label{ctilderint}
C_{\tilde{R}}\sim \left({R_\Moon\over a}\right)^{sn}\mid C_{nm}\mid^{s},~~\mbox{or}~~
C_{\tilde{R}}\sim \left({R_\Moon\over a}\right)^{sn}\mid S_{nm}\mid^{s}
\end{equation}
where $s$ is the order at which the term is omitted ($s=2$ in the above examples). On the other hand, for the relegation terms $n_\Moon C_{nm}$ or $n_\Moon S_{nm}$ we have the estimate:
\begin{equation}\label{ctilderintrel}
C_{\tilde{R}}\sim 
m{n_\Moon\over n_s}\left({R_\Moon\over a}\right)^{n}\mid C_{nm}\mid,~~\mbox{or}~~
C_{\tilde{R}}\sim m{n_\Moon\over n_s}\left({R_\Moon\over a}\right)^{n}\mid S_{nm}\mid~~,
\end{equation}
the ratio $n_\Moon/n_s$ varying in order of magnitude between $n_\Moon/n_s\sim 10^{-3}$ for trajectories close to the surface of the Moon, and $n_\Moon/n_s\sim 10^{-2}$ for trajectories around $a\sim 5000~km$. Thus, as shown in Fig.\ref{fig:c22relerror}, the growth coefficient of the error caused by the omission of the $n_\Moon C_{22}$ relegation terms prevails over the one due to the omission of the $C_{22}$ terms by a coefficient $n_\Moon/(n_s|C_{22}|) ~ 10-100$.\\
\\
\textbf{$C_{nm}^2$ and $C_{nm}-$relegation corrections for terms of higher degree:} finally, Fig.\ref{fig:c3031relerror} shows a comparison similar as in Figs. \ref{fig:j2sqerror} or \ref{fig:c22relerror} but for an example of higher order zonal harmonic $C_{30}$, as well as of tesseral harmonic $C_{31}$. As regards zonal harmonics, we find that second order corrections lead to a reduction of the errors well below the required limit, i.e. of the order of metters/year, thus they can be safely ignored for all zonal terms except for $J_2^2$. On the other hand, similarly to the $C_{22}$ case, the relegation terms $n_\Moon C_{nm}$ or $n_\Moon S_{nm}$ on the SPCs provide the leading second order corrections for all tesseral terms $m\neq 0$, by a factor of order $n_\Moon/(n_s|C_{nm}|)$ or $n_\Moon/(n_s|S_{nm}|)$.

\subsubsection{Relative importance of second order and relegation external terms}
\label{sssec:secondrelint}
\begin{figure}
\centering
\includegraphics[width=0.9\textwidth]{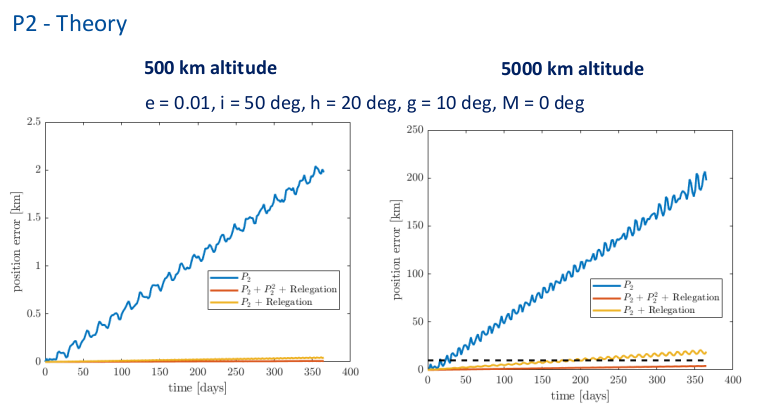}
\includegraphics[width=0.7\textwidth]{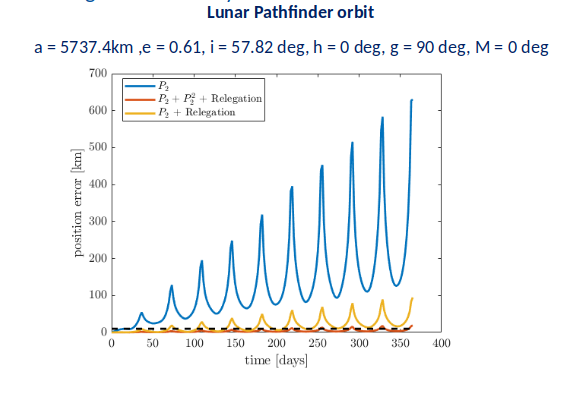}
\caption{\small Tests of the the distance error growth with the inclusion, or not, in the averaged theory of the Earth's $P_2^2$ and/or $n_\Moon P_2$ relegation term. The various curves show the growth in time of the error with various terms included, as indicated in the panels. (Top) Test for two trajectories with the same orbital elements as indicated above the figures, except for the semi-major axis, which is equal to (top left) $a=R_\Moon+500~km$, (top right) $a=R_\Moon+5000~km$. (Bottom) The same precision test for a lunar pathfinder-like trajectory with high initial value of the eccentricity $e=0.61$ exhibits important monthly oscillations around the mean linear trend of the error growth, related to periodic passages from the apocenter or pericenter of the trajectory.  }
\label{fig:p2sqrelerror}
\end{figure}

The influence of higher order corrections due to the Earth's external terms ($P_2^2$, $n_\Moon P_2$) can be assessed with the help of Fig.\ref{fig:p2sqrelerror}, showing the comparison of the error growth by omission of one or more of the Earth's $P_2^2$ and $n_\Moon P_2$ terms for two nearly circular trajectories with semi-major axis $a=R_\Moon+500~km$ (left), or $a=R_\Moon+5000~km$ (right). As expected, higher order corrections in the averaged theory (equations of motion or SPCs) become important at altitudes beyond $a=3000$. The corresponding coefficients $C_{\tilde{R}}$ in Eq.(\ref{errorlin2}) can be estimated as:
\begin{eqnarray}\label{ctilderext}
C_{\tilde{R}} &\sim & \left({M_\Earth R_\Moon^2\over M_\Moon r_\Earth^2}\right)^s
\left({a\over R_\Moon}\right)^{2s}\approx 2^s\times 10^{-3s}\left({a\over R_\Moon}\right)^{2s}~~~
\mbox{for the terms $P_2^s$, $s=1,2$} \nonumber\\
C_{\tilde{R}} &\sim & \left({M_\Earth R_\Moon^3\over M_\Moon r_\Earth^3}\right)^s
\left({a\over R_\Moon}\right)^{3s}\approx 9^s\times 10^{-6s}\left({a\over R_\Moon}\right)^{3s}~~~
\mbox{for the terms $P_3^s$, $s=1,2$} \nonumber\\
~&~&~\\
C_{\tilde{R}} &\sim & {n_\Moon\over n_s}\left({M_\Earth R_\Moon^2\over M_\Moon r_\Earth^2}\right)
\left({a\over R_\Moon}\right)^{2}
\approx 2\times 10^{-3} {n_\Moon\over n_s}\left({a\over R_\Moon}\right)^{2}~
\mbox{for the rel.terms $n_{moon}P_2$} \nonumber\\
C_{\tilde{R}} &\sim & {n_\Moon\over n_s}\left({M_\Earth R_\Moon^3\over M_\Moon r_\Earth^3}\right)
\left({a\over R_\Moon}\right)^{3}
\approx 9\times 10^{-6} {n_\Moon\over n_s}\left({a\over R_\Moon}\right)^{3}~
\mbox{for the rel.terms $n_{moon}P_3$} \nonumber
\end{eqnarray}
As a rough estimate, for an orbit at $a=3R_\Moon$ we have $n_\Moon/n_s\approx 3\times 10^{-3}$, thus: $C_{\tilde{R}}\sim 3\times 10^{-4}$ for the omission of the $P_2^2$ term, while $C_{\tilde{R}}\sim 5\times 10^{-5}$ for the omission of the $n_\Moon P_2$ term. Thus, the error growth factor due to omission of the $P_2^2$ term is about an order of magnitude larger than the one due to the omission of the $n_\Moon P_2$ relegation term, in agreement with what is observed in Fig.\ref{fig:p2sqrelerror}. In fact, substituting the above numbers to Eq.(\ref{errorlin2}) yields an order of magnitude correct estimate of the linear trend observed in both trajectories of the top panels of Fig.\ref{fig:p2sqrelerror}. 

On the other hand, working with highly eccentric trajectories, as the one of the bottom panel of Fig.\ref{fig:p2sqrelerror} (corresponding to the Lunar Pathfinder mission), we observe that, besides an overall linear trend in the factor of the error growth, there are also oscillations in the error curves generated by the large difference in the size of the tidal force between the moments when the trajectory is at apocenter or pericenter. 
Thus, these oscillations have a monthly periodicity. In this case, an upper bound for the error growth can be obtained substituting the value of the semi-major axis $a$ in Eqs.(\ref{ctilderext}) with the apocentric distance $r_a=a(1+e)$. 

\subsection{The SELENA final default model: precision tests}
\label{ssec:selenafinal}
\begin{figure}[h]
\centering
\includegraphics[width=0.8\textwidth]{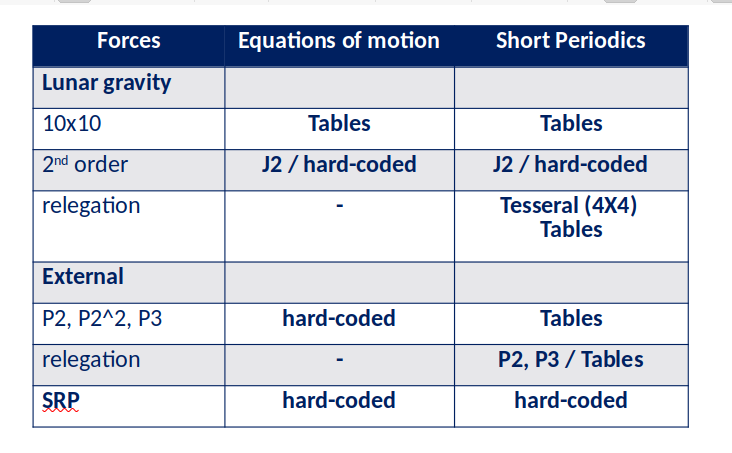}
\caption{\small The SELENA final default model adopted in the integration of the averaged equations of motions and computation of the associated SPCs.}
\label{fig:selena}
\end{figure}

In view of all the above trial-and-error tests, the final averaged theory adopted by SELENA is based on the integration of the averaged equations of motion under the Hamiltonian (\ref{hamsecfinal}), with short periodic corrections linking the osculating to mean elements as described in subsection \ref{ssec:eqmotrafinal}. Figure \ref{fig:selena} summarizes the structure of the adopted default SELENA propagator. Note that, in the numerical procedure for the computation of the right-hand side of the equations of motion or the transformations associated with SPCs, the greatest computational load comes from the internal terms associated with the Lunar potential. In the SELENA integrator, these terms are loaded to the equations of motion through Tables, i.e., ascii files yielding each trigonometric polynomial term in the series of the average theory via a symbolic representation. For few terms, instead, the corresponding equations of motion or SPC transformations are hard-coded into the propagator. The entire set of equations of motion and SPC transformations are provided in a symbolic form, along with the symbolic notebooks used in the production of this file, in SELENAs deliverable folder D5 (see the appendix, subsection \ref{ssec:seldeliv}). 

\begin{figure}
\centering
\includegraphics[width=0.9\textwidth]{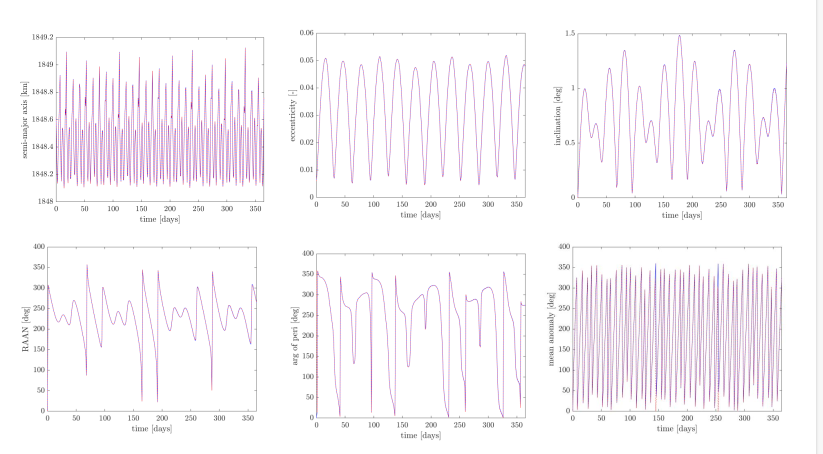}
\caption{\small The evolution of all six osculating orbital elements for a nearly circular orbit of the test set 1. Two curves are included in each plot: one computed with the SELENA propagator of the averaged equations of motions + SPCs and another with the full AUTh-Cartesian propagator of the equations of motion. However, the two curves are indistinguishable at the scale of the figures.}
\label{fig:orbtestset1}
\end{figure}

\begin{figure}
\centering
\includegraphics[width=0.9\textwidth]{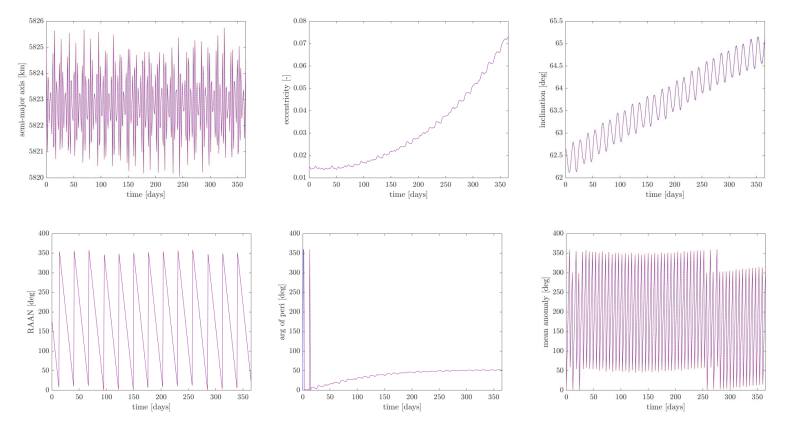}
\caption{\small Same as in Fig.\ref{fig:orbtestset1}, but for one of the orbits of the test data set 2a.}
\label{fig:orbtestset2a}
\end{figure}

\begin{figure}
\centering
\includegraphics[width=0.9\textwidth]{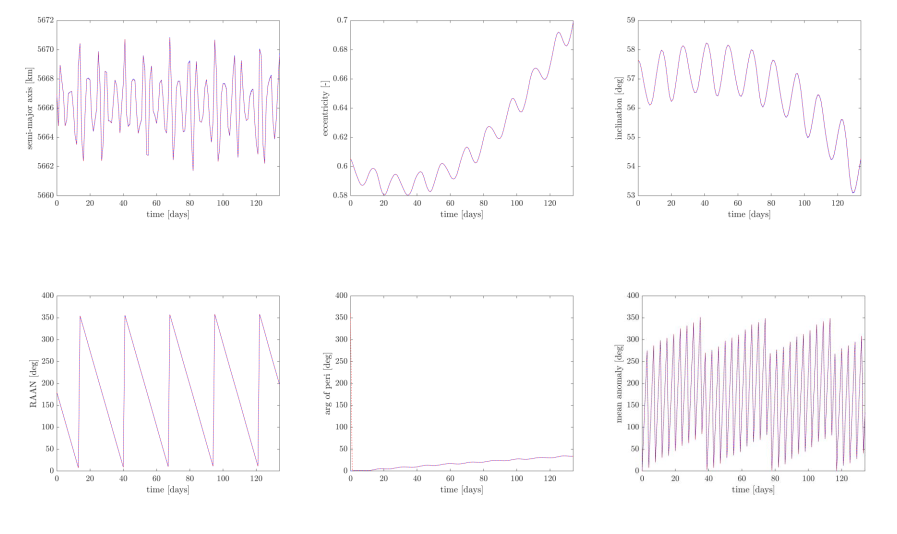}
\caption{\small Same as in Fig.\ref{fig:orbtestset1}, but for one of the orbits of the test data set 2b.}
\label{fig:orbtestset2b}
\end{figure}

Figures \ref{fig:orbtestset1}, \ref{fig:orbtestset2a}, \ref{fig:orbtestset2b} show examples of three trajectories from the test sets 1, 2a and 2b respectively (subsection \ref{ssec:testdatainte}), in which the evolution of all six osculating orbital elements over a period of one year is compared by two different propagation methods, i.e., SELENA and the full AUTh-Cartesian integration. In all cases the superposed curves showing the evolution for all elements by the two methods are indistinguishable at the scale of the figures, and relative errors per datum in the integration time series are typically of the order of $10^{-6} - 10^{-5}$, except for the curves on the mean anomaly, in which the relative error for the test set 2b (highly eccentric orbits at high altitude) can reach the level of $10^{-4}$ after one year. 

\begin{figure}
\centering
\includegraphics[width=0.8\textwidth]{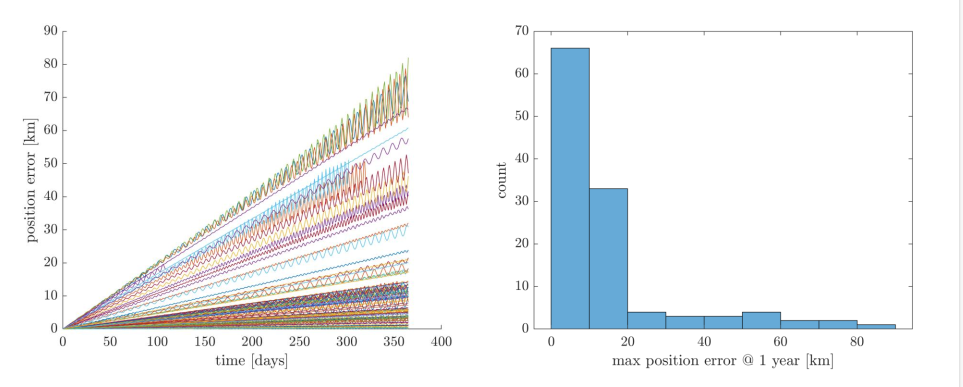}
\caption{\small (Left) Growth in time of the distance error $D$ between the SELENA and full AUTh-Cartesian integrated trajectory, for all the orbits of the test data set 1 (subsection \ref{ssec:testdatainte}). (Right) The number of orbits in the set with a distance error $D$ in kilometers after one year within each of the bins indicated in the abscissa.}
\label{fig:errorset1}
\end{figure}

Figure \ref{fig:errorset1} collects the information about the growth in time of the distance error $D$ computed by the comparison between the SELENA and AUTh-Cartesian propagators for the entire set of 120 trajectories of the test data set 1. All trajectories exhibit a linear trend for the growth of $D$ with time, with linear coefficients predicted at order of magnitude by the sum of the equations (\ref{ctilderint}) - (\ref{ctilderext}) for all the terms included in the averaged theory. Note that, as shown in the right panel of Fig.\ref{fig:errorset1}, about 60\% of these orbits yield a growth rate below the nominal threshold of $10~km/year$, while about 90\% are below $20~km/year$. Only the most distant trajectories overpass this limit, a fact, as explained in subsection \ref{ssec:selena}, due to the trajectories having the same angular error in the computation of the mean anomaly, but larger overall size, hence, larger growth of the error in terms of the distance $D$.

\begin{figure}
\centering
\includegraphics[width=0.75\textwidth]{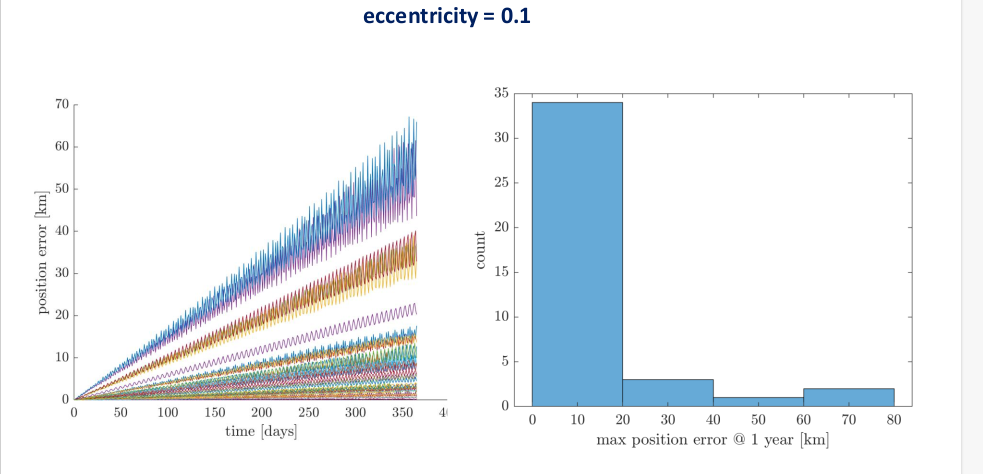}
\includegraphics[width=0.75\textwidth]{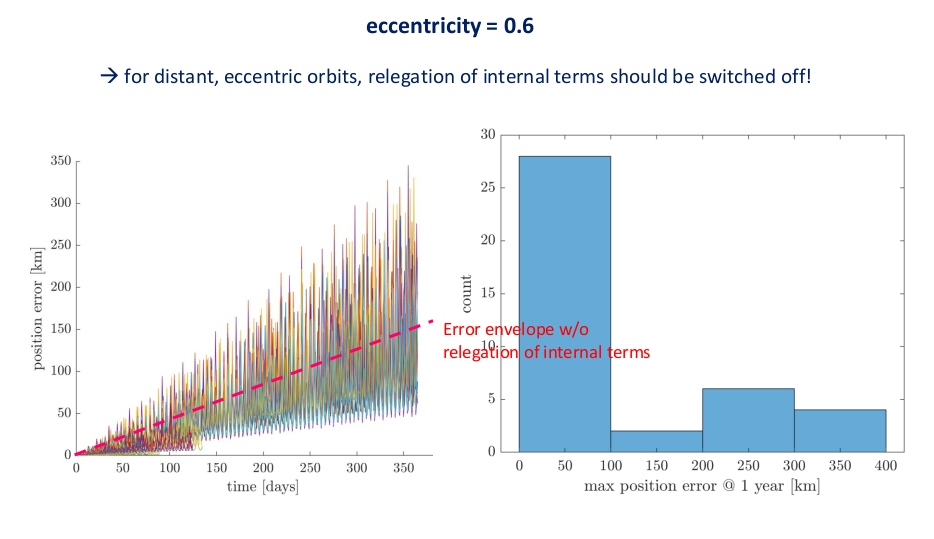}
\caption{\small Same as in Fig.\ref{fig:errorset1}, but for the set of orbits of the test data subsets (Top) 2a, and (Bottom) 2b.}
\label{fig:errorset2}
\end{figure}
\begin{figure}
\centering
\includegraphics[width=0.7\textwidth]{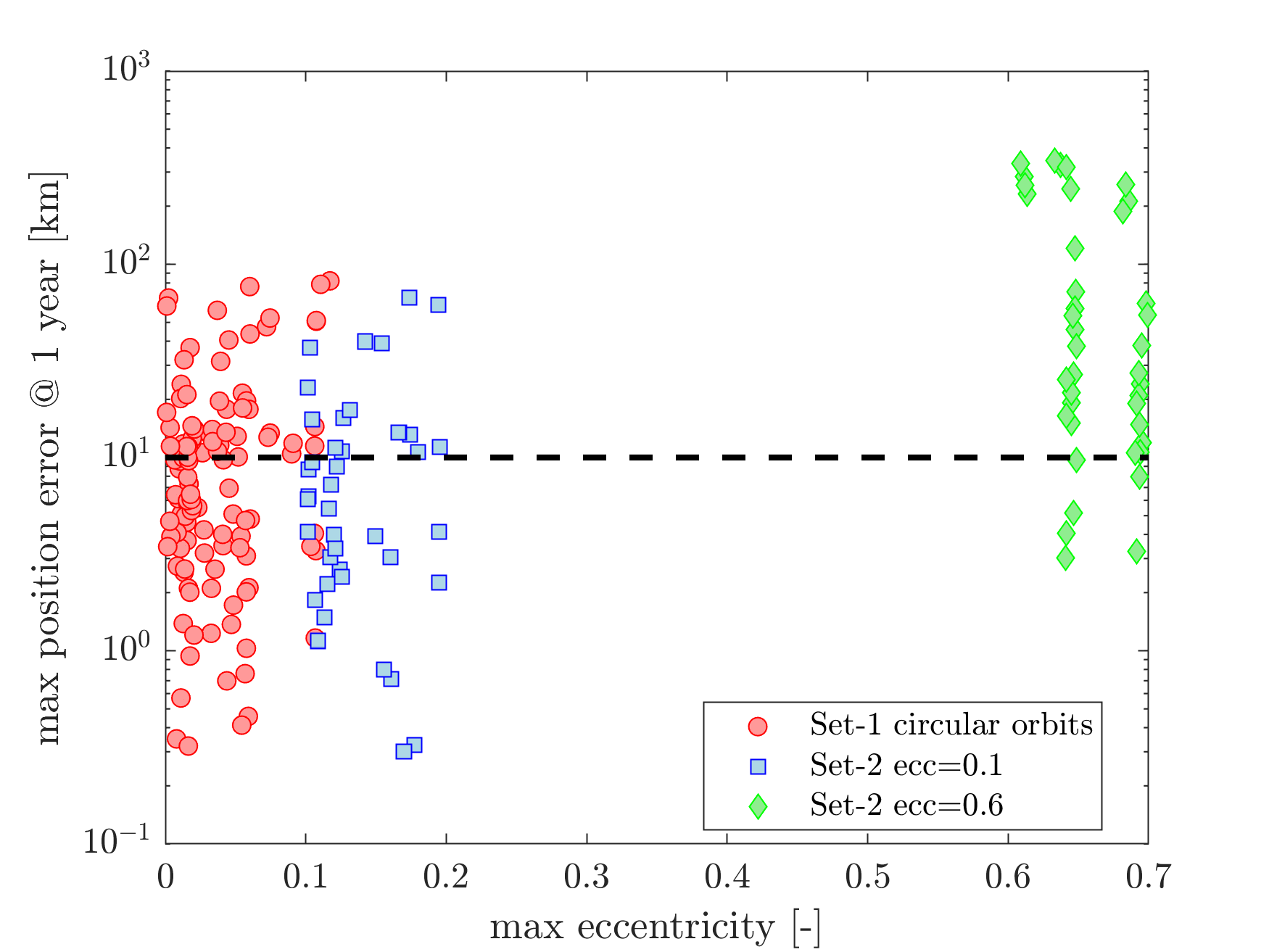}
\caption{\small The distance error $D$ at $t=1~year$ for the trajectories of all three groups against the maximum value of the eccentricity attained throughout the one-year integration for each individual trajectory.}
\label{fig:errorecc}
\end{figure}

Figure \ref{fig:errorset2}, now, yields the same information as in Fig.\ref{fig:errorset1}, but for the orbits of the test data set 2, divided in the subsets 2a (mildly eccentric orbits close to critical inclination values) or 2b (highly eccentric orbits; see subsection \ref{ssec:testdatainte}). We note that for mild eccentricities the behavior of the error is essentially identical to the one of the circular orbits, independently of the value of the inclination. However, a set of highly eccentric orbits exhibit a larger error growth factor, by nearly one order of magnitude. The fact that the high value of the eccentricity is the main factor contributing to the larger error can be substantiated with the help of Figure \ref{fig:errorecc}, in which the distance error $D$ at $t=1~year$ is plotted for the trajectories of all three groups against the maximum value of the eccentricity attained throughout the one-year integration for each individual trajectory. We note the overall trend for the error to increase with the maximum eccentricity attained by the trajectories. 

\begin{figure}
\centering
\includegraphics[width=0.45\textwidth]{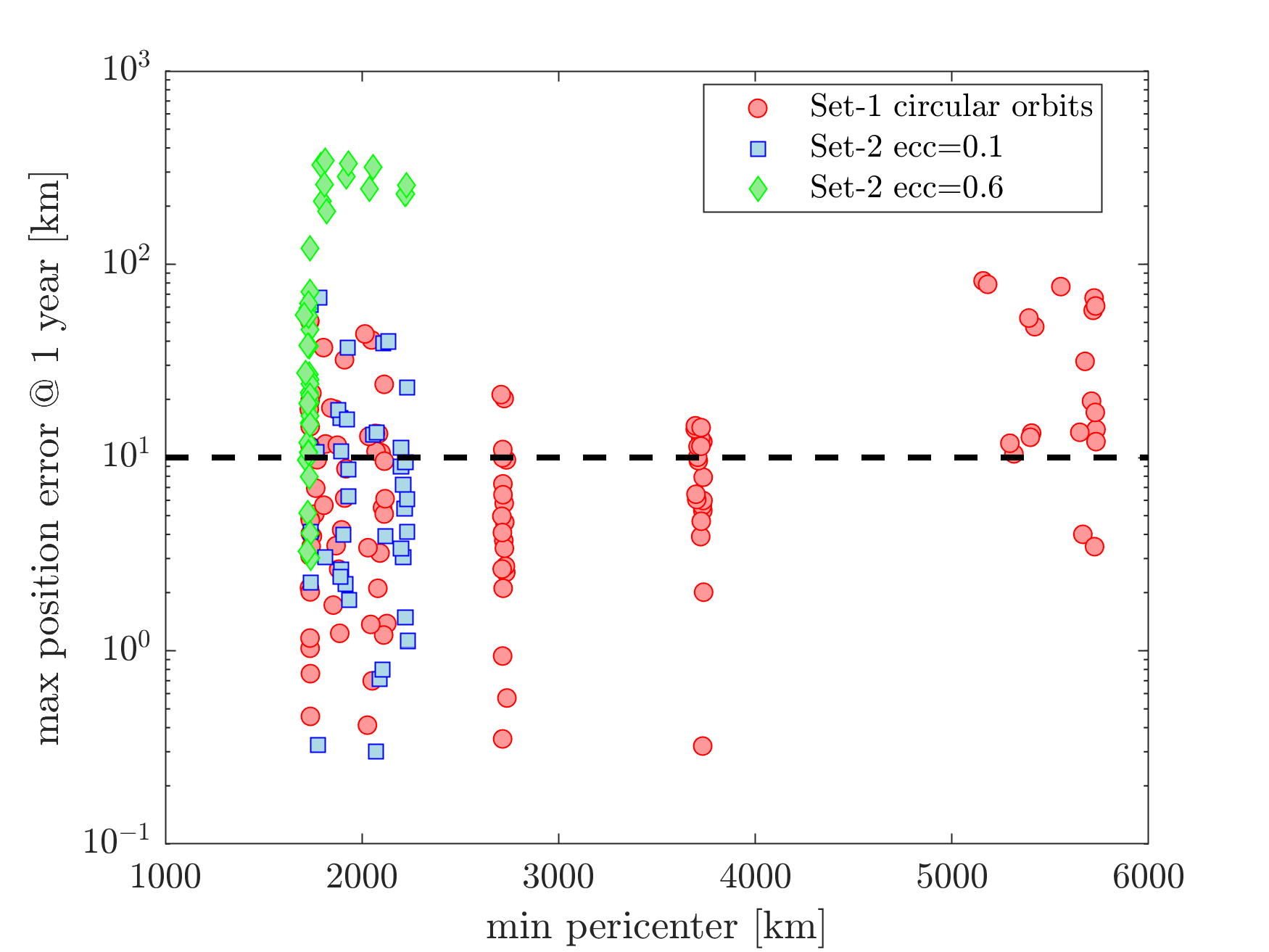}
\includegraphics[width=0.45\textwidth]{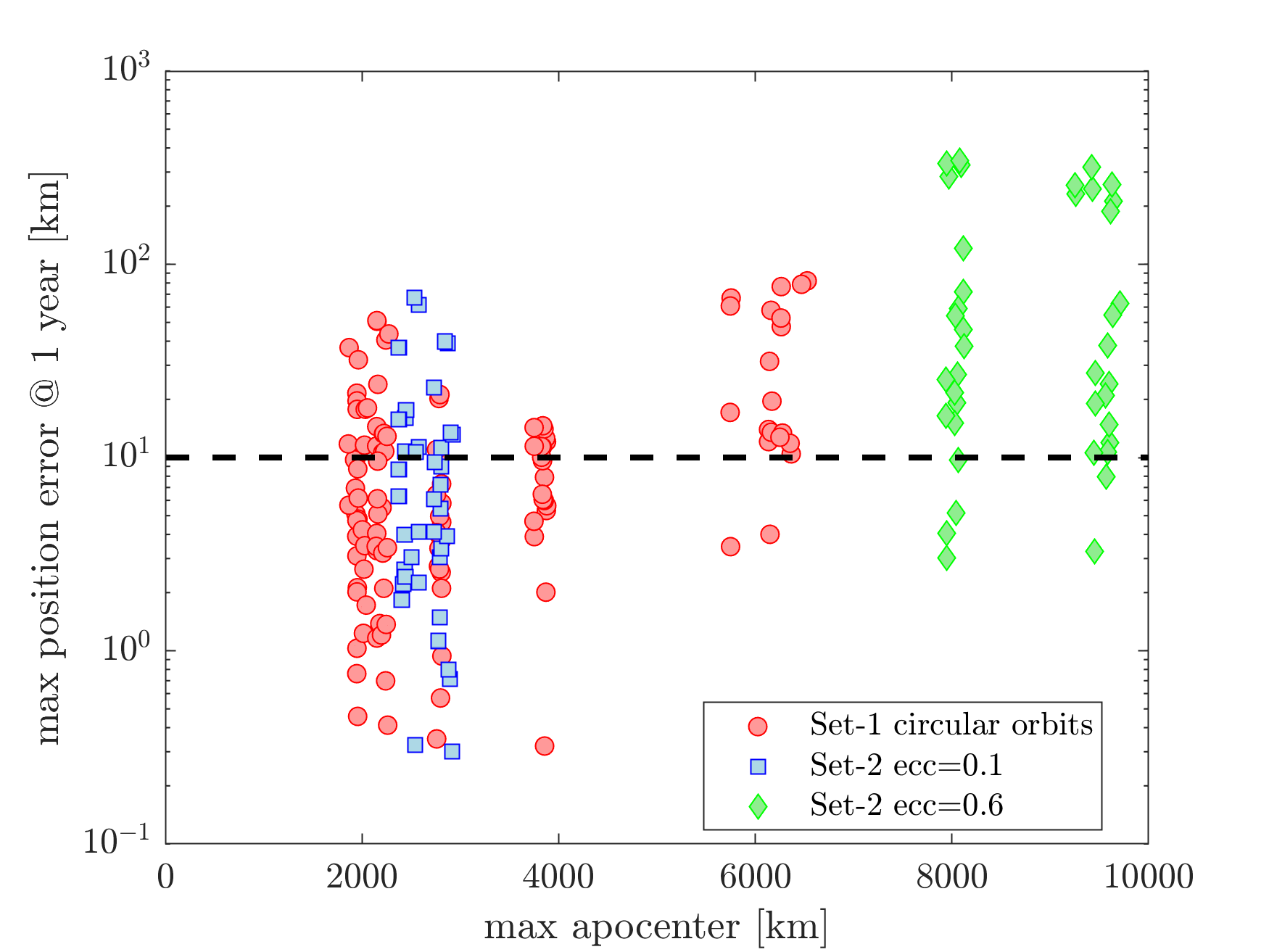}
\caption{\small The distance error $D$ at $t=1~year$ for the trajectories of all three groups against (left) the minimum attained pericentric distance $r_p$, or (right) the maximum attained apocentric distance $r_a$ throughout the one-year integration for each individual trajectory.}
\label{fig:errorrpra}
\end{figure}

The behavior exhibited in Fig.\ref{fig:errorecc} implies that the error is not affected by the mean lunicentric distance (value of the semi-major axis) along a trajectory as much as by the large difference between minimum pericentric distance $r_p=a(1-e)$ and maximum apocentric distance $r_a=a(1+e)$ for a trajectory. This conclusion is supported by the data in Fig.\ref{fig:errorrpra}, showing again the distance error $D$ at $t=1~year$ for the trajectories of all three groups against the smallest attained pericentric distance (left) or the largest attained apocentric distance (right). 
Technically, the fact that trajectories with a second order terms $\tilde{R}^{(1)}_4=\{CS[RM[V_{\Earth,P2}]],\chi^{(1)}\}$ neglected in the final model (see subsection \ref{sssec:step1}). Such terms express a second order correction in the averaged theory caused by the product of the most important internal terms with the Earth's $P_2$ external term. Physically, such terms become relevant only for trajectories with pericenter very close to the Moon's surface and apocenter reaching a distance $\sim 10^4~km$ where the Earth's $P_2$ tidal term is the leading perturbation. Introducing a closed-form representation for such terms is possible by the method proposed in \cite{caveft2022}, however the extra cost in the computation of the corresponding SPCs does not justify the gain in precision, which is anyway due, for the most part, to the overall growth in size of the trajectory, rather than the growth of the angular error in the evolution of the mean anomaly. As shown, however, in the bottom panel of Fig.\ref{fig:errorset2}, the control of the error is better when the Lie transformation at step 7 of the averaged theory (subsection \ref{sssec:step7}) is excluded from the computation of the SPCs for the trajectories of subset 2b. We attribute this to the fact that the computation of this particular transformation (rule $S_{HR}[\cdot,N_{rel}]$) is the only point where a series development (Eq.(\ref{phiexp})) instead of a closed-form is invoked. Since the series (\ref{phiexp}) are produced by the usual series inversion of Kepler's equation, their convergence is limited for eccentricities $e>0.6$. In fact, a closed-form replacement of this computation would likely increase SELENA's precision for the highly eccentric orbits, and it is proposed as a future improvement.   

\subsection{Performance of the SELENA-Mean propagation}
\label{ssec:selenamean}

\begin{figure}
\centering
\includegraphics[width=0.6\textwidth]{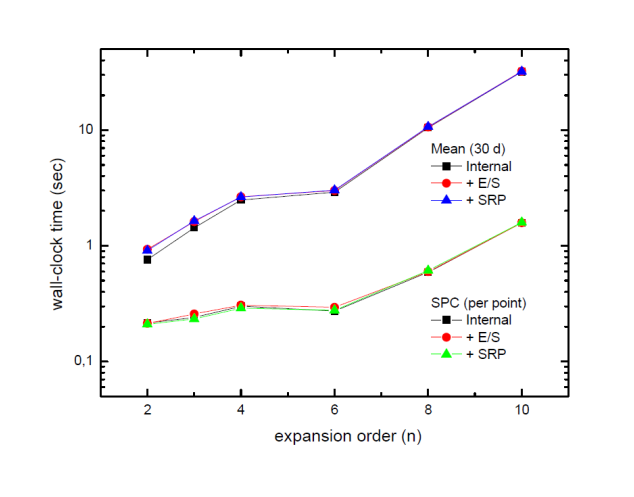}
\caption{\small The top curves show the (wall-clock) computation time $T_C$ in seconds required for a propagation of the mean elements for each group of terms in the averaged equations of motion, for a total orbital time of $t=30~days$. The time $T_C$ is estimated as a mean among all trajectories of the dataset 1. The lower curves give a mean estimate, for the same trajectories, of the physical (wall-clock) time $T_{SPC}$ required to perform a single SPC transformation of the type mean $\rightarrow$ osculating element for a single data point along the time series.}
\label{fig:exectime}
\end{figure}

All the above error analysis was based on the complete chain of the SELENA operations on the averaged theory, i.e., integration of the averaged equations of motion and the full chain of SPC corrections introduced in the chain `initial osculating elements $\rightarrow$ initial mean elements $\rightarrow$ propagation of mean elements $\rightarrow$ back-transform of the mean element time series to osculating element time series. In the present subsection we discuss the performance (in computational cost and accuracy) of the \textit{SELENA-Mean} chain of operations (subsection \ref{sssec:integrdef}), in which the last part of the chain is omitted, i.e., mean orbital elements are mapped to fictitious osculating elements by the identity instead of the SPC transformation. Figure \ref{fig:exectime} explains the obvious benefits in terms of computational cost out of such an approach.  
\begin{figure}
\centering
\includegraphics[width=0.7\textwidth]{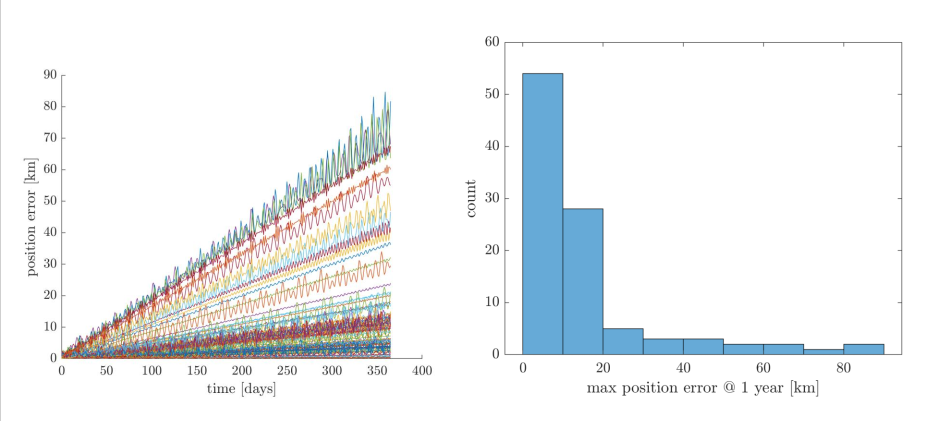}
\caption{\small Same as in Fig.\ref{fig:errorset1}, for the trajectories of the test dataset 1, but with the SELENA trajectories computed via the SELENA-Mean procedure rather than the full SELENA procedure. Excluding the SPCs computation at all timesteps except for the one at the begining of the integration ($t=0$) reduces dramatically the computational cost with practically no consequences for the accuracy of the propagation.}
\label{fig:errorset1mean}
\end{figure}

\begin{figure}
\centering
\includegraphics[width=0.7\textwidth]{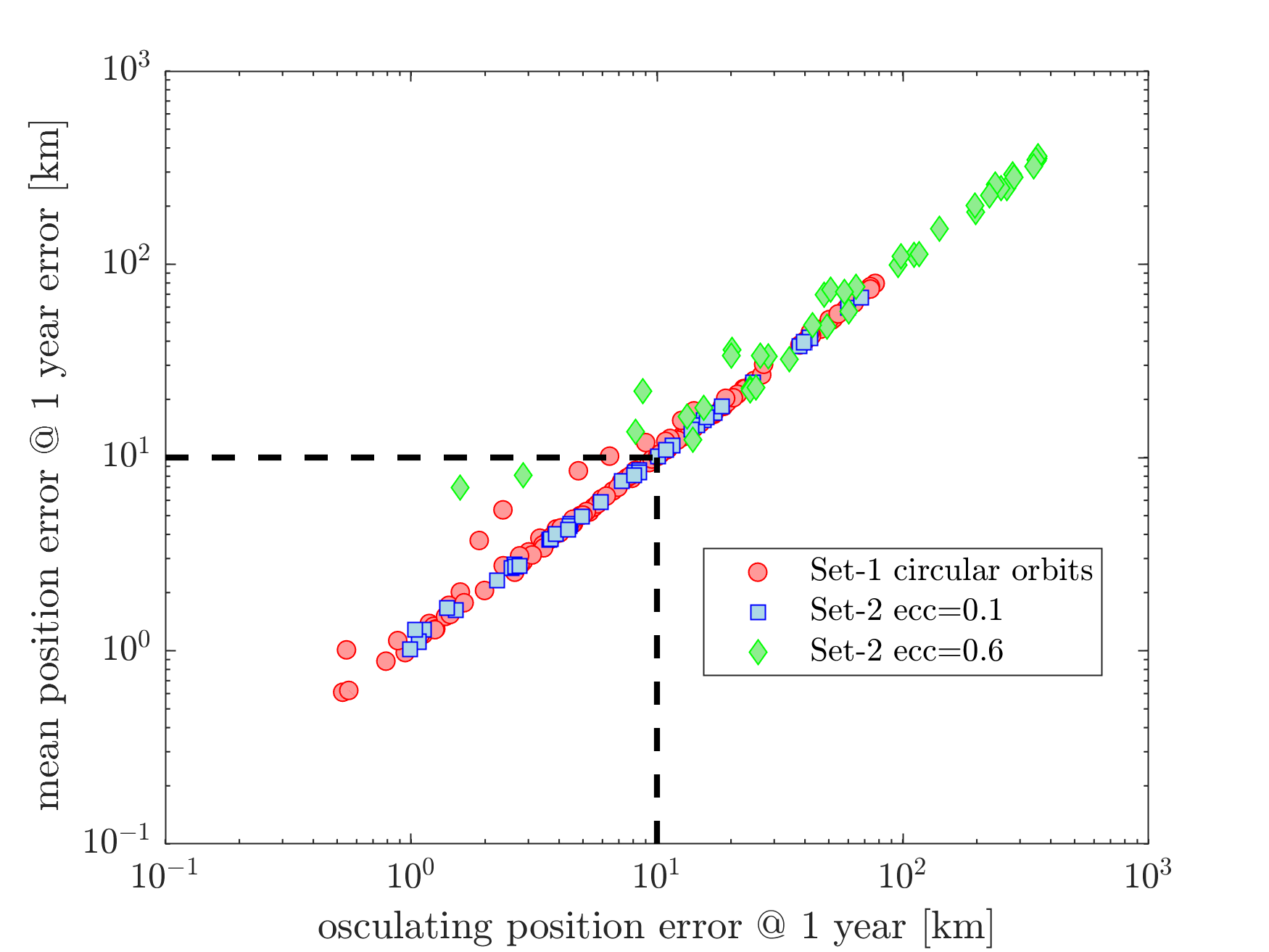}
\caption{\small Comparison of the distance error $D$ after one year as computed with the full SELENA procedure (abscissa) or with the SELENA-Mean procedure (ordinate), for the entire set of trajectories in the test datasets 1 and 2.}
\label{fig:errorcomp}
\end{figure}

The top curves in Fig.\ref{fig:exectime} show the total physical (wall-clock) computation time $T_C$ in seconds required for a propagation of the mean elements for each group of terms in the averaged equations of motion, for a total orbital time of $t=30~days$. All benchmarking tests are performed with the same, AUTh-based, computer unit. The physical computation time $T_C$ is given as a function of the order $n$ of the lunar potential harmonics included in the averaged equations of motion. As expected,  $T_C$ grows with $n$, as the number of internal terms produced in the averaged equations of motion grows, roughly as $\sim n^3$. The lower curves in the same figure show, now, the physical (wall-clock) time $T_{SPC}$ required to perform \textit{a single SPC transformation} of the type mean $\rightarrow$ osculating element along the same integration. We observe that the ratio $T_C/T_{SPC}$ remains nearly constant with $n$, $T_C/T_{SPC}\approx 10$. This implies that a sampling of the SELENA-produced time series of \textit{osculating} elements by (one datum)/(3 days) renders the computational time $T_{SPC}$ for computing the short-periodic corrections about equal to the total time $T_C$ taken to compute the averaged equations of motion. From this sampling frequency on, the time needed to compute the SCPs dominates over the time taken to integrate the averaged equations of motion, by a factor 3 at the sampling rate $1~datum/day$, and by a factor $\sim 50$ at the sampling rate $1~datum/(orbital period)$ of the trajectories. \\
\\
Yet, as shown in Fig.\ref{fig:errorset1mean}, such an extra cost can be avoided without consequences in the accuracy when the SELENA-Mean propagation procedure is selected. In fact, a comparison of the growth of the distance error $D$ among SELENA-Mean and full AUTh-Cartesian propagated trajectories shows that the time behavior of the error in the SELENA-Mean propagation is essentially identical to the one of the full SELENA procedure (cf. Figs.\ref{fig:errorset1mean} and \ref{fig:errorset1}). In fact, notwithstanding the dramatic gain of speed by the SELENA-Mean propagation over the full SELENA propagation, as shown in Fig.\ref{fig:errorcomp} the distance error $D$ after one year is practically identical for the two procedures over the entire set of trajectories included in the testing datasets 1 and 2. 

This behavior, which highlights the benefits of using an integrator based on the averaged equations of motion even when frequently-sampled osculating element time series of the trajectories are sought for, is based on the fact that the only \textit{systematic}, i.e. time-growing error in the trajectories is essentially the one due to the wrong evaluation of the \textit{fast} frequency $\dot{M}$, as analysed in subsection \ref{ssec:selena}. As discussed, the error in the evaluation of $\dot{M}$ is affected by two factors: i) the order of the averaged theory, and ii) the precision by which the \textit{initial} value of the mean semi-major axis is computed. Thus, the crucial SPC computation to be performed is the one concerning the passing from osculating to mean elements at the beginning of the integration process. All other corrections introduced by the SPCs at later times are, in fact, of first order in the perturbations included in the averaged theory. However, these corrections are oscillatory and introduce no systematic linear trend in the growth of the error with time. Of course, at any point of the mean time series, one can reconstruct the appropriate osculating 'initial condition' and switch to computing osculating data, for the time length desired. 

\clearpage


\clearpage
\section{Appendix}
\label{sec:appendix}

\subsection{List of SELENA deliverables and short description of the folders D1 to D5}
\label{ssec:seldeliv}
\begin{table}[!hb]
\centering
\begin{tabular}{|c|p{0.7\textwidth}|}
\hline
\textit{file/folder name} & \textit{description} \\
\hline\hline
\multicolumn{2}{|c|}{ \bf{Folder D1}}  \\
\hline\hline
Midterm report.pdf & The SELENA midterm report. \\
\hline
Final report.pdf & The SELENA final report. \\
\hline
finale\_presentation.pdf & The slides of the final presentation given to CNES. \\
\hline\hline
\multicolumn{2}{|c|}{ \bf{Folder D2}}  \\
\hline\hline
api\_examples & Sample python scripts that to allow to run single/multiple propagations with SELENA. \\
\hline
SELENA-propagator & The main code directory. Contains Python/C++ source code of SELENA. \\
\hline
SELENA-Tests-Data & The data of the SELENA verification campaign.  \\
\hline\hline
\multicolumn{2}{|c|}{ \bf{Folder D3}}  \\
\hline\hline
D3.pdf & Code documentation report. Contains the software architecture and the description of the functions used in the C++ and Python implementation.   \\
\hline\hline
\multicolumn{2}{|c|}{ \bf{Folder D4}}  \\
\hline\hline
D4.pdf & User's guide. Contains information to guide the user through installation and basic usage of SELENA. \\
\hline\hline
\multicolumn{2}{|c|}{ \bf{Folder D5}}  \\
\hline\hline
Mean-Theory & Contains the all the derived theory produced from the above Mathematica notebooks. The terms are
organized in ASCII files with the same exponent ordering as the one produced by Mathematica.  \\
\hline
Notebooks-Theory & Contains the Mathematica notebooks used to implement the analytical theory behind the 
SELENA propagator. They are organized by functionality (1st/2nd-order transformations, relegation) and force component 
(internal = Moon's potential, external = Earth/Sun gravity, SRP = solar radiation pressure). \\
\hline
\end{tabular}
\caption{\small Short description of deliverables: folders D1 to D5}
\end{table}

\clearpage

\begin{table}[!hb]
\centering
\begin{tabular}{|l|p{0.5\textwidth}|}
\hline
\textit{file name} & \textit{description} \\
\hline\hline
closedformnodulesfinalelpar4.nb & Collection of auxilary routines used in the theory computations. \\
\hline
external-P2-firstorder.nb & Derives the theory up to first order for the second order multipolar expansion of the third body perturbations. \\
\hline
external-P2-secondorder-relegation.nb & Derives the theory up to second order for the second order multipolar expansion of the third body perturbations. Computes also the relegation transformation. \\
\hline
external-P3-firstorder.nb & Derives the theory up to first order for the third order multipolar expansion of the third body perturbations. \\
\hline
internal-firstorder.nb & Derives the first order theory of the lunar gravitational attraction perturbation. \\
\hline
internal-relegation.nb &  Derives the relegation terms for the tesseral harmonics of the lunar gravitational attraction perturbation. \\
\hline
internal-secondorder.nb & Derives the second order theory of the lunar gravitational attraction perturbation.\\
\hline
srp.nb & Derives the first order theory for the solar radiation pressure perturbation.\\
\hline
\end{tabular}
\caption{\small Short description of the notebooks containing all the symbolic computations of  the averaged theory employed in the SELENA propagator}
\end{table}

~~~
\clearpage 

\begin{figure}
\centering
\includegraphics[width=0.4\textwidth]{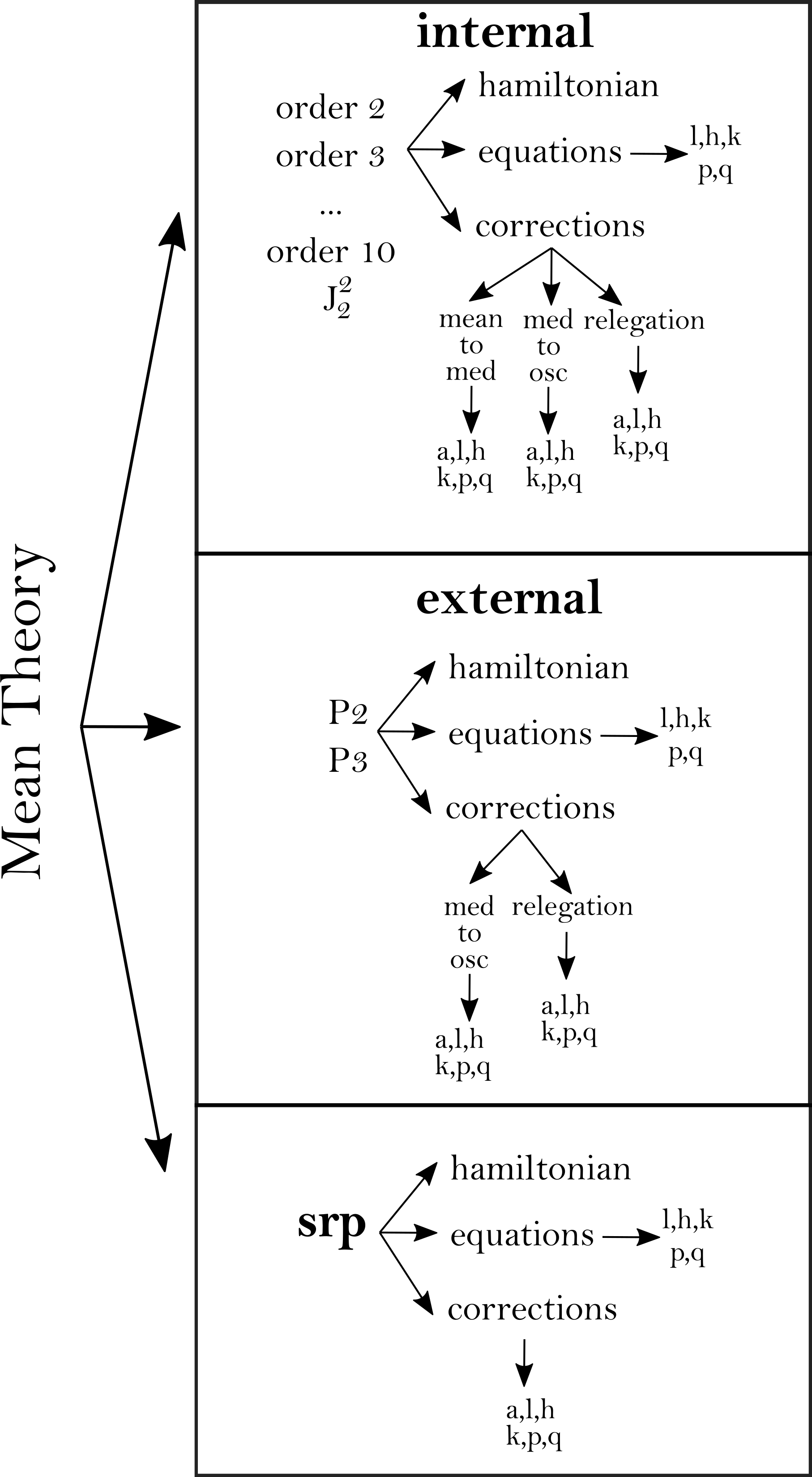}
\caption{Structure of the D5/Mean-Theory folder}
\end{figure}

~~~
\clearpage

\subsection{GRAIL Constants and $10\times 10$ Lunar Potential Coefficients}
\label{ssec:grailcoef}
\noindent
Equatorial radius of the Moon: $R_\Moon=1738.0~km$.\\
\\
Gravitational constant of the Moon: ${\cal G}M_L=0.490280012616\times~10^4 km^3/sec^2$\\
\\
\begin{table}[h]
\centering\small
\begin{tabular}{|c|c|c|c|} 
\hline
~~~n~~~ &~~~m~~~ 
&~~~~~~~~~~~\mbox{Coefficient $C_{nm}$}~~~~~~~~~~~ 
&~~~~~~~~~~~\mbox{Coefficient $S_{nm}$}~~~~~~~~~~~\\
\hline\hline
    1 & 0 &   0.0000000000000000E+00 &   0.0000000000000000E+00       \\    
\hline
    1 & 1 &   0.0000000000000000E+00 &   0.0000000000000000E+00       \\   
\hline
    2 & 0 &  -0.9087974694316000E-04 &   0.0000000000000000E+00       \\   
\hline
    2 & 1 &   0.4804858187034000E-11 &   0.8319976179256000E-09       \\   
\hline
    2 & 2 &   0.3467157070685000E-04 &  -0.2080905319450000E-09       \\   
\hline
    3 & 0 &  -0.3197483172669000E-05 &   0.0000000000000000E+00       \\     
\hline
    3 & 1 &   0.2636797758064000E-04 &   0.5454528124967000E-05       \\     
\hline
    3 & 2 &   0.1417154761018000E-04 &   0.4877983761921000E-05       \\     
\hline
    3 & 3 &   0.1227503842700000E-04 &  -0.1774392494042000E-05       \\     
\hline
    4 & 0 &   0.3234787544589000E-05 &   0.0000000000000000E+00       \\     
\hline
    4 & 1 &  -0.6013462538980000E-05 &   0.1664327383234000E-05       \\     
\hline
    4 & 2 &  -0.7116173124950000E-05 &  -0.6777040203801000E-05       \\     
\hline
    4 & 3 &  -0.1350004578974000E-05 &  -0.1344499562042000E-04       \\     
\hline
    4 & 4 &  -0.6007218421231000E-05 &   0.3926844078609000E-05       \\     
\hline
    5 & 0 &  -0.2237895028963000E-06 &   0.0000000000000000E+00       \\     
\hline
    5 & 1 &  -0.1011612026117000E-05 &  -0.4118918590128000E-05       \\     
\hline
    5 & 2 &   0.4399527730258000E-05 &   0.1057126381069000E-05       \\     
\hline
    5 & 3 &   0.4661799530086000E-06 &   0.8698891186531000E-05       \\     
\hline
    5 & 4 &   0.2754160543604000E-05 &   0.6762927601539000E-07       \\     
\hline
    5 & 5 &   0.3110802396520000E-05 &  -0.2754562706853000E-05       \\     
\hline
    6 & 0 &   0.3818429731721000E-05 &   0.0000000000000000E+00       \\     
\hline
    6 & 1 &   0.1528269259667000E-05 &  -0.2599593504407000E-05       \\     
\hline
    6 & 2 &  -0.4397305532799000E-05 &  -0.2167691353650000E-05       \\     
\hline
    6 & 3 &  -0.3317543579200000E-05 &  -0.3427424034693000E-05       \\     
\hline
    6 & 4 &   0.3411544820608000E-06 &  -0.4057987871277000E-05       \\     
\hline
    6 & 5 &   0.1454377510090000E-05 &  -0.1034179409900000E-04       \\     
\hline
    6 & 6 &  -0.4684302102779000E-05 &   0.7229676026112000E-05       \\     
\hline
\end{tabular}   
\end{table}

\begin{table}
\centering\small
\begin{tabular}{|c|c|c|c|} 
\hline
~~~n~~~ &~~~m~~~ 
&~~~~~~~~~~~\mbox{Coefficient $C_{nm}$}~~~~~~~~~~~ 
&~~~~~~~~~~~\mbox{Coefficient $S_{nm}$}~~~~~~~~~~~\\
\hline\hline
    7 & 0 &   0.5593395521180000E-05 &   0.0000000000000000E+00       \\     
\hline
    7 & 1 &   0.7471679448922000E-05 &  -0.1197372248415000E-06       \\     
\hline
    7 & 2 &  -0.6501386277161000E-06 &   0.2411114548482000E-05       \\     
\hline
    7 & 3 &   0.5994270540752000E-06 &   0.2357326252726000E-05       \\     
\hline
    7 & 4 &  -0.8437053566645000E-06 &   0.7565179395171000E-06       \\     
\hline
    7 & 5 &  -0.2068114534889000E-06 &   0.1069299864304000E-05       \\     
\hline
    7 & 6 &  -0.1065341843530000E-05 &   0.1100465135969000E-05       \\ 
\hline
    7 & 7 &  -0.1820298178535000E-05 &  -0.1600040313298000E-05       \\     
\hline
    8 & 0 &   0.2346831025052000E-05 &   0.0000000000000000E+00       \\     
\hline
    8 & 1 &   0.4172552758326000E-08 &   0.1098040161291000E-05       \\     
\hline
    8 & 2 &   0.3009317424280000E-05 &   0.1930559600363000E-05       \\     
\hline
    8 & 3 &  -0.1889046873321000E-05 &   0.9544856234862001E-06       \\     
\hline
    8 & 4 &   0.3408665044948000E-05 &  -0.5282366567240999E-06       \\     
\hline
    8 & 5 &  -0.1248041915031000E-05 &   0.2918563973395000E-05       \\     
\hline
    8 & 6 &  -0.1660485526106000E-05 &  -0.2114745802343000E-05       \\     
\hline
    8 & 7 &  -0.1509638443524000E-05 &   0.3268877315657000E-05       \\     
\hline
    8 & 8 &  -0.2485687875411000E-05 &   0.2116381872934000E-05       \\     
\hline
    9 & 0 &  -0.3530908226176000E-05 &   0.0000000000000000E+00       \\     
\hline
    9 & 1 &   0.1866966711820000E-05 &   0.8103920415300000E-07       \\     
\hline
    9 & 2 &   0.1927806207769000E-05 &  -0.1387568154334000E-05       \\     
\hline
    9 & 3 &  -0.1992407183919000E-05 &   0.2201757910190000E-05       \\     
\hline
    9 & 4 &  -0.1884446627709000E-05 &  -0.1425791526555000E-05       \\     
\hline
    9 & 5 &  -0.1562514670072000E-05 &  -0.3524685176376000E-05       \\     
\hline
    9 & 6 &  -0.2127263414440000E-05 &  -0.3002637322559000E-05       \\     
\hline
    9 & 7 &  -0.3914777590918000E-05 &  -0.1068148458108000E-06       \\     
\hline
    9 & 8 &  -0.1312026278751000E-05 &  -0.2203374191835000E-05       \\     
\hline
    9 & 9 &  -0.9384258636469000E-06 &   0.2488174834497000E-05       \\     
\hline
   10 & 0 &  -0.1069297690343000E-05 &   0.0000000000000000E+00       \\     
\hline
   10 & 1 &   0.8415984243707000E-06 &  -0.9540746227385000E-06       \\     
\hline
   10 & 2 &   0.3572480338553000E-06 &  -0.2651167020141000E-06       \\     
\hline
   10 & 3 &   0.4841979498210000E-06 &   0.6688279867555000E-06       \\     
\hline
   10 & 4 &  -0.3572977017844000E-05 &   0.1578945238596000E-05       \\     
\hline
   10 & 5 &   0.6996909768799000E-06 &  -0.3145796496684000E-06       \\     
\hline
   10 & 6 &  -0.1273180601437000E-06 &  -0.2095335437909000E-05       \\     
\hline
   10 & 7 &  -0.3998641877237000E-05 &  -0.9107313784394000E-06       \\     
\hline
   10 & 8 &  -0.3559447594232000E-05 &   0.2848851002686000E-05       \\     
\hline
   10 & 9 &  -0.4753134677431000E-05 &  -0.5153017437100000E-07       \\     
\hline
   10 & 10 &  0.9478676858446000E-06 &  -0.1719399182547000E-05      \\
\hline
\end{tabular}   
\end{table}

\end{document}